\documentclass[11pt]{article}
\pdfoutput=1
\usepackage{amsmath,amssymb,amsfonts}
\usepackage{graphicx}
\usepackage{changepage}
\usepackage{subfig}
\usepackage[usenames, dvipsnames]{color}
\usepackage[hidelinks]{hyperref}
\usepackage[dvipsnames]{xcolor}
\usepackage{cite}
\usepackage{multirow}
\usepackage{slashed}
\usepackage{verbatim} 
\usepackage{enumitem}
\usepackage{rotating}
\usepackage{xcolor}
\usepackage{multirow}
\usepackage{bm}
\usepackage[normalem]{ulem}
\usepackage{cleveref}
\usepackage{booktabs} 
\usepackage{tikz} 
\usepackage{float}
\usepackage{cancel}
\usepackage{xcolor}
\usepackage{relsize}
\usepackage{physics}
\usepackage{cleveref}

\usepackage{geometry,mathtools,pdflscape,lscape,rotating,empheq,mathrsfs,longtable}

\graphicspath{ {figures/} }

\usepackage{tikz}
\usepackage{diagbox}
\usetikzlibrary{positioning}
\usetikzlibrary{calc}
\usetikzlibrary{decorations.pathreplacing,calligraphy}

\usepackage{xstring}
\usetikzlibrary{decorations.pathmorphing} 
\usetikzlibrary{decorations.markings} 
\usetikzlibrary{arrows} 
\usetikzlibrary{shapes} 
\usetikzlibrary{matrix} 
\usetikzlibrary{positioning} 
\usepackage[english]{babel} 
\usepackage[autostyle]{csquotes}
\usepackage{pifont}
\usetikzlibrary{shapes.multipart}

\tikzset{->-/.style={decoration={
  markings,
  mark=at position .5 with {\arrow{>}}},postaction={decorate}}}

\tikzstyle{none} = []

\tikzstyle{cyan arrow}=[<->, solid, draw= black, dashed, line width=0.5mm]
\tikzstyle{orange arrow}=[<->, solid, draw=cyan, line width=0.5mm]
\tikzstyle{red arrow}=[<->, solid, draw=blue, line width=0.5mm]
\tikzstyle{blue arrow}=[->, solid, draw= black, line width=0.5mm]

\newcommand{\bea}{\begin{eqnarray}}
\newcommand{\eea}{\end{eqnarray}}
\newcommand{\be}{\begin{equation}}
\newcommand{\ee}{\end{equation}}
\newcommand{\ba}{\begin{aligned}}
\newcommand{\ea}{\end{aligned}}
\newcommand{\bit}{\begin{itemize}}
\newcommand{\eit}{\end{itemize}}
\newcommand{\ben}{\begin{enumerate}}
\newcommand{\een}{\end{enumerate}}
\newcommand{\nn}{\nonumber}

\newcommand{\Bsym}{\mathfrak{B}^{\text{sym}}}

\newcommand{\lb}{\left( }
\newcommand{\rb}{\right) }

\newcommand\Z{{\mathbb{Z}}}
\newcommand{\R}{{\mathbb{R}}}

\newcommand{\Spin}{\text{Spin}}
\newcommand{\cA}{\mathcal{A}}

\newcommand{\cC}{\mathcal{C}}
\newcommand{\cD}{\mathcal{D}}

\newcommand{\cL}{\mathcal{L}}

\newcommand{\cN}{\mathcal{N}}

\newcommand{\cT}{\mathcal{T}}

\newcommand{\cX}{\mathcal{X}}
\newcommand{\cY}{\mathcal{Y}}
\newcommand{\cZ}{\mathcal{Z}}

\newcommand{\mf}{\mathfrak}
\newcommand{\fT}{\mathfrak{T}}
\newcommand{\fB}{\mathfrak{B}}

\newcommand{\so}{\mathfrak{so}}

\newcommand{\up}[1]{^{({#1})}}
\newcommand{\dw}[1]{_{({#1})}}
\newcommand{\id}{\text{id}}

\newcommand{\LD}{\text{LD}}
\newcommand{\PD}{\text{PD}}
\newcommand{\de}{\partial}
\newcommand{\eps}{\epsilon}
\newcommand{\ol}{\overline}
\newcommand{\gs}{\geqslant}
\newcommand{\chiM}[1]{\chi_{(M_6{#1},\,\Z_2)}}
\newcommand{\chiiM}[1]{\chi_{(M_6{#1},\,\Z_2)}^{-1}}
\newcommand{\tpdf}{\texorpdfstring}
\newcommand{\gz}{^{\geqslant 0}}
\newcommand{\gep}{^{\geqslant \eps}}
\newcommand{\ze}{^{[0,\eps]}}
\newcommand{\GS}{\text{GS}}

\newcommand{\bD}{\mathbb{D}}

\renewcommand{\mod}{\text{mod}\;}
\usepackage{etoolbox}
\makeatletter
\patchcmd{\@sect}{#8}{\boldmath #8}{}{}
\let\ori@chapter\@chapter
\def\@chapter[#1]#2{\ori@chapter[\boldmath#1]{\boldmath#2}}
\makeatother

\begin{document}

\baselineskip=18pt  
\numberwithin{equation}{section}  

\thispagestyle{empty}

\vspace*{0.8cm} 
\begin{center}
{\Huge Non-Invertible Symmetries in 6d from \\
\smallskip
\smallskip
Green-Schwarz Automorphisms
}

\vspace*{0.8cm}
Fabio Apruzzi$^\star$, Sakura Sch\"afer-Nameki$^*$, Alison Warman$^*$\\

\vspace*{.4cm} 

{\it $^\star$ Dipartimento di Fisica e Astronomia “Galileo Galilei”, Università di Padova,\\ Via Marzolo 8, 35131 Padova, Italy  }\\

{\it $^\star$ INFN, Sezione di Padova, Via Marzolo 8, 35131 Padova, Italy   }\\

{\it ${}^{*}$ Mathematical Institute, University of Oxford, \\
Andrew Wiles Building,  Woodstock Road, Oxford, OX2 6GG, UK}

\vspace*{1.5cm}

\begin{adjustwidth}{1cm}{1cm}
\noindent
We construct non-invertible symmetries in 6d $\mathcal{N}=(2,0)$ superconformal field theories arising from Green-Schwarz (GS) automorphisms, which form abelian or non-abelian groups. Specifically, $\mathbb{Z}_2$, $\mathbb{Z}_3$ and $S_3$ GS automorphisms give rise to 
non-invertible duality, triality and $S_3$-ality defects, respectively, once combined with stacking symmetry protected topological phases (SPTs) and gauging 2-form symmetries. We derive the defects and their fusion rules from two distinct perspectives: from half-space gauging as well as from the Symmetry Topological Field Theory (SymTFT).
This is the first concrete construction of such symmetry defects in 6d, that are codimension 1, and have non-invertible and non-abelian fusion rules. This structure forms a central part of the underlying fusion 5-category.

\end{adjustwidth}

 \end{center}

\newpage

\tableofcontents

\section{Introduction}

One of the most surprising insights in the past few years is the existence of a new type of symmetry, called non-invertible or higher-categorical symmetries \cite{Kaidi:2021xfk, Choi:2021kmx, Bhardwaj:2022yxj} (for reviews of non-invertible symmetries see  \cite{Schafer-Nameki:2023jdn, Shao:2023gho}). As the name indicates, these symmetries may not admit inverses, and are therefore strictly beyond the realm of standard symmetries that form groups. 
One of the constructions of non-invertible or categorical symmetries in higher dimensional quantum field theories (QFTs) is inspired by the Kramers-Wannier (KW) duality symmetry in the 2d critical Ising model \cite{Frohlich:2004ef}. 
Namely, a theory $\fT$ which has a duality to a theory $\fT^\vee$, such that another topological operation, such as stacking a TQFT and/or gauging maps back to $\fT$, will admit a  non-invertible symmetry. 
In the 2d case this is the map $g\to 1/g$,  in the transverse field Ising model 
\be
H_{\text{Ising}} = -\sum_{j} \sigma^z_j \sigma^z_{j+1}- g \sum_j \sigma^x_j \,,
\ee
which becomes a symmetry at $g=1$, i.e. the critical Ising conformal field theory (CFT).
In $d$ spacetime dimensions we generically expect a fusion $(d-1)$-category as the most general (internal, finite) symmetry structure. 
In even $d=2n$ we expect that a generalization of the KW dualities exists, which is obtained by combining dualities and gauging of $(d/2-1)$-form symmetries. Indeed, examples are the non-invertible KW defects in 4d \cite{Choi:2021kmx, Choi:2022zal,Kaidi:2021xfk,Kaidi:2022cpf}. This construction works  in any even dimension, and generalizations of such duality defects were subsequently studied in various contexts: 
in 2d, beyond the KW duality defects there are generalizations to triality  and  $G$-ality defects for a finite group $G$ \cite{Lu:2022ver, Lu:2024lzf}. 
In 4d, using the Symmetry Topological Field Theory  (SymTFT) \cite{Gaiotto:2020iye, Apruzzi:2021nmk, Freed:2022qnc} or holography in \cite{Kaidi:2022cpf, Antinucci:2022vyk}, from  branes and string compactifications \cite{Apruzzi:2023uma, Apruzzi:2022rei, Heckman:2022xgu} and in 
class S theories \cite{Bashmakov:2022uek, Antinucci:2022cdi}.   In particular the latter have the feature that the duality symmetry can form a non-abelian finite group. Using a more mathematical approach the possibilities of duality defects and generalizations thereof were studied in the context of fusion 3-categories in \cite{Bhardwaj:2024xcx}.

{The relevant Green-Schwarz automorphisms are automorphisms of the BPS string charge lattice, whose role in 6d SCFTs was developed in \cite{Apruzzi:2017iqe}.} Finally in 6d $\mathcal{N}=(2,0)$  theories it was suggested that using Green-Schwarz dualities {combined with stacking TQFTs and/or gauging 2-form symmetries} one can construct non-invertible duality and triality defects in \cite{Lawrie:2023tdz}. The purpose of the present paper is to concretely construct the full $S_3$-ality defects and crucially to compute their fusion. Achieving this will  act as a proper starting point for the  exploration of these fusion 5-categories. 

{Throughout this paper, when referring to a fusion 5-category we use this as shorthand for the higher-categorical symmetry structure expected in a 6d QFT.} {Our explicit results determine the codimension-one topological defects generated by the GS automorphisms, 2-form gauging and SPT stacking, together with their fusion rules. We do not attempt here to construct the full higher category, including all higher morphisms and associated coherence data.}

More generally we expect that in any even dimension $d=2n$ such non-invertible defects will exist that are $G$-ality defects for some duality symmetry $G$. 
With the huge proliferation of non-invertible symmetries, a fair question is why these $G$-ality defects are particularly interesting. 
Non-invertible symmetries are broadly characterized by either being group-theoretical or not. {\bf Group-theoretical} ({also} referred to as ``non-intrinsically non-invertible") means that the non-invertible symmetry is related by gauging to an invertible, i.e. group-like, symmetry. This may be a $p$-form symmetry or a higher-group symmetry. A non-group-theoretical one is such that there is no gauging that maps it to an invertible symmetry. We should emphasize that whether or not a non-invertible 
symmetry is group-theoretical does not diminish or increase its appeal, or the non-invertible specific physical properties. In fact non-invertible symmetries of any type have interesting physical implications (e.g. predict new phases, new order parameters). 

Nevertheless it is interesting -- and challenging in general -- to construct {\bf non-group-theoretical} non-invertible symmetries. From general considerations, such (non-group-theoretical) duality defects will occur only when the spacetime dimension $d$ is even. On the other hand, a reasonable conjecture is that in $d=(2n+1)$ generalized symmetries are  (mostly) related by gauging to invertible symmetries -- a fact that seems to be true in 3d bosonic symmetries \cite{decoppet2022drinfeld, TianWen, LanKongWen}. With duality defects very well known in 4d and 2d, it is interesting in light of the existence of non-trivial 6d QFTs (and strongly-coupled UV fixed points in particular) to determine non-group-theoretical non-invertible symmetries in 6d. 

Concretely we will use two approaches to construct these non-invertible symmetries: \vspace{-1mm}
\begin{itemize}
\item Half-space gauging, \vspace{-1mm}
\item Gauging an outer automorphism of the SymTFT.
\end{itemize}
The first approach was first introduced for duality and their $\Z_3$ generalization, triality, defects and appeared in \cite{Choi:2022zal}, which constructs the defects and fusion: pictorially we start with a half-space gauging interface (for a $d$-dimensional theory, this can be the gauging of a $(d/2-1)$-form symmetry) of codimension 1 
\be 
     \begin{split}
    &\;{\color{red}\cD(M_{d-1})}\\
    \fT|_{M_d^{<0}} \qquad &{\color{red} \quad\bigg|} \qquad  (\fT / G)|_{M_d^{\gs 0}}
\end{split}
\ee 
On the left hand side, in the half-space $M_d^{<0}$ we have the original theory $\fT$, on the right hand side, after the insertion of a half-space gauging defect the theory $\fT/G$. Combining this with a duality transformation  that maps the theory $\fT/G$ back to $\fT$ we obtain a non-invertible {\it duality defect}. 
We will generalize this to triality and $S_3$-ality defects in turn.

Our construction is applicable to a large class of 6d theories, but the richest setting, which has the largest Green-Schwarz duality symmetry, is that of the 6d $\mathcal{N}=(2,0)$ $\mf{so}(8)$ SCFT. The $S_3$-ality defects result from (twisted) gauging of the 2-form symmetry and Green-Schwarz (GS) automorphisms. 

We should note that compactification of these theories on a two-torus gives rise to 4d $\mathcal{N}=4$ Super Yang-Mills, which only has duality or triality defects. The main reason for this is that the modulus $\tau$ of the torus becomes part of the action of the duality symmetry, and there is no locus where both triality and duality defects can coexist. This is in turn possible for class S constructions, which generically are non-Lagrangian and can have $G$-ality defects. 

We start in section \ref{sec:twix} with the discussion of the 6d  SCFTs and their Green-Schwarz dualities as well as the possible choices of polarizations and consequently, absolute theories. Then we construct the non-invertible defects in 6d from half-space gauging in section \ref{sec:half-space}, and from the SymTFT point of view in section \ref{sec:SymTFT}. 

A  summary of defects and their fusion for $S_3$-ality defects in 6d is given in subsection \ref{sec:Bounty}. Note that for a special case, when the group is $\Z_2$, i.e. duality defects, this is a generalization of the Tambara-Yamagami construction that is well-known in 2d and 4d, now to 6d.

\section{6d $\mathcal{N}=(2,0)$ Theories: Polarizations and GS Automorphisms}
\label{sec:twix}
6d $\mathcal{N}=(2,0)$ SCFTs of ADE type are specified by an ADE Lie algebra $\mf{g}$, whose weight lattice, $\Lambda$, describes string charges under a two-form symmetry \cite{Gaiotto:2014kfa,DelZotto:2015isa,Monnier:2017klz,Heckman:2017uxe}. For 6d $\mathcal{N}=(1,0)$ SCFTs the lattice of BPS string charges can be more general. A necessary condition for the lattice is that its Dirac pairing must be positive definite. These 6d SCFTs are in general relative theories \cite{Witten:2009at,Freed:2012bs},
coupled to a 7d bulk: to get an absolute theory with a well-defined partition function a polarization must be chosen: this ensures mutual locality (trivial Dirac pairing) of the operators in the absolute theory \cite{Witten:1998wy,Monnier:2017klz,Monnier:2016jlo,Gukov:2020btk,Lawrie:2023tdz,Witten:1996hc,Witten:2009at,Freed:2012bs}. Generalized symmetries in this context were studied in  \cite{Cordova:2020tij, Bhardwaj:2020phs, Apruzzi:2017iqe,Gukov:2020btk,Apruzzi:2021mlh, Lawrie:2023tdz}.

{Before choosing a polarization, 6d SCFTs are coupled to a 7d bulk anomaly theory, whose action is that of a 3-form Chern-Simons theory  \cite{Witten:1998wy,Monnier:2017klz,Heckman:2017uxe,Gukov:2020btk}
\begin{equation} \label{eq:7d_CS}
    S_{\text{CS}(K)}=\frac{1}{4\pi} \int_{M_7}\,\sum_{I,J}\,K_{IJ}\;c_3\up{I}\wedge\dd c_3\up{J}\ \,,\quad I,J\in\{1,...,r\}\,,
\end{equation}
where $c_3\up{I},$ for $I\in\{1,...,r\}$ are 3-form $U(1)$ gauge fields on $M_7$. The matrix $K$ is integral, positive definite and symmetric, and governs the braiding statistics for strings in the boundary 6d system. 
It can be obtained geometrically from an F-theory compactification on an elliptically fibered Calabi-Yau threefold as the intersection pairing for the base \cite{Heckman:2017uxe,Apruzzi:2017iqe}.
If the diagonal
entries of $K$ are not all even, the manifold $M_7$ requires a ``Wu$^c$ structure", as explained in appendix A of \cite{Gukov:2020btk}.
For 6d SCFTs with $\cN=(2,0)$ of ADE type, $K$ is the Cartan matrix of the ADE Lie algebra $\mf{g}$ \cite{Monnier:2017klz,Gukov:2020btk}. }

\subsection{Absolute Theories with GS Automorphisms}\label{sec:Heroes}
Green-Schwarz (GS) automorphisms in 6d SCFTs were analyzed in \cite{Apruzzi:2017iqe}. They are automorphisms, $\eta$, of the BPS string lattice, $\Lambda$,
\begin{equation}
  \eta\, : \Lambda \rightarrow \Lambda,  \quad \text{s.t.} \quad \eta^T \, K \, \eta = K.
\end{equation}
We recall that for the $\mathcal{N}=(2,0)$ case $K$ always corresponds to the Cartan matrix of $\mf{g}$, $\text{CM}_{\mathfrak{g}}$, whereas only for certain $(1,0)$ theories this is true. These automorphisms decompose into two parts, $\mathcal O(\text{GS}) \ltimes \mathcal W(\text{GS})$:
\begin{enumerate}
    \item The inner automorphisms, $\mathcal W(\text{GS})$, which for the case of $K=\text{CM}_{\mathfrak{g}}$, correspond to the Weyl group $\mathcal W_{\mathfrak{g}}$ are redundancies of the tensor branch description. In particular they imply identifications of the tensor branch moduli space into Weyl chambers,
    \begin{equation}
        \mathcal{M}_T = \mathbb{R}^{(5)T}/\mathcal W \,,
    \end{equation}
    where the $(5)$ stands for the $(2,0)$ case, since every $(2,0)$ tensor multiplet has 5 scalars. 
    \item The outer automorphisms, $\mathcal{O}(\text{GS})$, instead, provide self-dualities\footnote{We use the word duality even though we will soon discover that this is really a $G$-ality.} of the 6d (relative) theory. We also recall that 6d SCFTs do not have marginal deformations or a conformal manifold \cite{Cordova:2016xhm, Apruzzi:2013yva, Louis:2015mka}. Therefore the self-duality will not change any coupling constant. This could be enough to declare that the self-duality is a symmetry, though 6d theories are generically relative theories. 
    Indeed we will see that when the 6d theory admits polarizations that lead to absolute theories, these dualities will in fact map one absolute theory to another. In order to clarify the nomenclature, $\mathcal{O}(\text{GS})$ are self-$G$-alities of the relative theory, and $G$-alities of absolute theories when such polarizations exist. 
\end{enumerate}
We recall that the GS automorphisms are read off from the low-energy tensor branch description where the Dirac pairing appears in the coupling among the various tensor multiplets. Moreover, in the $(1,0)$ case $K$ appears also in coupling tensors to vector multiplets. The low-energy tensor branch description of 6d $(1,0)$ theories consists of a quiver gauge theory coupled to tensor multiplets. The automorphism is preserved when it leaves the quiver structure unchanged.

We discuss the types of  $G$-alities that are possible in 6d SCFTs: in \cite{Apruzzi:2017iqe} a  strategy was implemented to  classify automorphisms based on the F-theory classification of 6d SCFTs \cite{Heckman:2013pva, Heckman:2015bfa, DelZotto:2014hpa}. We skip the details of the classification and refer to \cite{Apruzzi:2017iqe}. Non-trivial GS outer automorphisms $\mathcal O(\text{GS})$  fall into two main classes:
\begin{enumerate}
    \item $G=\mathbb{Z}_2$, i.e. duality, these are either reflection of the Dirac pairing $K$, or exchange of the outer automorphisms of the $K=C_{\mathfrak{so}(2N)}$ (with $N>4$) and $K=C_{\mathfrak{e}_6}$.
    \item $G=S_3$, i.e. $S_3$-ality, these are the automorphisms of $K=C_{\mathfrak{so}(8)}$.
\end{enumerate}

In this paper we will in particular construct the $S_3$-ality non-invertible symmetries for 6d SCFTs which have $K=C_{\mathfrak{so}(8)}$. This example admits polarizations that lead to absolute theories. In addition, the outer automorphism group provides the most general $G$-ality that can appear, therefore the defects that we construct are quite general and their construction can be directly applied to all the other examples that have $G$-ality and polarizations leading to absolute theories. For example $S_3$-ality will contain duality defects. They will also appear in 6d theories of $A_{N-1}$-type with $N=n^2$ where $n \in \mathbb N$.

\subsection{6d \tpdf{$\mathcal{N}=(2,0)$}{N2} \tpdf{$\mathfrak{so}(8)$}{so8} Theory}

A particularly interesting case is the 6d $\mathcal{N}=(2,0)$ SCFT associated to the Lie algebra $\mathfrak{so}(8)$, as its  automorphism group is the non-abelian group of permutations on three elements: $S_3$. In particular we will show that this implies that there are  non-invertible $S_3$-ality defects. These are known to exist in 2d QFTs, but so far not in higher dimensions (though in theory certain non-Lagrangian theories may have such $S_3$-ality defects in 4d, there are no completely explicit constructions including fusions known to us). 

\subsubsection{Defect Group and Polarizations} 

The defect group for the 6d $\mathcal{N}=(2,0)$ $\mf{so}(8)$ relative theory is \cite{DelZotto:2015isa,Apruzzi:2017iqe,Gukov:2020btk,Lawrie:2023tdz}
\begin{equation} \label{eq:D_L_Lbar}
    \bD=\Z_2\oplus\Z_2 \,.
\end{equation}
This is easily derived from the Cartan matrix of $\mathfrak{so}(8)$, which identifies the Dirac pairing, and computing the Smith normal form thereof. 
We can derive three polarizations $L$, which satisfy $\mathbb{D}= L \oplus \overline{L}$.
 $L$ can be chosen, one for each of the three $\Z_2$ subgroups of the defect group \eqref{eq:D_L_Lbar}, which we label by their generators:
\begin{align}
    &\text{Polarization:} &&L_S =(1,0)  && L_C =(0,1) && L_V =(1,1)\\
    &\text{Absolute theory:} && Ss(8) && Sc(8) && SO(8)
\end{align}
However, as stressed in \cite{Gukov:2020btk,Lawrie:2023tdz}, fixing a subgroup $L$ of $\bD$ in equation \eqref{eq:D_L_Lbar} is not sufficient to fully specify an absolute theory: one ought to also specify $\ol{L}$. This can also be seen by similar reasoning to the 4d discussion of \cite{Aharony:2013hda}, with the difference that in 6d the Dirac pairing is symmetric and we must specify the weights not of lines but of 2d operators. This implies the existence of six absolute 6d $\mathcal{N}=(2,0)$ $\mf{so}(8)$ theories: 
\begin{align} \label{eq:abs_theories}
    Ss(8)_{+}\;&=\lb L_{S},\,\ol{L}_{C}\rb ,  
    &Sc(8)_{+}\;&=\lb L_{C},\,\ol{L}_{S}\rb , 
    & SO(8)_{+}\;&=\lb L_{V},\,\ol{L}_{S}\rb ,  \nn \\[2mm]
    Ss(8)_{-}&=\lb L_{S},\,\ol{L}_{V}\rb , 
    &Sc(8)_{-}&=\lb L_{C},\,\ol{L}_{V}\rb ,  
    &SO(8)_{-}&= \lb L_{V},\,\ol{L}_{C}\rb  \,.
\end{align}
These six theories are related by a web of topological manipulations (gauging, possibly after stacking with an SPT). Furthermore, they are related by Green-Schwarz dualities. 
This combination of facts allows us to construct non-invertible $G$-ality defects in these theories. 

\subsubsection{Green-Schwarz Automorphisms and Dualities}

A key ingredient for the construction of non-invertible defects is the existence of dualities. In this case we will see that there are dualities mapping between different polarizations. These dualities have closed orbits, and combined form a group $S_3 = \Z_3 \rtimes \Z_2$. 

Let us consider an absolute theory with polarization pair $\lb L_J,\,\ol{L}_I\rb,\;I\neq J\in\{S,C,V\} $, defined on the 6-dimensional spacetime $M_6$, coupled to classical background fields for the global 2-form  symmetry $\Z_2\up{I}$: 
\begin{equation}
    C_3\up{I}\in H^3(M_6,\,\Z_2)\,.
\end{equation}
We denote the partition function by $Z_{\fT_J}[C_3\up{I}]$, where the superscript $I$ denotes the global 2-form symmetry whereas the subscript $J$ specifies the polarization, which, as we will recall shortly, corresponds to the dual 2-form symmetry $\Z_2\up{J}$ one obtains after gauging $\Z_2\up{I}$. 

\begin{figure}
\centering
\begin{tikzpicture}[scale=3]
   \node [style=none, align=center] (6) at (-0.866*2, +1/2*2) {$SO(8)_{+}$\\[1mm] \footnotesize{$\lb L_{V},\,\ol{L}_{S}\rb$}};
   \node [style=none, align=center] (1) at (-0.866, +1/2) {$SO(8)_{-}$\\[1mm] \footnotesize{$\lb L_{V},\,\ol{L}_{C}\rb$}};
   \node [style=none, align=center] (4) at (+0.866*2, +1/2*2) {$Ss(8)_{+}$\\[1mm] \footnotesize{$\lb L_{S},\,\ol{L}_{C}\rb$}};
   \node [style=none, align=center] (3) at (0, -0.8) {$Ss(8)_{-}$\\[1mm] \footnotesize{$\lb L_{S},\,\ol{L}_{V}\rb$}};
   \node [style=none, align=center] (5) at (0, -0.8*2) {$Sc(8)_{-}$\\[1mm] \footnotesize{$\lb L_{C},\,\ol{L}_{V}\rb$}};
   \node [style=none, align=center] (2) at (+0.866, +1/2) {$Sc(8)_{+}$\\[1mm] \footnotesize{$\lb L_{C},\,\ol{L}_{S}\rb$}};
    \draw [style= blue arrow] (1) [bend left] to (2);
\draw [style=blue arrow] (2) [bend left] to (3);
    \draw [style=blue arrow] (3) [bend left] to (1);
    \draw [style=blue arrow] (5)  [bend right]to (4);
    \draw [style=blue arrow] (6) [bend right] to (5);
    \draw [style=blue arrow] (4) [bend right] to (6);

    \draw [style=cyan arrow, transform canvas={xshift=1.5mm,yshift=3mm}] (1)  to (6); 
    \draw [style=cyan arrow, transform canvas={xshift=-1.5mm,yshift=3mm}] (2)  to (4);
    \draw [style=cyan arrow] (3)  to (5);

     \draw [style=red arrow] (1) to (6);
	\draw [style=orange arrow] (1) [bend right] to (5);
	\draw [style=red arrow] (5) [bend right] to (2);
	\draw [style=orange arrow] (2) to (4);
	\draw [style=red arrow] (4) [bend left] to (3);
	\draw [style=orange arrow] (3) [bend left] to (6); 
\end{tikzpicture}
\caption{The Green-Schwarz automorphisms of the relative 6d $(2,0)$ SCFTs with algebra $\mathfrak{so}(8)$. 
$\GS 2$ (whose action is shown as black dashed arrows) is of order 2 and exchanges $S\leftrightarrow C$, whereas $\GS 3$ (shown in black arrows) is of order 3 and cyclically permutes $S,\,V,\,C$. The operations shown in cyan and blue are gauging the 2-form symmetry (cyan) and stacking a TQFT (blue).\label{fig:GS2_GS3}}
\end{figure}
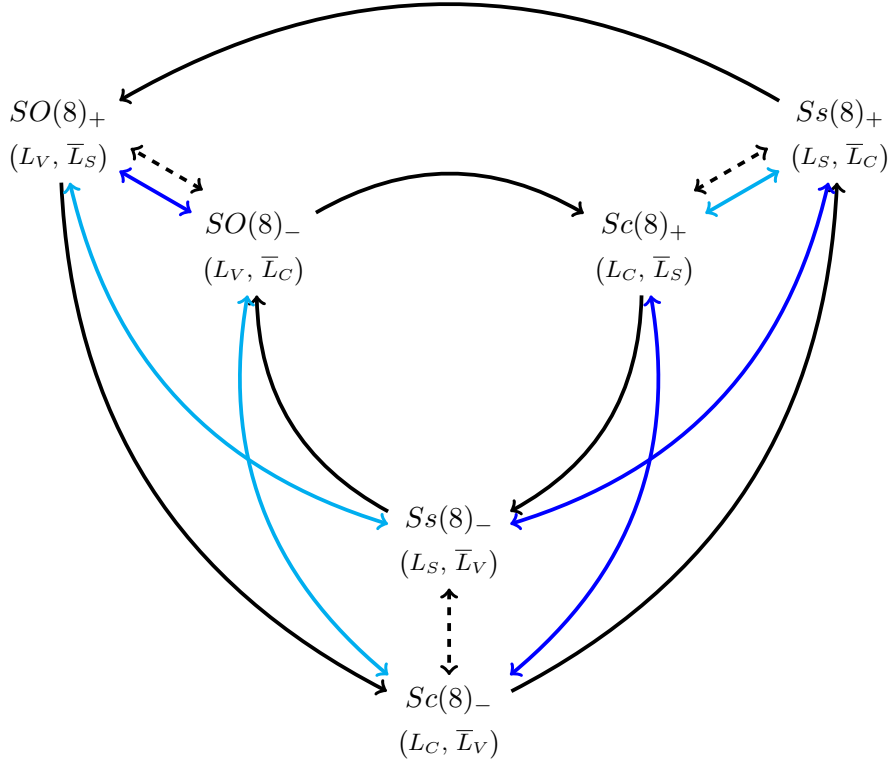

\paragraph{Green-Schwarz (GS) Dualities.} 
The GS automorphisms act on the polarization pairs and fall into two types: order 2 and order 3. The order 2 GS automorphisms act as follows:
\be  \label{eq:GS2_def1}
\Z_2:\qquad 
    \ba
         \GS 2\up{V}:& \qquad S\leftrightarrow C, \qquad V\leftrightarrow V \\
         \GS 2\up{S}:& \qquad C\leftrightarrow V, \qquad S\leftrightarrow S \\
         \GS 2\up{C}:& \qquad S\leftrightarrow V, \qquad C\leftrightarrow C 
    \ea
\ee
and those of order 3 are cyclic or anti-cyclic permutations 
\be \label{eq:GS3_def1}
\Z_3:\qquad 
 \ba
         {\GS 3}: &\qquad V\to C\to S\to V \\
         \overline{\GS 3}: &\qquad V\to S\to C\to V \,.
\ea
\ee
From this action on the polarizations we can infer the action on the absolute theories as shown in figure  \ref{fig:GS2_GS3} in terms of the black solid/dashed lines. 
It is important to note that the combination gives rise to an $S_3$ group of automorphisms, where 
\be
S_3 = \{\id, a,a^2, b, ab, a^2 b \} 
\ee
satisfying 
\be
a^3= b^2=\id \,,\qquad  ba^2 = ab  \,.
\ee
We can identify $\GS2$ with $b$ and $\GS3$ with $a$ and check the  third relation above:
\be
Sc(8)_+ 
\ \stackrel{\GS2}{\longrightarrow}\  
Ss(8)_+ 
\ \stackrel{\GS3}{\longrightarrow}\ 
SO(8)_+ 
\ \stackrel{\GS3}{\longrightarrow}\ 
Sc(8)_-
\ee
which equals 
\be
Sc(8)_+ 
\ \stackrel{\GS3}{\longrightarrow}\  
Ss(8)_- 
\ \stackrel{\GS2}{\longrightarrow}\ 
Sc(8)_-
\ee

More precisely,  the GS automorphisms define interfaces between theories with different polarizations. 
We denote the interfaces implementing the $\GS2\up{I},\,\GS3,\,\ol{\GS3}$ automorphisms (where $I\in\{S,C,V\}$) defined in equations \eqref{eq:GS2_def1}-\eqref{eq:GS3_def1} by $D_5\up{I}(M_5),\,T_5(M_5),\,\ol{T}_5(M_5)$ respectively: they are supported on a 5-manifold $M_5$ between two different absolute theories, which are related by the corresponding Green-Schwarz duality.

It follows from \eqref{eq:GS2_def1} that the corresponding interfaces $D_5\up{I}$ are invertible of order 2:
\begin{align}
    &D_5\up{V}\otimes D_5\up{V}=D_5\up{\id}, &&D_5\up{S}\otimes D_5\up{S}=D_5\up{\id}, &&D_5\up{C}\otimes D_5\up{C}=D_5\up{\id},
\end{align}
whereas the $\GS3$ imply that the 
interfaces $T_5,\,\ol{T}_5$ are invertible of order 3:
\begin{align}
    &T_5\otimes T_5\otimes T_5=D_5\up{\id}, && \ol{T}_5\otimes \ol{T}_5\otimes \ol{T}_5=D_5\up{\id}
\end{align}

The fusions of the GS defects follow the $S_3$ group multiplication, which we summarize in table \ref{tab:GS_S3_mult1}.

\begin{table}
    \centering
    \begin{tabular}{|c||c|c|c|c|c|}
    \hline
$\otimes$ & $D_5\up{S}$ & $D_5\up{V}$ & $D_5\up{C}$ & $T_5$ & $\ol{T}_5$ \\ \hline
 \hline 
$D_5\up{S}$ & $D_5\up{\id}$ & $\ol{T}_5$ & $T_5$ & $D_5\up{C}$ & $D_5\up{V}$ \\ 
 \hline 
$D_5\up{V}$ & $T_5$ & $D_5\up{\id}$ & $\ol{T}_5$ & $D_5\up{S}$ & $D_5\up{C}$ \\ 
 \hline 
$D_5\up{C}$ & $\ol{T}_5$ & $T_5$ & $D_5\up{\id}$ & $D_5\up{V}$ & $D_5\up{S}$ \\ 
 \hline 
$T_5$ & $D_5\up{V}$ & $D_5\up{C}$ & $D_5\up{S}$ & $\ol{T}_5$ & $D_5\up{\id}$ \\ 
 \hline 
$\ol{T}_5$ & $D_5\up{C}$ & $D_5\up{S}$ & $D_5\up{V}$ & $D_5\up{\id}$ & $T_5$ \\ 
    \hline
    \end{tabular}
    \caption{Fusion table of the invertible GS interfaces: $D$ denotes the order 2 duality interfaces, $T$ the order 3 triality ones. }
    \label{tab:GS_S3_mult1}
\end{table}

\paragraph{Gauging and stacking.} 
The construction of duality/triality and in this case $S_3$-ality defects requires further that we have operations that can be used to construct maps that ``invert" the action of the GS dualities. 
For this we define the topological operations: 
\be
\ba
\sigma: &\qquad \text{gauging of the global 2-form symmetry $\Z_2\up{I}$} \cr 
\tau :&\qquad \text{stacking a 6d $\Z_2$ SPT phase} \,.
\ea
\ee
More concretely: 
$\sigma$ acts on the  polarization pairs of the theory as 
\be
\sigma :\qquad \lb L_J,\,\ol{L}_I\rb \to  
\lb L_I,\,\ol{L}_J\rb\,, 
\ee
whereas $\tau$ consists of stacking a 6d SPT phase with $\Z_2\up{I}$ symmetry on the partition function of the $\lb L_J,\,\ol{L}_I\rb$ theory, giving a theory with $\lb L_J,\,\ol{L}_K\rb,\,K\neq J,I$ symmetry (stacking with an SPT does not change the polarization but changes the second element of the polarization pair). Gauging after stacking with an SPT is called ``twisted" gauging. These 6d topological manipulations were analyzed in \cite{Gukov:2020btk} and described in terms of polarization pairs in \cite{Lawrie:2023tdz}.

{In the present construction the relevant SPT stackings are the trivial one and the non-trivial quadratic-refinement stacking $\tau$ used below. For bosonic SPTs coupled to the $\Z_2$ 2-form symmetry background, we believe that the SPTs considered here exhaust the possibilities relevant to this construction. For fermionic SPTs, we consider the particular Arf--Kervaire/$v_4$-fermionic example that appears below; a complete examination of fermionic invertible phases is left to future work. Additional phases that are decoupled from the $\Z_2$ 2-form background do not change the  fusion rules of the codimension-one defects computed here.}

\paragraph{Non-Invertible Symmetry Defects.} Combining a Green-Schwarz automorphism and a topological manipulation (gauging, possibly with twist) gives rise to duality and triality defects, and in the present case to $S_3$-ality ones, within a single absolute theory. 
We will derive them both from the ``half-space" gauging approach in section \ref{sec:half-space} and from the SymTFT in section \ref{sec:SymTFT-defects}.
We will in particular compute the fusion rules of the defects implementing these symmetries, which are an important step in identifying the symmetry fusion 5-category.

\section{$S_3$-ality Defects in 6d from Half-space Gauging} 
\label{sec:half-space}

\subsection{Non-invertible Symmetries from Half-space Gauging}

Non-invertible duality and triality symmetry defects satisfy fusion rules of order two or three respectively with non-trivial condensation defects and TQFT coefficients. The half-space gauging approach to construct non-invertible duality defects in 4d was used in \cite{Choi:2021kmx,Choi:2022zal,Kaidi:2022uux,Kaidi:2022cpf}. Non-invertible triality defects from half-space twisted gauging (when the gauging occurs after stacking with a non-trivial SPT) were studied in 4d in \cite{Choi:2022zal}. Recently, this approach for non-invertible duality and triality symmetries has also been employed for 3d theories \cite{Cui:2024cav}. 

In this section, we will apply the half-space (twisted) gauging method to construct and analyze in detail duality, triality, and $S_3$-ality defects in the 6d $\cN=(2,0)\;\mf{so}(8)$ SCFT. For duality defects, we will firstly review the operation $\sigma$ of gauging the 2-form global symmetry of an absolute 6d theory. Then, following the above cited works in lower dimensions, we define a codimension-1 half-space gauging \emph{interface} $\sigma(M_5)$ between a 6d absolute theory and the theory obtained from it by means of the $\sigma$-operation. 

Denote by $x$ the coordinate on the 6d spacetime $M_6$ in the direction locally orthogonal to the 5-manifold $M_5$ supporting the half-space gauging interface $\sigma(M_5)$, and separate this into two regions $x<0$ and $x\geqslant 0$. Schematically this will be depicted as follows:
\begin{equation} 
     \begin{split}
    &\;{\color{red}\sigma(M_5)}\\
    \fT|_{M_6^{<0}} \qquad &{\color{red} \quad\bigg|} \qquad  (\fT / \Z_2)|_{M_6^{\gs 0}}
\end{split}
\end{equation}
On $M_6^{<0}$ the theory is $\fT$, whereas on $M_6\gz$ lies the theory obtained from it by gauging its $\Z_2$ 2-form global symmetry. The half-space gauging interface $\sigma(M_5)$ separates them at $M_5=\de M_6\gz$: it is obtained by gauging the $\Z_2$ 2-form global symmetry of $\fT$ in half of spacetime while imposing the Dirichlet boundary condition for the dynamical gauge field on the interface $M_5$; on the remaining half of spacetime we have the theory with the gauged symmetry $\fT/\Z_2$ and $M_5$ thus supports a codimension-1 interface $\sigma(M_5)$. 
Since we will concretely perform computations in terms of partition functions, we will schematically depict the interfaces between effective Lagrangian densities, like in reference \cite{Choi:2022zal}, rather than abstract theories, e.g. 
\begin{equation}
     \begin{split}
    \hspace{3cm}&\;{\color{red}\sigma(M_5)}\\
    \hspace{3cm}\cL_{\fT_J}[C_3\up{I}]  &{\color{red} \quad\bigg|_{c_3\up{I}|_{M_{5}}=0}} \quad  \cL_{\fT_J}[c_3\up{I}]\,+\pi(c_3\up{I}\cup C_3\up{J})
\end{split}
\end{equation}
The theory on the left-hand side has polarization $L_J$, with $J\in\{S,C,V\}$, so we denote it as $\fT_J$ and its effective Lagrangian density, that we use only to describe its $\Z_2$ 2-form symmetry, as $\cL_{\fT_J}$.\footnote{{Note that 6d SCFTs are generically Non-Lagrangian theories, i.e. they do not admit a description in terms of a local Lagrangian density. However, we can still use an effective Lagrangian to describe only their discrete 2-form symmetry.}} The effective Lagrangian is a function of background fields $C_3\up{I}$ for the $\Z_2\up{I}$ 2-form global symmetry of the theory, where $I\neq J\in \{S,C,V\}$. The interface $\sigma(M_5)$ gauges $\Z_2\up{I}$ by making $c_3\up{I}$ dynamical with the Dirichlet boundary condition for it imposed on $M_5$ and couples it to a background field $C_3\up{J}$ for the dual $\Z_2\up{J}$ global symmetry. The precise equation corresponding to the above schematic figure is \eqref{eq:sigma_6d_def_eq}.

Triality interfaces can instead be constructed by gauging in combination with stacking the partition function with a $\Z_2$ TQFT. This requires a quadratic refinement of the intersection pairing: we write it as the integral of a 6-form $\bar{q}$, which gives a $\Z_2$-valued quadratic function of the 3-form background fields for the 2-form $\Z_2$ symmetry. When these fields are classical backgrounds (i.e. the 2-form symmetry is global) we denote them by an uppercase letter, e.g. $C_3\up{I},\,I\in\{S,C,V\}$, whereas when they are dynamical (i.e. the symmetry is being gauged and the fields are summed-over in the path integral) we use lowercase. Therefore, if we perform twisted gauging (where we first stack and then gauge) of a $\Z_2\up{I}$ symmetry, we will encounter $\bar{q}(c_3\up{I})$, whereas if we stack after gauging, then $\bar{q}(C_3\up{K})$ will appear, where $\Z_2\up{K}$ is the global symmetry of the resulting theory. By performing these operations in half of spacetime we will define the \emph{interfaces} $\tau\sigma(M_5)$ and  $\sigma\tau(M_5)$ respectively. 

\paragraph{Defects.}
To define duality/triality/$S_3$-ality \emph{defects} in a given absolute theory, we combine:
\begin{itemize}
    \item an order-2 Green–Schwarz duality interface  with a half-space gauging interface $\sigma(M_5)$ to construct a non-invertible duality defect $\cD_5(M_5)$,
    \item an order-3 Green–Schwarz duality interface with a half-space twisted gauging interface $\tau\sigma(M_5)$ or gauging and stacking interface $\sigma\tau(M_5)$ to construct non-invertible triality defects $\cT_5(M_5)$ and $\ol{\cT}_5(M_5)$.
\end{itemize}

\paragraph{Fusion.}
To compute the (parallel) fusion of two codimension-1 defects, following \cite{Kaidi:2022cpf}, we place the first at $x=0$ and the second at $x=\eps$. Schematically this looks as follows:\footnote{For the effective Lagrangian densities see equation \eqref{eq:cD_cD_6d_fig}.}
\begin{align} 
    &\;{\color{red}\cD_5(M_{5|0})} &&\;{\color{red}\cD_5(M_{5|\eps})} \nonumber \\
    {\fT_J}|_{M_6^{<0}} \quad &{\color{red} \quad\bigg|}  \hspace{2cm} ({\fT_I}/\Z_2\up{J})|_{M_6\ze}  \hspace{-1cm}  &&{\color{red}\quad\bigg|} \hspace{1.7cm} ({\fT_J}/\Z_2\up{I}/\Z_2\up{J})|_{M_6^{>\eps}}\\
        &\;{x=0} &&\;{x=\epsilon} \nonumber
\end{align}
spacetime $M_6$ is divided into three regions:
\begin{itemize}
    \item $M_6^{<0}$ hosts the initial theory, 
    \item $M_6^{[0,\eps]}$ constitutes the region between the two defects whose fusion we are computing. We denote its boundaries by $M_{5|0}$ and $M_{5|\eps}$ to indicate that they are 5-manifolds located at $x=0$ and $x=\eps$ respectively. The theory on $M_6\ze$ is obtained from the starting one by acting only with the first defect.
    \item $M_6^{>\eps}$ hosts the final theory, given by acting on the initial one with both defects sequentially.
\end{itemize}
For these computations, we will denote the theories by their effective Lagrangian densities, since the action of the symmetry defects is given concretely on them. After performing appropriate manipulations\footnote{For example, integrating out dynamical fields, using properties of the quadratic refinement, cohomology isomorphisms and Poincaré duality.} we will take $\eps\to0$ and write the resulting fusion.

\subsection{Non-invertible Duality Defects}

We now consider the duality GS-dualities $\GS2$, and combine them with gauging/stacking operations of order two to obtain non-invertible duality symmetries. First we discuss the operation $\sigma$ of gauging the 2-form symmetry, which will be used in the construction. 

\subsubsection{Gauging 2-form Symmetries}

If one starts with the absolute 6d $\cN=(2,0)\;\mf{so}(8)$ theory with polarization pair $\lb L_J,\,\ol{L}_I\rb$ the operation $\sigma$ of gauging its global 2-form $\Z_2\up{I}$ symmetry without any twist gives rise to a theory with polarization pair $\lb L_I,\,\ol{L}_J\rb$ and a dual $\Z_2\up{J}$ global 2-form symmetry by acting as follows on the partition function:
\begin{align} \label{eq:sigma_6d}
     \lb L_J,\,\ol{L}_I\rb &\;\xrightarrow{\sigma}\; \lb L_I,\,\ol{L}_J\rb \nn\\[2mm]
    \ Z_{\fT_J}[C_3\up{I}] &\;\xmapsto{\sigma} N\dw{M_6,\,\Z_2}\sum_{c_3\up{I}\in H^3(M_6,\,\Z_2)} Z_{\fT_J}[c_3\up{I}] \,\exp(i\pi\int_{M_6}c_3\up{I}\cup C_3\up{J}) \,.
\end{align}
On the right-hand side of equation \eqref{eq:sigma_6d}, one makes the gauge field for the $\Z_2\up{I}$ 2-form symmetry dynamical and sums over all field configurations weighted by a term coupling the dynamical gauge field $c_3\up{I}$ to the classical background $C_3\up{J}$ for the dual ${\Z}_2\up{J}$ 2-form symmetry. The normalization factor
\begin{equation} \label{eq:N_Z2_M6}
     N\dw{M_6,\,\Z_2}=\frac{|H^1(M_6,\,\Z_2)|}{|H^2(M_6,\,\Z_2)|\,|H^0(M_6,\,\Z_2)|}
\end{equation}
takes into account the volume of the gauge redundancy of the 2-form symmetry being gauged on $M_6$. 

The resulting absolute theory, with polarization pair $\lb L_I,\,\ol{L}_J\rb$, has a global $\Z_2\up{J}$ 2-form symmetry: by gauging it, we return to the original theory $\lb L_J,\,\ol{L}_I\rb$. We can check this by applying the $\sigma$ operation (\ref{eq:sigma_6d}) twice:
\begin{align}
    Z_{\fT_J}[C_3\up{I}]\;\xmapsto{\sigma^2}\; N\dw{M_6,\,\Z_2}^2\sum_{\substack{c_3\up{I},\,c_3\up{J}\in H^3(M_6,\,\Z_2)}} Z_{\fT_J}[c_3\up{I}] \,\exp(i\pi\int_{M_6}\lb c_3\up{I}\cup c_3\up{J}+c_3\up{J}\cup C_3\up{I}\rb) \,.
\end{align}
In the above expression, we can integrate out $c_3\up{J}$: this enforces $c_3\up{I}=C_3\up{I}$ which trivializes the exponential and produces a factor of $|H^3(M_6,\,\Z_2)|$
showing that gauging twice maps the partition function to itself, up to an Euler counterterm:\footnote{For the normalization, recall the definition of $ N\dw{M_6,\,\Z_2}$ in eq. \eqref{eq:N_Z2_M6}, the property $|H^n(M_6,\,\Z_2)|=|H^{6-n}(M_6,\,\Z_2)|$ and the definition of the Euler counterterm
\begin{equation} \label{eq:chi6d}
    \chiM{}=\frac{|H^0(M_6,\,\Z_2)|\,|H^2(M_6,\,\Z_2)|\,|H^4(M_6,\,\Z_2)|\,|H^6(M_6,\,\Z_2)|}{|H^1(M_6,\,\Z_2)|\,|H^3(M_6,\,\Z_2)|\,|H^5(M_6,\,\Z_2)|} \,.
\end{equation} \label{foot:chi}}
\begin{equation} 
    Z_{\fT_J}[C_3\up{I}]\;\xmapsto{\sigma^2}\;\chiiM{}Z_{\fT_J}[C_3\up{I}] \,,
\end{equation}
which means that
\begin{equation} \label{eq:sigma_squared_6d}
    \sigma^2=\chiiM{}\,.
\end{equation}
The normalization $N\dw{M_6,\,\Z_2}$ in the gauging operation $\sigma$ could be re-defined by a power of the Euler counterterm $\chiM{}$: in particular multiplying the partition function in (\ref{eq:sigma_6d}) by $\chiM{}^{1/2}$ would eliminate the counterterm from (\ref{eq:sigma_squared_6d}) and give $\sigma^2=\id$. However, in the following, we will maintain the normalization of equation (\ref{eq:sigma_6d}), which enables us to keep track of the volume of gauge redundancies.

\subsubsection{Definition of Duality Interface and Defects} \label{sec:sigma_sigma}
\paragraph{Half-space gauging interface $\boldsymbol{\sigma(M_5)}$.} A codimension-1 half-space gauging interface $\sigma(M_5)$ between the 6d absolute theories with polarization pairs $\lb L_J,\,\ol{L}_I\rb$ and $\lb L_I,\,\ol{L}_J\rb$ can be constructed by gauging the $\Z_2\up{I}$ 2-form symmetry of the former theory, equation (\ref{eq:sigma_6d}), in half of spacetime while imposing the Dirichlet boundary condition for the dynamical gauge field $c_3\up{I}$ on the interface $M_5$.  Schematically:
\begin{equation} \label{eq:sigma_6d_def}
     \begin{split}
    &\quad{\color{red}\sigma(M_5)}\\
    \cL_{\fT_J}[C_3\up{I}]  &{\color{red} \qquad\bigg|_{c_3\up{I}|_{M_{5}}=0}} \quad  \cL_{\fT_J}[c_3\up{I}]\,+\pi(c_3\up{I}\cup C_3\up{J})
\end{split}
\end{equation}
which, in terms of partition functions, is written as:
\begin{align} \label{eq:sigma_6d_def_eq}
    &\quad{\color{red}\sigma(M_5)}\\
    Z_{\fT_J}[C_3\up{I}] &{\color{red} \qquad\bigg|_{c_3\up{I}|_{M_{5}}=0}} \quad  N\dw{M_6\gz,\,\Z_2}\sum_{c_3\up{I}\in H^3(M_6\gz,\,M_{5|0},\,\Z_2)} Z_{\fT_J}[c_3\up{I}]\,\exp(i\pi\int_{M_6\gz}c_3\up{I}\cup C_3\up{J})  
\end{align}
with the normalization factor\footnote{Recall that, when considering a manifold with boundary, cohomology becomes relative but homology is still absolute \cite{Kaidi:2022cpf}.}
\begin{equation}
    N\dw{M_6\gz,\,\Z_2}=\frac{|H^1(M_6\gz,M_{5|0},\,\Z_2)|}{|H^2(M_6\gz,M_{5|0},\,\Z_2)|\,|H^0(M_6\gz,M_{5|0},\,\Z_2)|} \,.\label{eq:N_Z2_M6gz}
\end{equation}

\paragraph{Duality defect $\boldsymbol{\cD_5(M_5)}$.}
In a fixed absolute theory with polarization pair $\lb L_J,\,\ol{L}_I\rb$, a non-invertible duality \emph{defect} $\cD_5(M_5)$ can be constructed by composing a GS2 duality and half-space gauging \cite{Lawrie:2023tdz}:
\begin{align} \label{eq:cD_5I_def_fus}
    \cD_5(M_5)=D_5\up{K}(M_5)\otimes \sigma(M_5) \,, \qquad I\neq J\neq K \in\{S,C,V\} \,.
\end{align}
First the $\GS2\up{K}$ duality (with $K\neq I,J$), realized by $D_5\up{K}(M_5)$, exchanges the labels $I\leftrightarrow J$ and then the $\sigma(M_5)$ interface implements gauging of the new global $\Z_2\up{J}$ 2-form symmetry:
\begin{align}
    \lb L_J,\,\ol{L}_I\rb \;\xrightarrow{D_5\up{K}(M_5)}\; &\lb L_I,\,\ol{L}_J\rb \;\xrightarrow{\sigma(M_5)}\; \lb L_J,\,\ol{L}_I\rb \nn\\[2mm]
    Z_{\fT_J}[C_3\up{I}]\;\xmapsto{D_5\up{K}(M_5)}\;  &Z_{\fT_I}[C_3\up{J}]\;\xmapsto{\sigma(M_5)}\; \chi^{-1/2}\,Z_{\fT_J}[C_3\up{I}]  \,.
\end{align}
$\cD_5(M_5)$ is thus a defect in a given absolute theory with polarization pair $\lb L_J,\,\ol{L}_I\rb$,
and as we shall see below, obeys non-invertible fusion rules.
Recalling the definition of the $\sigma(M_5)$ duality interface \eqref{eq:sigma_6d_def} and the fact that $D_5\up{K}(M_5)$ exchanges $I\leftrightarrow J$, the action of the duality defect $\cD_5(M_5)$ can be depicted as follows:
\begin{equation} \label{eq:cD5_6d_fig}
     \begin{split}
    &\quad{\color{red}\cD_5(M_5)}\\
    \cL_{\fT_J}[C_3\up{I}]  &{\color{red} \qquad\bigg|_{c_3\up{J}|_{M_{5}}=0}} \quad   \cL_{\fT_I}[c_3\up{J}]\,+\pi(c_3\up{J}\cup C_3\up{I})\,.
\end{split}
\end{equation}
It translates to the following equation:
\begin{align} \label{eq:cD5_6d_eq}
    Z_{\fT_J}[C_3\up{I};\,\cD_5(M_5)]=N\dw{M_6\gz,\,\Z_2}\sum_{c_3\up{J}\in H^3(M_6\gz,\,M_{5|0},\,\Z_2)} Z_{\fT_I}[c_3\up{J}]\,\exp(i\pi\int_{M_6\gz}c_3\up{J}\cup C_3\up{I})\,.
\end{align}
{where on the left hand side we denote the partition function with the insertion of the $\cD_5(M_5)$ defect.}

\subsubsection{Fusion of Duality Defects} \label{sec:D5_D5_fusion}
We now compute the fusion of two duality defects, defined in equations \eqref{eq:cD5_6d_fig}-\eqref{eq:cD5_6d_eq}, located at $x=0$ and $x=\eps$ respectively, and then take $\eps\to0$.  As described in the summary, the two interfaces divide spacetime $M_6$ into three regions, $M_6^{<0},\,M_6^{[0,\eps]},\,M_6^{>\eps}$. Schematically this is depicted as:
\begin{align} \label{eq:cD_cD_6d_fig}
    &\;{\color{red}\cD_5(M_{5|0})} &&\;{\color{red}\cD_5(M_{5|\eps})} \nonumber \\
    \cL_{\fT_J}[C_3\up{I}]  &{\color{red} \quad\bigg|_{c_3\up{J}|_{M_{5|0}}=0}}  \cL_{\fT_I}[c_3\up{J}]\;+\pi(c_3\up{J}\cup C_3\up{I})   &&{\color{red}\quad\bigg|_{\substack{c_3\up{J}|_{M_{5|\eps}}=0}}} \hspace{3mm}\cL_{\fT_J}[c_3\up{I}] \,+\pi\lb c_3\up{I}\cup c_3\up{J}+c_3\up{J}\cup C_3\up{I}\rb\,.
\end{align}
From its definition in equation \eqref{eq:cD_5I_def_fus}, the duality defect $\cD_5(M_{5|\eps})$ first acts with a Green-Schwarz interface $D_5\up{K}$ at $M_{5|\eps}$: when $c_3\up{J}$ passes through $D_5\up{K}(M_{5|\eps})$, it gets converted to $c_3\up{I}$,
\begin{equation} \label{eq:GSK_JI_eps}
   c_3\up{J}|_{M_{5|\eps}}\, D_5\up{K}(M_{5|\eps})=D_5\up{K}(M_{5|\eps})\,c_3\up{I}|_{M_{5|\eps}}\,.
\end{equation}
We then impose the Dirichlet boundary condition on $c_3\up{J}$ when performing half-space gauging by means of $\sigma(M_{5|\eps})$: we can therefore split it into ${c}_3\up{J}\in H^3(M_6^{[0,\eps]},\,M_{5|0}\cup M_{5|\eps},\,\Z_2)$ and $c_3\up{J} \in H^3(M_6^{\gs\eps},\,M_{5|\eps},\,\Z_2)$. Since $D_5\up{K}$ is invertible, equation \eqref{eq:GSK_JI_eps} implies that we must also impose the Dirichlet boundary condition for $c_3\up{I}$ on $M_{5|\eps}$. The equation corresponding to \eqref{eq:cD_cD_6d_fig} is thus:
\begin{align} \label{eq:cD_cD_6d_eq}
    Z_{\fT_J}[C_3\up{I};\,\cD_5(M_{5|0}),\, \cD_5(M_{5|\eps})]= N\dw{M_6\ze,\,\Z_2} \sum_{{c}_3\up{J} \in H^3(M_6^{[0,\eps]},\,M_{5|0}\cup M_{5|\eps},\,\Z_2)} 
    \exp(i\pi\int_{M_6\ze}{c}_3\up{J}\cup C_3\up{I})\,\times\nn\\
    \times N\dw{M_6\gep,\,\Z_2}^2\sum_{c_3\up{I},\,c_3\up{J}\in H^3(M_6^{\gs\eps},\,M_{5|\eps},\,\Z_2)}
    Z_{\fT_J}[c_3\up{I}]\,\exp(i\pi\int_{M_6\gep}\lb c_3\up{I}\cup c_3\up{J}+c_3\up{J}\cup C_3\up{I}\rb)\,.
\end{align}
Integrating out $c_3\up{J}$ on $M_6\gep$ enforces $c_3\up{I}=C_3\up{I}$ and leaves the partition function $Z_{\fT_J}[C_3\up{I}]$ on $M_6\gep$.\footnote{For the normalization, we have defined
\begin{equation}
    N\dw{M_6\ze,\,\Z_2}=\frac{|H^1(M_6\ze,M_{5|0}\cup M_{5|\eps},\,\Z_2)|}{|H^2(M_6\ze,M_{5|0}\cup M_{5|\eps},\,\Z_2)|\,|H^0(M_6\ze,M_{5|0}\cup M_{5|\eps},\,\Z_2)|}\,.
\end{equation}
Integrating out $c_3\up{J}$ yields a factor of $|H^3(M_6^{\gs\eps},\,M_{5|\eps},\,\Z_2)|$, which when combined with $N\dw{M_6\gep,\,\Z_2}^2$, produces the Euler counterterm $\chiiM{\gep}$, defined in equation \eqref{eq:chi6d}.
}
\begin{align}
    Z_{\fT_J}[C_3\up{I};\,\cD_5(M_{5|0}),\, \cD_5(M_{5|\eps})]=&N\dw{M_6\ze,\,\Z_2}
    \sum_{{c}_3\up{J}\in H^3(M_6^{[0,\eps]},\,M_{5|0}\cup M_{5|\eps},\,\Z_2)} \,\exp(i\pi\int_{M_6\ze}{c}_3\up{J}\cup C_3\up{I})\,\times\nn\\
    &\times\,\chiiM{\gep}\,Z_{\fT_J}[ C_3\up{I}]\,.
\end{align}
We now use the isomorphism (which is described for $n$-cohomologies in appendix B of \cite{Kaidi:2021xfk})
\begin{equation} \label{eq:iso_H^3_H^2}
    H^3(M_6^{[0,\eps]},\,M_{5|0}\cup M_{5|\eps},\,\Z_2)\to H^2(M_5,\,\Z_2)
\end{equation}
to map ${c}_3\up{J}\mapsto b_2\in H^2(M_5,\,\Z_2)$ and write in the $\eps\to 0$ limit:\footnote{For the normalization we use the properties
    \begin{align}
        |H^0(M_6^{[0,\eps]},\,M_{5|0}\cup M_{5|\eps},\,\Z_2)|=1, \qquad
        |H^n(M_6^{[0,\eps]},\,M_{5|0}\cup M_{5|\eps},\,\Z_2)|=|H^{n-1}(M_5,\,\Z_2)|\,,
    \end{align}
    which imply
    \begin{align}
         N\dw{M_6\ze,\,\Z_2}=\frac{|H^1(M_6\ze,M_{5|0}\cup M_{5|\eps},\,\Z_2)|}{|H^2(M_6\ze,M_{5|0}\cup M_{5|\eps},\,\Z_2)|\,|H^0(M_6\ze,M_{5|0}\cup M_{5|\eps},\,\Z_2)|} \quad\rightarrow\quad N\dw{M_5,\,\Z_2}=\frac{|H^0(M_5,\,\Z_2)|}{|H^1(M_5,\,\Z_2)|}\,.
    \end{align} \label{foot:norm2}}
\begin{align}
   \cD_5(M_5)\otimes \cD_5(M_5)=&\chiiM{\gz}\,N\dw{M_5,\,\Z_2}\sum_{b_2\in H^2(M_5,\,\Z_2)}\exp(i\pi\int_{M_5} b_2 \cup C_3\up{I})\,.
\end{align}
We then use Poincaré duality on $M_5$ to convert the sum over $b_2\in H^2(M_5,\,\Z_2)$ to one over $M_3=\PD(b_2)\in H_3(M_5,\,\Z_2)$ and obtain:
\begin{align} 
    \cD_5(M_5)\otimes \cD_5(M_5)&=\chiiM{\gz}\;N\dw{M_5,\,\Z_2}\sum_{M_3\in H_3(M_5,\,\Z_2)}\,\exp(i\pi\int_{M_3}C_3\up{I})\,,
\end{align}
whose right hand side is a condensation defect (we include the Euler counterterm $\chiiM{\gz}$ in its normalization)
\begin{equation} \label{eq:C5I_def}
    \cC_5\up{0}(M_5)=\chiiM{\gz}\;N\dw{M_5,\,\Z_2}\sum_{M_3\in H_3(M_5,\,\Z_2)}\,\exp(i\pi\int_{M_3}C_3\up{I})
\end{equation}
of the 2-form symmetry defect
\begin{equation}
    D_3^{(I)}({M}_3)=\exp\lb i\pi\int_{M_3} C_3^{(I)}\rb 
\end{equation}
without discrete torsion (hence the $0$ superscript in $\cC_5\up{0}(M_5)$). Its normalization is (up to the Euler counterterm)
\begin{equation} \label{eq:Norm_M5}
    N\dw{M_5,\,\Z_2}=\frac{|H^0(M_5,\,\Z_2)|}{|H^1(M_5,\,\Z_2)|}\,,
\end{equation}
(derived in \cref{foot:norm2}) which correctly takes into account the gauge redundancy from condensing $D_3^{(I)}(M_3)$ on $M_5$, on which it generates a 1-form symmetry. 
The fusion of two duality defects is therefore:
\begin{equation} \label{eq:cD5_cD5_fusion_6d}
    \cD_5(M_5)\otimes \cD_5(M_5)=\cC_5\up{0}(M_5)\,.
\end{equation}

\subsection{Non-invertible Triality Defects} \label{eq:sec_triality}

We now consider the triality GS-dualities $\GS3$, and combine them with order three gauging/stacking operations to obtain non-invertible triality symmetries. 

\subsubsection{Twisted Gauging} \label{sec:twisted_gauging}
A 3-form background gauge field $C_3$ for a 2-form global symmetry has odd form degree, therefore:
\begin{equation}
    C_3\cup C_3=-C_3\cup C_3 \quad\Rightarrow\quad 2\,C_3\cup C_3=0\,.
\end{equation}
This means that $\int_{M_6}C_3\cup C_3$ is zero unless one takes the coefficients to be in a ring without a multiplicative inverse of 2, such as $\Z_2=\{0,1\}$ \cite{Lawrie:2023tdz}. A possible 6d counterterm one can stack onto the partition function before gauging can be defined by means of a quadratic refinement of the intersection pairing, notions which we will now review, following \cite{Gukov:2020btk,Bhardwaj:2020ymp}.
Recall that, if $M_6$ is an oriented 6-manifold, the intersection form:
\begin{equation}
     H^3(M_6,\,\Z_2)\times H^3(M_6,\,\Z_2)\to \Z_2: \qquad (C_3,\Tilde{C}_3)\mapsto\int_{M_6}C_3\cup \Tilde{C}_3 \quad(\mod 2)
\end{equation}
can have a quadratic refinement\footnote{Note that the existence of a quadratic refinement requires specific dimension-dependent spacetime structures: on general manifolds it is $\Z_4$-valued but for orientable manifolds, it is $\Z_2$-valued (see e.g. appendix B of \cite{Gukov:2020btk}). We will follow the conventions in e.g. section 4.4 and appendix B of \cite{Gukov:2020btk} and section 2.4 of \cite{Lawrie:2023tdz} by taking a $\Z_2$-valued quadratic refinement. For the $\Z_4$-valued convention and the relation with the $\Z_2$-valued one, see e.g. section 2.1 of \cite{Bhardwaj:2020ymp} or appendix B of \cite{Hsin:2021qiy}.}
which we write as the integral of a density $\bar{q}$
\begin{equation} \label{eq:q_def}
    \int_{M_6}\bar{q}: \quad H^3(M_6,\,\Z_2)\to \Z_2
\end{equation}
such that
\begin{equation} \label{eq:q_prop}
        \int_{M_6}\bar{q}(C_3+\Tilde{C}_3)=\int_{M_6}\lb \bar{q}(C_3)+\bar{q}(\Tilde{C}_3)+C_3\cup\Tilde{C}_3\rb \qquad (\mod 2)\,.
\end{equation}
Note that $\bar{q}$ is a special case of a $\Z_4$-valued quadratic refinement $q$ that satisfies:
\begin{equation}
            \int_{M_6}{q}(C_3+\Tilde{C}_3)=\int_{M_6}\lb {q}(C_3)+{q}(\Tilde{C}_3)+2\,C_3\cup\Tilde{C}_3\rb \qquad (\mod 4)\,,
\end{equation}
by setting $\Tilde{C}_3=0$ one learns that $q(0)=0$ and by then taking $\Tilde{C}_3=C_3$, one has $q(C_3)=C_3\cup C_3\ (\mod 2)$. The relation between the $\Z_4$-valued ${q}$ and $\Z_2$-valued $\bar{q}$ is ${q}=2\bar{q}$.

From the $\Z_2$-valued quadratic refinement one can define a 3-form $\Z_2$ gauge theory with the following topological action \cite{Gukov:2020btk}:
\begin{equation} \label{eq:q_action}
    \pi \int_{M_6}\bar{q}(C_3)
\end{equation}
and the Arf--Kervaire invariant - AK\cite{Arf:1941,Kervaire:1960,Browder:1969,Brown:1972}, that is valued in $\Z_2=\{0,1\}$, can be expressed as 
\begin{equation}
    \pi \text{AK}=\text{Arg}\sum_{c_3\in H^3(M_6,\,\Z_2)}(-1)^{\int_{M_6}\bar{q}(c_3)}\,.
\end{equation}

If one starts with the 6d absolute theory specified by the polarization pair $\lb L_J,\,\ol{L}_I\rb$ and corresponding partition function $Z_{\fT_J}[C_3\up{I}]$, one can define the operation $\tau$ \cite{Bhardwaj:2020ymp,Choi:2022zal} which acts on the partition function by stacking with  \eqref{eq:q_action}. This operation does not affect the polarization $L$ but changes $\ol{L}$ (the second entry of the polarization pair) and hence the global symmetry of the theory. This follows from appendix B of \cite{Lawrie:2023tdz} and the earlier discussion in section 4.4 of \cite{Gukov:2020btk}.
\begin{align} \label{eq:tau_def_6d}
     \lb L_J,\,\ol{L}_I\rb &\;\xrightarrow{\;\tau\;}\; \lb L_J,\,\ol{L}_K\rb,\qquad K\neq I,J\in \{S,C,V\}\,, \nn\\[2mm]
    \ Z_{\fT_J}[C_3\up{I}] &\;\xmapsto{\;\tau\;}  Z_{\fT_J}[C_3\up{I}]\, \exp(i\pi\int_{M_6}\bar{q}(C_3\up{I}))
    =Z_{\fT_J}[C_3\up{K}]\,.
\end{align}

The combined ``twisted gauging" operation $\tau\sigma$ of stacking \eqref{eq:tau_def_6d} and then gauging \eqref{eq:sigma_6d} acts as follows:
\begin{align}  \label{eq:tau_sigma_6d}
     \lb L_J,\,\ol{L}_I\rb \;\xrightarrow{\tau\sigma}
     &\;\lb L_K,\,\ol{L}_J\rb,\qquad K\neq I,J\in \{S,C,V\}\,, \nn\\[2mm]
     Z_{\fT_J}[C_3\up{I}] \;\xmapsto{\tau\sigma}  &\;N\dw{M_6,\,\Z_2}\sum_{c_3\up{I}\in H^3(M_6,\,\Z_2)} Z_{\fT_J}[c_3\up{I}] \,\exp(i\pi \int_{M_6}\lb \bar{q}(c_3\up{I})+c_3\up{I}\cup C_3\up{J}\rb)\\
     =&\;\chiM{}^{-1/2}\;Z_{\fT_K}[C_3\up{J}]\,.
\end{align}
We have gauged the $\Z_2\up{I}$ global symmetry of the starting theory after having stacked it with a non-trivial SPT: this gives rise to a theory with a  global $\Z_2\up{J}$ symmetry and a dual $\Z_2\up{K}$ symmetry (which is not $\Z_2\up{I}$ due to the twisted gauging). $N\dw{M_6,\,\Z_2}$ is the normalization factor given in equation \eqref{eq:N_Z2_M6}. 

Starting from the theory with partition function $Z_{\fT_J}[C_3\up{I}]$ and gauging \eqref{eq:sigma_6d} then stacking with \eqref{eq:q_action} defines the operation $\sigma\tau$:
\begin{align}  \label{eq:sigma_tau_6d}
     \lb L_J,\,\ol{L}_I\rb \;\xrightarrow{\sigma\tau}
     &\; \lb L_I,\,\ol{L}_K\rb,\qquad K\neq I,J\in \{S,C,V\}\,, \nn\\[2mm]
     Z_{\fT_J}[C_3\up{I}] \;\xmapsto{\sigma\tau}
     &\;  N\dw{M_6,\,\Z_2}\sum_{c_3\up{I}\in H^3(M_6,\,\Z_2)} Z_{\fT_J}[c_3\up{I}] \,\exp(i\pi\int_{M_6}\lb c_3\up{I}\cup C_3\up{K}+\bar{q}(C_3\up{K})\rb)\,\\
     =&\;\chiM{}^{-1/2}\;Z_{\fT_I}[C_3\up{K}]\,.
\end{align}
We note that  $\sigma\tau$ and $\tau\sigma$ are inverses of each other. This can be seen in terms of polarization pairs
\begin{align}
    \lb L_J,\,\ol{L}_I\rb
     &\;\xrightarrow{\sigma\tau}\; \lb L_I,\,\ol{L}_K\rb\;\xrightarrow{\tau\sigma}\; \lb L_J,\,\ol{L}_I\rb \,,\nn\\
     \lb L_J,\,\ol{L}_I\rb &\;\xrightarrow{\tau\sigma}\; \lb L_K,\,\ol{L}_J\rb\;\xrightarrow{\sigma\tau}\; \lb L_J,\,\ol{L}_I\rb\,,
\end{align}
as well as from the partition functions
\begin{align}
    Z_{(\sigma\tau\tau\sigma)\,\fT_J}[C_3\up{I}]&=N\dw{M_6,\,\Z_2}^2\sum_{c_3\up{I},\,c_3\up{K}\in H^3(M_6,\,\Z_2)} Z_{\fT_J}[c_3\up{I}] \,e^{i\pi\int_{M_6}\lb c_3\up{I}\cup c_3\up{K}+\bar{q}(c_3\up{K})+\bar{q}(c_3\up{K})+c_3\up{K}\cup C_3\up{I}\rb}\,,\\[2mm]
    Z_{(\tau\sigma\sigma\tau)\,\fT_J}[C_3\up{I}]&=N\dw{M_6,\,\Z_2}^2\sum_{c_3\up{I},\,c_3\up{J}\in H^3(M_6,\,\Z_2)} Z_{\fT_J}[c_3\up{I}] \,e^{i\pi \int_{M_6}\lb \bar{q}(c_3\up{I})+c_3\up{I}\cup c_3\up{J}+c_3\up{J}\cup C_3\up{I}+\bar{q}(C_3\up{I})\rb}\,.
\end{align}
Using the fact that $\int_{M_6}\bar{q}$ is $\Z_2$-valued \eqref{eq:q_def} and integrating out $c_3\up{K}$ in the first case and $c_3\up{J}$ in the second sets $c_3\up{I}=C_3\up{I}$ giving: \vspace{2mm}
\begin{equation}
    Z_{(\sigma\tau\tau\sigma)\,\fT_J}[C_3\up{I}]=Z_{(\tau\sigma\sigma\tau)\,\fT_J}[C_3\up{I}]=\chiiM{}\;Z_{\fT_J}[C_3\up{I}]
\end{equation}

Both $\tau\sigma$ and $\sigma\tau$ are order-3 operations (with opposite cyclicity). This can be seen both in terms of polarization pairs \cite{Lawrie:2023tdz}:
\begin{align}
    \lb L_J,\,\ol{L}_I\rb \;\xrightarrow{\tau\sigma}\; \lb L_K,\,\ol{L}_J\rb \;\xrightarrow{\tau\sigma}\; \lb L_I,\,\ol{L}_K\rb \;\xrightarrow{\tau\sigma}\; \lb L_J,\,\ol{L}_I\rb \\
    \lb L_J,\,\ol{L}_I\rb \;\xrightarrow{\sigma\tau}\; \lb L_I,\,\ol{L}_K\rb \;\xrightarrow{\sigma\tau}\; \lb L_K,\,\ol{L}_J\rb \;\xrightarrow{\sigma\tau}\; \lb L_J,\,\ol{L}_I\rb 
\end{align}
and from partition functions \cite{Bhardwaj:2020ymp}. Indeed, the action of $(\tau\sigma)^2$ is:
\begin{align} \label{eq:tau_sigma_squared}
    Z_{(\tau\sigma)^2\,\fT_J}[C_3\up{K}] &= N\dw{M_6,\,\Z_2}^2\sum_{c_3\up{I},\,c_3\up{J}\in H^3(M_6,\,\Z_2)} Z_{\fT_J}[c_3\up{I}] \,e^{i\pi \int_{M_6}\lb \bar{q}(c_3\up{I})+c_3\up{I}\cup c_3\up{J}+\bar{q}(c_3\up{J})+c_3\up{J}\cup C_3\up{K}\rb}=\nn\\
    &=N\dw{M_6,\,\Z_2}^2\sum_{c_3\up{I},\,c_3\up{J}\in H^3(M_6,\,\Z_2)} Z_{\fT_J}[c_3\up{I}] \,e^{i\pi \int_{M_6}\lb \bar{q}(c_3\up{I}+c_3\up{J}+C_3\up{K})+c_3\up{I}\cup C_3\up{K}+\bar{q}(C_3\up{K}) \rb}=\nn\\
    &=\lb N\dw{M_6,\,\Z_2}\sum_{\Tilde{c}_3\up{J}\in H^3(M_6,\,\Z_2)} e^{i\pi \int_{M_6}\bar{q}(\Tilde{c}_3\up{J})}\rb
   N\dw{M_6,\,\Z_2}\sum_{c_3\up{I}\in H^3(M_6,\,\Z_2)} Z_{\fT_J}[c_3\up{I}] \,e^{i\pi\int_{M_6}\lb c_3\up{I}\cup C_3\up{K}+\bar{q}(C_3\up{K}) \rb},
\end{align}
where we used the property of $\bar{q}$, eq. \eqref{eq:q_prop}, the fact that the fields are $\Z_2$ cocycles (so their sign is irrelevant) and we replaced the sum over $c_3\up{J}$ with one over $\Tilde{c}_3\up{J}=c_3\up{I}+c_3\up{J}+C_3\up{K}$. In the second factor we recognize the action of $\sigma\tau$, therefore
\begin{align}
    Z_{(\tau\sigma)^2\,\fT_J}[C_3\up{K}]=\cZ\dw{Y,M_6}\; Z_{(\sigma\tau)\fT_J}[C_3\up{K}]\,,
\end{align}
where 
\begin{equation} \label{eq:Z_Y}
    \cZ\dw{Y,M_6}=N\dw{M_6,\,\Z_2}\,\sum_{c_3\in H^3(M_6,\,\Z_2)}\,e^{i\pi\, \int_{M_6} \bar{q}(c_3)}
\end{equation}
is (up to normalization) the exponential of the Arf--Kervaire invariant of the quadratic refinement of the intersection form \cite{Bhardwaj:2020ymp,Gukov:2020btk}. \\
Similarly, for $(\sigma\tau)^2$ we have:
\begin{align}
    Z_{(\sigma\tau)^2\,\fT_J}[C_3\up{J}] &=N^2\dw{M_6,\,\Z_2}\sum_{c_3\up{I},\,c_3\up{K}\in H^3(M_6,\,\Z_2)} Z_{\fT_J}[c_3\up{I}] \,e^{i\pi\int_{M_6}\lb c_3\up{I}\cup c_3\up{K}+ \,\bar{q}(c_3\up{K})+c_3\up{K}\cup C_3\up{J}+\bar{q}(C_3\up{J})\rb}= \nn\\
    &=N\dw{M_6,\,\Z_2}^2\sum_{c_3\up{I},\,c_3\up{K}\in H^3(M_6,\,\Z_2)} Z_{\fT_J}[c_3\up{I}] \,e^{i\pi\int_{M_6}\lb  \bar{q}(c_3\up{I}+c_3\up{K}+C_3\up{J})+\bar{q}(c_3\up{I})+c_3\up{I}\cup C_3\up{J} \rb}=\nn\\
    &=\lb N\dw{M_6,\,\Z_2}\sum_{\Tilde{c}_3\up{K}\in H^3(M_6,\,\Z_2)} e^{i\pi \int_{M_6}\bar{q}(\Tilde{c}_3\up{K})}\rb
   N\dw{M_6,\,\Z_2}\sum_{c_3\up{I}\in H^3(M_6,\,\Z_2)} Z_{\fT_J}[c_3\up{I}] \,e^{i\pi \int_{M_6}\lb \bar{q}(c_3\up{I})+c_3\up{I}\cup C_3\up{J} \rb}\,,
\end{align}
where $\Tilde{c}_3\up{K}=c_3\up{I}+c_3\up{K}+C_3\up{J}$. 
The second factor is the action on the partition function of $\tau\sigma$, therefore:
\begin{align}
    Z_{(\sigma\tau)^2\,\fT_J}[C_3\up{J}]=\cZ\dw{Y,M_6}\; Z_{(\tau\sigma)\fT_J}[C_3\up{J}]
\end{align}
with $\cZ\dw{Y,M_6}$ defined in equation \eqref{eq:Z_Y}.

\subsubsection{Definition of Triality Interfaces and Defects}

\paragraph{Half-space twisted gauging interface $\boldsymbol{\tau\sigma(M_5)}$.} 
A codimension-1 interface $\tau\sigma(M_5)$ between the 6d absolute theories with polarization pairs $\lb L_J,\,\ol{L}_I\rb$ and $\lb L_K,\,\ol{L}_J\rb$ can be defined by performing twisted gauging, equation \eqref{eq:tau_sigma_6d}, in half of spacetime and imposing the Dirichlet boundary condition for the dynamical gauge field $c_3\up{I}$ on the interface $M_5$. Schematically:
\begin{equation} \label{eq:tau_sigma_6d_fig}
     \begin{split}
    &\quad{\color{red}\tau\sigma(M_5)}\\
    \cL_{\fT_J}[C_3\up{I}]  &{\color{red} \qquad\bigg|_{c_3\up{I}|_{M_{5}}=0}} \quad \cL_{\fT_J}[c_3\up{I}] \,+\pi\lb  \bar{q}(c_3\up{I})+c_3\up{I}\cup C_3\up{J}\rb
\end{split}
\end{equation}
which, in terms of partition functions, is written as:
\begin{align} \label{eq:tau_sigma_6d_eq}
    &\;{\color{red}\tau\sigma(M_5)}\\
    Z_{\fT_J}[C_3\up{I}] &{\color{red} \quad\bigg|_{c_3\up{I}|_{M_{5}}=0}\quad} N\dw{M_6\gz,\,\Z_2}\sum_{c_3\up{I}\in H^3(M_6\gz,\,M_{5|0},\,\Z_2)} Z_{\fT_J}[c_3\up{I}] \,\exp(i\pi\int_{M_6\gz}\lb  \bar{q}(c_3\up{I})+c_3\up{I}\cup C_3\up{J}\rb)
\end{align}
with the normalization factor $N\dw{M_6\gz,\,\Z_2}$ defined in equation \eqref{eq:N_Z2_M6gz}.

\paragraph{Triality defect $\boldsymbol{\cT_5(M_5)}$.}  By combining a $\GS3$ automorphism and a triality interface $\tau\sigma(M_5)$ we can construct a non-invertible triality defect $\cT_5(M_5)$ in a given absolute theory with polarization pair $\lb L_J,\,\ol{L}_I\rb$
\begin{align} \label{eq:cT5_def_fus}
    \cT_5(M_5)={T}_5(M_5)\otimes \tau\sigma(M_5) \,.
\end{align}
First the ${\GS3}$ automorphism  performs an anti-cyclic permutation of $I,J,K$ and then the $\tau\sigma(M_5)$ interface implements twisted gauging of the new global $\Z_2\up{K}$ 2-form symmetry:
\begin{align}\label{T5Def}
     \lb L_J,\,\ol{L}_I\rb \;\xrightarrow{T_5(M_5)}\; &\lb L_I,\,\ol{L}_K\rb \;\xrightarrow{\tau\sigma(M_5)}\; \lb L_J,\,\ol{L}_I\rb \nn\\[2mm]
    Z_{\fT_J}[C_3\up{I}]\;\xmapsto{T_5(M_5)}\;  &Z_{\fT_I}[C_3\up{K}]\;\;\xmapsto{\tau\sigma(M_5)}\; \chi^{-1/2}\,Z_{\fT_J}[C_3\up{I}] 
\end{align}
$\cT_5(M_5)$ is thus a defect in a given absolute theory with polarization pair $\lb L_J,\,\ol{L}_I\rb$
and, as we shall see below, obeys non-invertible fusion rules.
Recalling the fact that ${\GS3}$ anti-cyclically permutes $I,J,K$ and the definition of the $\tau\sigma(M_5)$ triality interface \eqref{eq:tau_sigma_6d_fig}, the triality defect $\cT_5(M_5)$ can be depicted as follows:
\begin{equation} \label{eq:cT5_6d_fig}
      \begin{split}
    &\quad{\color{red}\cT_5(M_5)}\\
    \cL_{\fT_J}[C_3\up{I}]  &{\color{red} \qquad\bigg|_{c_3\up{K}|_{M_{5}}=0}} \quad \cL_{\fT_I}[c_3\up{K}] \,+\pi\lb  \bar{q}(c_3\up{K})+c_3\up{K}\cup C_3\up{I}\rb
    \end{split}
\end{equation}
with corresponding equation:
\begin{align} \label{eq:cT5_6d_eq}
    Z_{\fT_J}[C_3\up{I};\,\cT_5(M_5)]=N\dw{M_6\gz,\,\Z_2}\sum_{c_3\up{K}\in H^3(M_6\gz,\,M_{5|0},\,\Z_2)} Z_{\fT_I}[c_3\up{K}] \,\exp(i\pi\int_{M_6\gz}\lb  \bar{q}(c_3\up{K})+c_3\up{K}\cup C_3\up{I}\rb)
\end{align}
with the normalization factor $N\dw{M_6\gz,\,\Z_2}$ defined in equation \eqref{eq:N_Z2_M6gz}.\\

\paragraph{Gauging and stacking interface $\boldsymbol{\sigma\tau(M_5)}$.}
A codimension-1 interface $\sigma\tau(M_5)$ between the 6d absolute theories with polarization pairs $\lb L_J,\,\ol{L}_I\rb$ and $\lb L_I,\,\ol{L}_K\rb$ can be defined by performing the $\sigma\tau$ operation \eqref{eq:sigma_tau_6d} in half of spacetime and imposing the Dirichlet boundary condition for the dynamical gauge field $c_3\up{I}$ on the interface $M_5$. We depict it as follows:
\begin{equation} \label{eq:sigma_tau_6d_fig}
     \begin{split}
    &\quad{\color{red}\sigma\tau(M_5)}\\
    \cL_{\fT_J}[C_3\up{I}]  &{\color{red} \qquad\bigg|_{c_3\up{I}|_{M_{5}}=0}} \quad  \cL_{\fT_J}[c_3\up{I}] \,+\pi\lb c_3\up{I}\cup C_3\up{K}+\bar{q}(C_3\up{K})\rb
\end{split}
\end{equation}
and write it in terms of partition functions as:
\begin{align} \label{eq:sigma_tau_6d_eq}
    &\quad{\color{red}\sigma\tau(M_5)}\\
     Z_{\fT_J}[C_3\up{I}] &{\color{red} \qquad\bigg|_{c_3\up{I}|_{M_{5}}=0}} \quad N\dw{M_6\gz,\,\Z_2}\sum_{c_3\up{I}\in H^3(M_6\gz,\,M_{5|0},\,\Z_2)} Z_{\fT_J}[c_3\up{I}] \,\exp(i\pi\int_{M_6\gz}\lb c_3\up{I}\cup C_3\up{K}+\bar{q}(C_3\up{K})\rb).
\end{align}

\paragraph{Triality defect $\boldsymbol{\ol{\cT}_5(M_5)}$.}  By combining a $\ol{\GS3}$ automorphism and a triality interface $\sigma\tau(M_5)$ we can construct another non-invertible triality defect $\ol{\cT}_5(M_5)$ in a given absolute theory with polarization pair $\lb L_J,\,\ol{L}_I\rb$
\begin{align} \label{eq:cT5bar_def_fus}
    \ol{\cT}_5(M_5)=\ol{T}_5(M_5)\otimes \sigma\tau(M_5) \,.
\end{align}
First the $\ol{\GS3}$ automorphism performs a cyclic permutation of $I,J,K$ and then the $\sigma\tau(M_5)$ interface implements gauging of $\Z_2\up{J}$ and stacking with an SPT for the new $\Z_2\up{I}$ global 2-form symmetry:
\begin{align}
     \lb L_J,\,\ol{L}_I\rb \;\xrightarrow{\ol{T}_5(M_5)}\; &\lb L_K,\,\ol{L}_J\rb \;\xrightarrow{\sigma\tau(M_5)}\; \lb L_J,\,\ol{L}_I\rb \,,\nn\\[2mm]
    Z_{\fT_J}[C_3\up{I}]\;\xmapsto{\ol{T}_5(M_5)}\;  &Z_{\fT_K}[C_3\up{J}]\;\;\xmapsto{\sigma\tau(M_5)}\; \chi^{-1/2}\,Z_{\fT_J}[C_3\up{I}] \,.
\end{align}
$\ol{\cT}_5(M_5)$ is thus a defect in a given absolute theory with polarization pair $\lb L_J,\,\ol{L}_I\rb$
and, as we shall see below, obeys non-invertible fusion rules.
Recalling the fact that $\ol{\GS3}$ cyclically permutes $I,J,K$ and the definition of the $\sigma\tau(M_5)$ triality interface \eqref{eq:sigma_tau_6d_fig} the action of the triality defect $\ol{\cT}_5(M_5)$ can be depicted as:
\begin{equation} \label{eq:cT5bar_6d_fig}
      \begin{split}
    &\quad{\color{red}\ol{\cT}_5(M_5)}\\
    \cL_{\fT_J}[C_3\up{I}]  &{\color{red} \qquad\bigg|_{c_3\up{J}|_{M_{5}}=0}} \quad  \cL_{\fT_K}[c_3\up{J}] \,+\pi\lb c_3\up{J}\cup C_3\up{I}+\bar{q}(C_3\up{I})\rb
    \end{split}
\end{equation}
with corresponding equation:
\begin{align} \label{eq:cT5bar_6d_eq}
    Z_{\fT_J}[C_3\up{I};\,\ol{\cT}_5(M_5)]=N\dw{M_6\gz,\,\Z_2}\sum_{c_3\up{J}\in H^3(M_6\gz,\,M_{5|0},\,\Z_2)} Z_{\fT_K}[c_3\up{J}] \,\exp(i\pi\int_{M_6\gz}\lb c_3\up{J}\cup C_3\up{I}+\bar{q}(C_3\up{I})\rb)\,,
\end{align}
where the normalization factor $N\dw{M_6\gz,\,\Z_2}$ is defined in equation \eqref{eq:N_Z2_M6gz}.

The fusion of the triality defects is derived in appendix \ref{app:Twix}, and summarized in the next subsection, alongside the duality and condensation defect fusions.

\subsection{Summary of the $S_3$-ality Defects and Fusions}
\label{sec:Bounty}

After having defined the duality and triality defects, the remaining structure in this fusion 5-category is to compute the fusions -- at least at the level of objects. These are provided in detail in appendix \ref{app:Twix}. Here we simply summarize them. 

The  5d topological defects, that form the simple objects of this fusion 5-category are: 
\begin{itemize}
\item duality defect $\mathcal{D}_5(M_5)$ defined in (\ref{eq:cD5_6d_fig}), 
\item triality defects $\mathcal{T}_5(M_5)$ and $\overline{\mathcal{T}_5} (M_5)$, defined in (\ref{eq:cT5_6d_fig}) and (\ref{eq:cT5bar_6d_fig}). 
\end{itemize}
The objects in this higher category are obtained by taking fusions of these. 
The lower dimensional topological defects are $k$-morphisms in terms of the category. In the present case we also have 
$D_3^{(J)}$, which generate a $\Z_2^{(I)}$ 2-form symmetry and are invertible $D_3^{(J)} \otimes D_3^{(J)}= D_3^{(\id)}$. In terms of the fusion 5-category they are 2-morphisms. In particular we will need to include the condensation defects of these, which are denoted by $\cC_5$.

Note that further topological defects -- invertible and non-invertible -- can be obtained completely analogously by starting with a fixed absolute theory in figure \ref{fig:GS2_GS3}, and following the arrows in a closed loop. If this involves GS automorphisms and stacking, i.e. $\tau$, only, the defect is invertible, if closing the loop involves also  gauging, then it is non-invertible. The computation of fusions is entirely analogous to the ones we provided in detail. 
{These are $D_5^{(J)} \, \tau$ and $D_5^{(I)} \, \tau \sigma \tau$.} {The former, invertible defect, appears as the $e^{i\pi\int_{M_6\gz}\bar{q}(C_3\up{I})}$ factor in equations \eqref{eq:6d_DT_res1}-\eqref{eq:6d_DT_res2}, whereas the latter, non-invertible defect is what appears on both sides of equation \eqref{eq:6d_DT_res3}: the concrete expression for its insertion in the partition function is \eqref{eq:same1}.}

\paragraph{Condensation Defects and TQFT Coefficients.}
A central type of defect in any fusion higher category are condensation defects. For this particular theory we will have condensation defects of the 2-form symmetry 
\be\ba
    \cC_5\up{0}(M_5)&=\chiiM{\gz}\;N\dw{M_5,\,\Z_2}\sum_{M_3\in H_3(M_5,\,\Z_2)}\,\exp(i\pi\int_{M_3}C_3\up{I})\,,\\
    \cC_5\up{1}(M_5)&= \chiiM{\gz}\,N\dw{M_5,\,\Z_2}\sum_{M_3\in H_3(M_5,\,\Z_2)}\,\exp(i\pi\,Q(M_5,M_3)+i\pi\int_{M_3}C_3\up{I})\,,
\ea\ee
where\footnote{Here $\beta:H^2(M_5,\,\Z_2)\to H^{3}(M_5,\,\Z_2)$ is the Bockstein homomorphism associated to the short exact sequence $$1\to\Z_2\to\Z_4\to\Z_2\to1.$$We denote the Poincaré dual of $M_3\in H_3(M_5,\Z_2)$ by $\PD(M_3)\in H^2(M_5,\Z_2)$.}
\be
 Q(M_5,M_3)=\int_{M_5}\PD(M_3)\cup\beta(\PD(M_3)) \,.
\ee
Furthermore, some fusions will have TQFT coefficients which we abbreviate by
\be\ba \label{eq:ZY_def}
     \cZ\dw{Y,M_6\gz}&=\chiM{\gz}\,N\dw{M_6\gz,\,\Z_2}\,\sum_{c_3\in H^3(M_6\gz,\,M_5,\,\Z_2)}\,e^{i\pi\, \int_{M_6\gz} \bar{q}(c_3)}\cr 
    (\cZ_2)_{2\ell}\up{I}(M_5)&= \chiiM{\gz}
    \,N\dw{M_5,\,\Z_2}^2\,\sum_{b_2\up{I},\,b_2\up{J}\in H^2(M_5,\,\Z_2)}\exp(i\pi \int_{M_5}\ell \,b_2\up{I}\cup\beta(b_2\up{I})+b_2\up{J}\cup(\delta\,b_2\up{I}-C_3\up{I}))\,.
\ea\ee
for $\ell\in\{0,1\}$.

We now summarize all the fusions of duality and triality defects. It is clear from these that the defects are non-invertible, as e.g. the duality squares to a condensation defect (not the identity). 
The duality and triality defects fused among themselves just generalize the known duality and triality fusions. They will be relevant equally for the theories discussed in section \ref{sec:Heroes}.

Furthermore, both duality and triality defects are symmetries of the theory, so that we can also fuse them and generate the full symmetry category, which then has generically $S_3$-ality defects -- they arise from the $S_3=\Z_3\rtimes \Z_2$ invertible defects dressed by the stacking with SPT and gauging operation. Concretely we will see the full structure of the $S_3$-ality defects from the fusion of duality with triality defects.

\paragraph{Duality Defect Fusions.}
\begin{align} \label{eq:6d_DD_res}
     \cD_5(M_5)\otimes \cD_5(M_5)&=\cC_5\up{0}(M_5)\,.
\end{align}

\paragraph{Triality Defect Fusions.}
\begin{align}
    \ol{\cT}_5(M_5)\otimes \cT_5(M_5)&=\cC_5\up{0}(M_5)\,, \label{eq:6d_TT1}\\
     \cT_5(M_5)\otimes \ol{\cT}_5(M_5)&= \cC_5\up{1}(M_5)\,,\\
      \cT_5(M_5)\otimes \cT_5(M_5)&=\cZ\dw{Y,M_6\gz}\, \cC_5\up{1}(M_5)\otimes \ol{\cT}_5(M_5)\,,\\
      \ol{\cT}_5(M_5)\otimes \ol{\cT}_5(M_5)&=
    \cZ\dw{Y,M_6\gz}\, \cC_5\up{0}(M_5)\otimes{\cT}_5(M_5)\,. \label{eq:6d_TT4}
\end{align}

\paragraph{Fusions between Duality and Triality 
Defects: $S_3$-ality Structure}
\begin{align} 
   \cD_5(M_5)\otimes \ol{\cT}_5(M_5)&=\cC_5\up{0}\;(M_5)\;e^{i\pi\int_{M_6\gz}\bar{q}(C_3\up{I})}\,, \label{eq:6d_DT_res1}\\
    \cT_5(M_5)\otimes \cD_5(M_5)&=\cC_5\up{1}(M_5)\;e^{i\pi\int_{M_6\gz}\bar{q}(C_3\up{I})}\,,
    \label{eq:6d_DT_res2}\\
     \cD_5(M_5)\otimes {\cT}_5(M_5)&=\ol{\cT}_5(M_5)\otimes \cD_5(M_5)\,. \label{eq:6d_DT_res3}
\end{align}
The last equation is precisely the relation expected from the $S_3$-ality nature of these defects.

\paragraph{Condensation Defect Fusions.}
For completeness we compute the fusion of condensation defects as well. 
Let $\cX_5(M_5)\in\{\cD_5(M_5),\cT_5(M_5),\ol{\cT}_5(M_5)\}$, the resulting fusions are:
\begin{align}
   \cC_5\up{\ell}(M_5) \otimes \cX_5(M_5)&=(\cZ_2)_{2\ell}\up{I}(M_5)\;\cX_5(M_5)\,,\\
   \cX_5(M_5)\otimes \cC_5\up{\ell}(M_5)&=(\cZ_2)_{2 \delta_{(\cX,\cT)}}\up{I}(M_5)\;\cX_5(M_5)\,,\\
   \cC_5\up{\ell}(M_5)\otimes \cC_5\up{\ell'}(M_5)&=(\cZ_2)_{2\ell}\up{I}(M_5)\;\cC_5\up{\ell'}(M_5)\,.
\end{align}
{These equations show closure of the codimension-one sector considered in this paper, generated by $\cD_5,\cT_5,\ol{\cT}_5$ and $\cC_5\up{0,1}$, up to the TQFT coefficients displayed above. We do not claim here to have established Karoubi or condensation completeness of the full fusion 5-category, nor the complete action on all operators and extended defects of the SCFT.}
We should note that a complete description of the full fusion 5-category is beyond the scope of this paper -- even for fusion 3-categories the full structure is currently not entirely worked out. It should be possible to generalize the mathematical analysis of  \cite{Bhardwaj:2024xcx}  for fusion 3-categories to the present case.

\section{SymTFT Perspective on $S_3$-ality Defects}
\label{sec:SymTFT}

An alternative way to construct duality and triality defects is using the SymTFT approach: we consider the 7d 3-form Chern-Simons theory with Lie algebra $\mathfrak{g}=\so(8)$. The $S_3$ GS automorphisms act as outer automorphisms of the SymTFT. We construct the condensation defects implementing these invertible GS transformations and then the non-invertible twist defects: combining them we obtain duality, triality, and $S_3$-ality defects in the boundary 6d theory.

\subsection{SymTFT for the 6d \tpdf{$\mathcal{N}=(2,0)$}{N2} \tpdf{$\mathfrak{so}(8)$}{so8} theory} \label{sec:SymTFT_1}

\subsubsection{7d Chern-Simons Theory} 
As reviewed in section \ref{sec:twix}, before specifying a polarization, 6d SCFTs are in general relative theories, coupled to a 7d bulk. The action for a 3-form Chern-Simons (CS) theory (in 7d) depends on a $K$-matrix and takes the form \cite{Witten:1998wy,Monnier:2017klz,Heckman:2017uxe,Gukov:2020btk, Tian:2024dgl}
\begin{equation} \label{eq:7d_CS_K}
    S_{\text{CS}(K)}=\frac{1}{4\pi} \int_{M_7}\,\sum_{I,J}\,K_{IJ}\;c_3\up{I}\wedge\dd c_3\up{J}\ \,,\quad I,J\in\{1,...,r\} \,,
\end{equation}
where $c_3\up{I},$ for $I\in\{1,...,r\}$ are 3-form $U(1)$ gauge fields on $M_7$ and $K$ is an integral, positive definite, symmetric matrix. For 6d SCFTs with $\cN=(2,0)$ of ADE type, $K$ is the Cartan matrix of the ADE Lie algebra $\mf{g}$.\footnote{This is reminiscent of the 3d Chern-Simons theories describing Abelian topological phases characterized by ADE Lie algebras\cite{PhysRevB.90.235149,Teo:2015xla}.}

\paragraph{Topological operators.} The 7d Chern-Simons theory \eqref{eq:7d_CS_K} has topological
operators of dimension 3 given by the holonomy of the 3-form gauge fields \cite{Witten:1998wy,Gaiotto:2014kfa,Gukov:2020btk}. 
\begin{equation} \label{eq:UI}
    U_3\up{I}(M_3)=\exp(i\int_{M_3}c_3\up{I})
\end{equation}
with fusions corresponding to the defect group of the 6d theory 
\begin{equation}
    \bD=\Z^r/K\Z^r\cong \cZ(\Tilde{G})
\end{equation}
which is isomorphic to the center of the simply-connected group $\Tilde{G}$ with ADE Lie algebra $\mf{g}$ whose Cartan matrix is $K$. Properties of the operators
\begin{equation}
    U_3\up{\lambda_I}(M_3)=\exp(i \int_{M_3}\lambda_I\,c_3\up{I})\,\qquad\quad \lambda_I\in\bD
\end{equation}
can be determined from a quadratic function $s$, written in terms of the Cartan matrix $K$: the ``spins" $s(\lambda)$ and braidings $\langle \lambda,\lambda'\rangle$ are \cite{Gukov:2020btk}:
\begin{align} \label{eq:spin_braiding_CS}
    s(\lambda)&=\tfrac{1}{2}\lambda^T\,K^{-1}\,\lambda \,,\\
    \langle \lambda,\,\lambda'\rangle&=\exp[-2\pi i \lb s(\lambda+\lambda')-s(\lambda)-s(\lambda')\rb ]\,.
\end{align}

\paragraph{$K_{\mathfrak{so}(8)}$ Chern-Simons Theory.} 
The 7d CS theory of interest is the one with $K$-matrix given by the Cartan matrix of $\mf{g}=\mf{so}(8)$ 
\begin{equation}
    K_{\mathfrak{so}(8)}= \text{CM}_{\mathfrak{so}(8)}= 
    \begin{pmatrix}
        2 & -1 & -1 & -1 \\
        -1 & 2 & 0 & 0 \\
        -1 & 0 & 2 & 0 \\
        -1 & 0 & 0  & 2 
    \end{pmatrix}
\end{equation}
The 3-dimensional operators \eqref{eq:UI} satisfy \cite{Gukov:2020btk}:
\begin{align}
    U_3\up{1}=U_3\up{\id},\qquad U_3\up{I}U_3\up{J}=U_3\up{K},\qquad I,J,K\in\{S,C,V\}
\end{align}
indeed the defect group for the 6d $\cN=(2,0)\;\mf{so}(8)$ theory is $\bD=\Z_2\times\Z_2$. Their ``spins" follow from \eqref{eq:spin_braiding_CS} and are:
\begin{equation} \label{eq:spin1/2}
    s(U_3\up{1})=0,\qquad s(U_3\up{S})=s(U_3\up{C})=s(U_3\up{V})=\frac{1}{2}\,.
\end{equation}
The braiding can be expressed by the commutation relation:
\begin{equation}  \label{eq:braid}
    U_3\up{I}(M_3)U_3\up{J}(M_3')=(-1)^{\langle M_3,M_3'\rangle}\, U_3^{(J)}(M_3')\,U_3^{(I)}(M_3),\quad I\neq J\in\{S,C,V\}.
\end{equation}
showing that in any basis the bilinear pairing on $\bD$ is non-trivial only between the two different $\Z_2$ factors.
The 3-form Chern-Simons action \eqref{eq:7d_CS_K} whose $K$-matrix is the Cartan matrix of $\mathfrak{so}(8)$ is \cite{Gukov:2020btk}:
\begin{equation} \label{eq:7dCSso8}
    S_{\text{CS} (K_{\mathfrak{so}(8)})}=\int_{M_7}\left[\frac{2}{4\pi}\lb c_3\up{1}\wedge\dd c_3\up{1}+c_3\up{S}\wedge\dd c_3\up{S}+c_3\up{C}\wedge\dd c_3\up{C}+c_3\up{V}\wedge\dd c_3\up{V}\rb -\frac{1}{2\pi}c_3\up{1}\wedge\dd(c_3\up{S}+c_3\up{C}+c_3\up{V})\right]
\end{equation}
which has an $S_3$ symmetry permuting $c_3\up{S},c_3\up{C},c_3\up{V}$. We note that, from equation \eqref{eq:spin1/2}, the operators $U_3\up{S},U_3\up{C},U_3\up{V}$  all have ``spin" $\frac{1}{2}$ so it is consistent to permute them.

The equations of motion, in $\R/2\pi\Z$, which follow from the action \eqref{eq:7dCSso8} are:
\begin{equation}
    \begin{cases}
        \;2\,c_3\up{1}-c_3\up{S}-c_3\up{C}-c_3\up{V}=0 \\[1mm]
        \;2\,c_3\up{S}-c_3\up{1}=0 \\[1mm]
        \;2\,c_3\up{C}-c_3\up{1}=0 \\[1mm]
        \;2\,c_3\up{V}-c_3\up{1}=0 \\[1mm]
    \end{cases}
\end{equation}
which imply (in $\R/2\pi\Z$)
\begin{equation} \label{eq:CinZ2}
    \begin{cases}
        c_3\up{1}=0\\[2mm]
        2\,c_3\up{S}=2\,c_3\up{C}=2\,c_3\up{V}=0 
    \end{cases}
\end{equation}
in addition to 
\begin{equation} \label{eq:Csum}
    c_3\up{I}=c_3\up{J}+c_3\up{K},
\end{equation}
where $(I,J,K)$ is any permutation of $(S,C,V)$. Using the above equation, one can eliminate either $c_3\up{S}$ or $c_3\up{C}$ or $c_3\up{V}$ when writing the on-shell actions: these are three equivalent choices as a consequence of the $S_3$ permutation symmetry of \eqref{eq:7dCSso8}. 

There are three possible boundary conditions which are obtained by imposing Dirichlet boundary conditions on a gauge field, i.e. fixing it to a classical background field on the boundary.
They correspond to each one of the three choices in \eqref{eq:Csum} evaluated on a 6d boundary of the 7d bulk:
\begin{equation} \label{eq:bc}
    \fB\up{I}:\qquad c_3\up{I}|_{\fB\up{I}}=(c_3\up{J}+c_3\up{K})|_{\fB\up{I}}=C_3\up{I}, \\
\end{equation}
where, again, $(I,J,K)$ is any permutation of $(S,C,V)$ and we use uppercase letters for non-dynamical background fields. 

Equation \eqref{eq:CinZ2} implies that we can use $\Z_2$-valued gauge fields, which are related to their $U(1)$-valued counterparts by a factor of $\pi$ as follows: $c_{3,U(1)}=\pi \, c_{3,\Z_2}$. If we then use the boundary condition \eqref{eq:bc}, we can write the boundary action in terms of ($\Z_2$-valued) gauge fields as:\footnote{Our conventions for the quadratic refinement of the intersection pairing are reviewed in subsection \ref{sec:twisted_gauging}.}
\begin{equation} \label{eq:S_boundary}
    S|_{\fB\up{I}}=\frac{\pi}{2}\int_{\fB\up{I}}\,\lb\bar{q}(c_3\up{J})+\bar{q}(C_3\up{I})\rb\qquad\text{or}\qquad S|_{\fB\up{I}}=\frac{\pi}{2}\int_{\fB\up{I}}\,\lb\bar{q}(c_3\up{K})+\bar{q}(C_3\up{I})\rb.\\
\end{equation} 
The case in which $C_3\up{I}=0$ was discussed in \cite{Gukov:2020btk}.

\paragraph{Alternative presentation of the SymTFT.} The $\Z_2$ 3-form Dijkgraaf-Witten (DW) theory
\begin{equation} \label{eq:Z2_DW_spin0}
    \pi\int_{M_7} c_3\up{I}\cup\delta c_3\up{J},
\end{equation}
where $c_3\up{I},c_3\up{J}$ for $I\neq J\in\{S,C,V\}$ are $\Z_2$-valued 3-forms, has 4 dimension-3 operators 
\be
\ba
     U_3\up{\id}(M_3)&, &\quad U_3\up{I}(M_3)&=\exp(i\pi\int_{M_3}c_3\up{I}),\\
     U_3\up{J}(M_3)&=\exp(i\pi\int_{M_3}c_3\up{J}), &\quad U_3\up{IJ}(M_3)&=\exp(i\pi\int_{M_3}(c_3\up{I}+c_3\up{J}))
\ea
\ee
whose ``spins" are
\begin{equation}
    s(U_3\up{\id})=s(U_3\up{I})=s(U_3\up{J})=0,\qquad s(U_3\up{IJ})=\frac{1}{2},
\end{equation}
which differ from those in the 7d CS($K_{\so(8)}$) theory \eqref{eq:spin1/2}. To obtain this latter theory starting from the $\Z_2$ DW \eqref{eq:Z2_DW_spin0} one can add a coupling between $(c_3\up{I}+c_3\up{J})$ and the 4th Wu class $v_4$:\footnote{It can be written in terms of the Stiefel–Whitney classes $w_i$ of the tangent bundle on orientable manifolds as: $v_4=w_4+w_2^2$ \cite{Gukov:2020btk}.}
\begin{equation} \label{eq:SIJ}
    S\up{I,J}=\pi\int_{M_7}\lb c_3\up{I}\cup\delta c_3\up{J}+(c_3\up{I}+c_3\up{J})\cup v_4\rb.
\end{equation}
The coupling to $v_4$ changes the ``spins" of $U_3\up{I}$ and $U_3\up{J}$ to $\frac{1}{2}$:
\begin{equation}
    s(U_3\up{\id})=0,\qquad s(U_3\up{I})=s(U_3\up{J})=s(U_3\up{IJ})=\frac{1}{2},
\end{equation}
matching those of the 7d CS($K_{\so(8)}$), \eqref{eq:spin1/2}.
The braidings and fusions also match those of the 7d CS($K_{\so(8)}$), so \eqref{eq:SIJ} is an alternative presentation of the theory, as discussed in Appendix F of \cite{Gukov:2020btk}.

\subsubsection{GS-duality Web and Topological Manipulations}

To construct the defects in the SymTFT picture we need to map the dualities and topological manipulations into the language of background fields for the two-form symmetries. 

\begin{figure}
\centering
\begin{tikzpicture}[scale=3]
    \node [style=none] (6) at (-0.866*2, +1/2*2) {$ S\up{V,S}$};
    \node [style=none] (1) at (-0.866, +1/2) {$ S\up{V,C}$};
    \node [style=none] (4) at (+0.866*2, +1/2*2) {$ S\up{S,C}$};
    \node [style=none] (3) at (0, -0.8) {$ S\up{S,V}$};
    \node [style=none] (5) at (0, -0.8*2) {$ S\up{C,V}$};
    \node [style=none] (2) at (+0.866, +1/2) {$ S\up{C,S}$};
    
    \draw [style=blue arrow] (1) [bend left] to (2);
    \draw [style=blue arrow] (2) [bend left] to (3);
    \draw [style=blue arrow] (3) [bend left] to (1);
    \draw [style=blue arrow] (5)  [bend right]to (4);
    \draw [style=blue arrow] (6) [bend right] to (5);
    \draw [style=blue arrow] (4) [bend right] to (6);

    \draw [style=cyan arrow, transform canvas={xshift=1.5mm,yshift=3mm}] (1)  to (6); 
    \draw [style=cyan arrow, transform canvas={xshift=-1.5mm,yshift=3mm}] (2)  to (4);
    \draw [style=cyan arrow] (3)  to (5);

         \draw [style=red arrow] (1) to (6);
	\draw [style=orange arrow] (1) [bend right] to (5);
	\draw [style=red arrow] (5) [bend right] to (2);
	\draw [style=orange arrow] (2) to (4);
	\draw [style=red arrow] (4) [bend left] to (3);
	\draw [style=orange arrow] (3) [bend left] to (6); 
\end{tikzpicture}
\caption{The Green-Schwarz Automorphisms of the SymTFT actions $S\up{I,J}$ defined in equation \eqref{eq:SIJ}, for the 6d $(2,0)$ SCFTs with Lie algebra $\mathfrak{so}(8)$: 
$\GS 2$ is of order 2 (black dashed arrows) and exchanges $c_3^{(S)}\leftrightarrow c_3^{(C)}$, whereas $\GS 3$ is of order 3 and cyclically permutes $c_3^{(S)},\, c_3^{(V)},\,c_3^{(C)}$ (black arrows). 
The topological manipulations are the $\sigma$ operation, eq. \eqref{eq:sigma_SymTFT} in cyan, and the $\tau$ operation, eq. \eqref{eq:tau_SymTFT} in blue arrows. 
This reproduces, in the SymTFT formalism, the web of figure \ref{fig:GS2_GS3}.
The notation is such that the first and second entries refer respectively to Neumann and to Dirichlet boundary conditions for the corresponding gauge fields. }
\label{fig:GS2_GS3_SymTFT}
\end{figure}
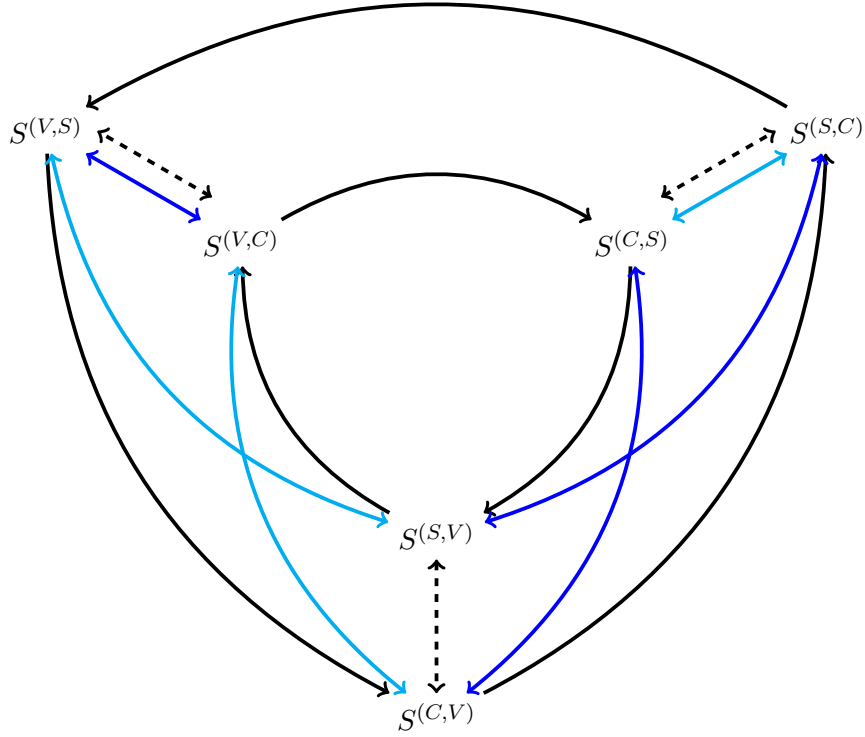

In terms of the background fields  the action of the 
 GS dualities \eqref{eq:GS2_def1}-\eqref{eq:GS3_def1} is 
\begin{subequations}
    \begin{align}
         \GS 2\up{V}:& \qquad c_3^{(S)}\leftrightarrow c_3^{(C)}, \qquad c_3^{(V)}\leftrightarrow c_3^{(V)} \label{eq:GS2V}\\
         \GS 2\up{S}:& \qquad c_3^{(C)}\leftrightarrow c_3^{(V)}, \qquad c_3^{(S)}\leftrightarrow c_3^{(S)} \label{eq:GS2S}\\
         \GS 2\up{C}:& \qquad c_3^{(S)}\leftrightarrow c_3^{(V)}, \qquad c_3^{(C)}\leftrightarrow c_3^{(C)} \label{eq:GS2C}
    \end{align}
\end{subequations}
\vspace{-1cm}
\begin{subequations}
    \begin{align}
         {\GS 3}: &\qquad c_3^{(V)}\to c_3^{(C)}\to c_3^{(S)}\to c_3^{(V)} \label{eq:GS3} \\
         \overline{\GS 3}: &\qquad c_3^{(V)}\to c_3^{(S)}\to c_3^{(C)}\to c_3^{(V)} \label{eq:GS3bar}
    \end{align}
\end{subequations}
of order 2 and 3. We show in figure \ref{fig:GS2_GS3_SymTFT} how the SymTFT actions \eqref{eq:SIJ} are related by the GS automorphisms. 

Furthermore in order to construct the non-invertible symmetries we need to define topological manipulations, which ``undo" the action of the GS dualities. These map the SymTFT actions \eqref{eq:SIJ} as follows:\footnote{The $\sigma$ and $\tau$ operations are reminiscent of the $S$ and $T$ actions on 3d abelian Chern-Simons theories of \cite{Witten:2003ya}.}
\begin{align}
    \sigma: \quad S\up{I,J} \;\;&\mapsto\quad S\up{I,J}+\pi\int_{M_7}  \delta\lb c_3\up{I}\cup c_3\up{J}\rb =S\up{J,I} \label{eq:sigma_SymTFT} \\
    \tau: \quad S\up{I,J} \;\;&\mapsto\quad S\up{I,J}+\pi\int_{M_7} \lb c_3\up{I}\cup \delta c_3\up{I}+c_3\up{I}\cup v_4\rb= S\up{I,K} \label{eq:tau_SymTFT}
\end{align}
where we used the fact that $c_3\up{I}+c_3\up{J}=c_3\up{K}$ for $(I,J,K)$ any permutation of $(S,C,V)$.
It is important to note that, since we are considering $\Z_2$ gauge fields, $2\,c_3^{(I)}=0,\;\forall\;I\in\{S,C,V\}$.
We show in figure \ref{fig:GS2_GS3_SymTFT} how the SymTFT actions  $S\up{I,J}$ (eq. \eqref{eq:SIJ} for $I\neq J\in\{S,C,V\}$) are related by the GS and $\sigma,\tau$ operations.

\subsection{SymTFT Construction of Duality and Triality Defects} \label{sec:SymTFT-defects}

We now construct the topological defects as condensation defects in the SymTFT, which in our case is the 7d 3-form Chern-Simons theory $\text{CS} (K_{\mathfrak{so}(8)})$, equation \eqref{eq:7dCSso8}. Subsequently, we will construct the twist defects and push them to the symmetry boundaries to obtain the symmetries of the absolute theories. {It is also possible to construct the codimension-one topological defect in the 7d $\mathbb{Z}_2$ DW theory, where the $S_3$ symmetry is not manifest, see appendix \ref{app:bosint} for the details of this construction.}

\subsubsection{Invertible Topological Defects}

\paragraph{Defects of dimension 3.}
There are 4 dimension-3 invertible topological defects, which we denote as\footnote{These are reminiscent of the line defects in the 3d $\Spin(8)$ theory \cite{Bhardwaj:2022yxj,PhysRevB.90.235149,Teo:2015xla,BoyleSmith:2024qgx}.}
\be
\ba
        D_3^{(\id)}(M_3)&,
        &\qquad D_3^{(S)}(M_3)&=\exp\lb i\pi\int_{M_3} c_3^{(S)}\rb ,\\
        D_3^{(C)}(M_3)&=\exp\lb i\pi\int_{M_3} c_3^{(C)}\rb ,  
        &\qquad D_3^{(V)}(M_3)&=\exp\lb i\pi\int_{M_3} c_3^{(V)}\rb  \,,
\ea
\ee
which satisfy the fusions
\begin{align}
    D_3^{(I)}(M_3)\otimes D_3^{(I)}(M_3)&=D_3^{(\id)}(M_3), \quad I\in\{S,C,V\} \,,\\[1mm]
    D_3^{(S)}(M_3)\otimes D_3^{(C)}(M_3) \otimes D_3^{(V)}(M_3)&=D_3^{(\id)}(M_3), \label{eq:D3_fusions}
\end{align}
and exchange relations 
\begin{equation}
     D_3^{(I)}(M_3) \, D_3^{(J)}(M_3')=(-1)^{\langle M_3,M_3'\rangle}\, D_3^{(J)}(M_3')\,D_3^{(I)}(M_3),\quad I\neq J\in\{S,C,V\} \,, \label{eq:comm_rel}
\end{equation}
equivalent to \eqref{eq:braid}. Furthermore, 
\begin{equation}
    D_3^{(I)}(M_3+M_3')=(-1)^{\langle M_3,M_3'\rangle}\,D_3^{(I)}(M_3)\,D_3^{(I)}(M_3'),\quad I\in\{S,C,V\} \,. \label{eq:quantum_torus}
\end{equation}

\paragraph{Codimension-1 GS2 defects.} Condensation defects are constructed by higher-gauging a subgroup of the 
higher-form symmetry on a submanifold \cite{Roumpedakis:2022aik}. By higher-gauging a $\Z_2$ subgroup of the $\Z_2^{(S)}\times\Z_2^{(C)}$ 3-form symmetry of the SymTFT on a codimension-1 manifold $M_6$, on which it is a 2-form symmetry, we obtain: 
\begin{equation} \label{eq:D6I_def}
    D_6^{(I)}(M_6)= N\dw{M_6,\,\Z_2}\sum_{M_3\in H_3(M_6,\,\Z_2)}D_3^{(I)}(M_3), \quad I\in\{S,C,V\} \,,
\end{equation}
where the normalization factor $ N\dw{M_6,\,\Z_2}$ was defined in \eqref{eq:N_Z2_M6}.

When crossing $D_6^{(I)}(M_6)$, the 3-dimensional defect $D_3^{(I)}(M_3')$ is left unchanged:
\begin{align}
     D_3^{(I)}(M_3')\,D_6^{(I)}(M_6)&= N\dw{M_6,\,\Z_2}\sum_{M_3\in H_3(M_6,\,\Z_2)} D_3^{(I)}(M_3')\,D_3^{(I)}(M_3)= \nn\\
     &= N\dw{M_6,\,\Z_2}\sum_{M_3\in H_3(M_6,\,\Z_2)} D_3^{(I)}(M_3)\,D_3^{(I)}(M_3')= \nn\\
     &=D_6^{(I)}(M_6)\,D_3^{(I)}(M_3')\,. \label{eq:D3I_D6I}
\end{align}

\begin{figure}
\centering
    \begin{tikzpicture}
    \begin{scope}
        \draw [purple, thick, fill=purple, opacity=1]
    (0,-1.5) -- (0,4) -- (1.5,5) -- (1.5,-0.5) -- (0,-1.5);
     \draw [purple, thick, fill=purple, opacity=1]
     (0,-1.5) -- (0,4) -- (1.5,5) -- (1.5,-0.5) -- (0,-1.5);
     \draw [blue, thick] (-2.3,3) -- (0, 3) ;
     \draw [blue, thick, dashed] (0, 3) -- (0.7,3) ;
     \draw [blue, thick] (0.7,3) -- (3.6, 3) ;
    \node [blue] at (-1.5, 3.4) {$D_3^{(J)}(M_3')$} ;
    \node[purple] at (1.3,5.4) {$D_6^{(J)}(M_6)$};
    \node [blue] at (3, 3.4) {$D_3^{(J)}(M_3')$} ;
    \draw [black,fill=black] (0.7,3) ellipse (0.03 and 0.03);
      \draw [blue, thick] (-2.3,1.5) -- (0, 1.5) ;
     \draw [blue, thick, dashed] (0, 1.5) -- (0.7,1.5) ;
     \draw [blue, thick] (0.7,1.5) -- (3.6, 1.5) ;
    \node [blue] at (-1.5, 1.9) {$D_3^{(I)}(M_3')$} ;
    \node [blue] at (3, 1.9) {$D_3^{(K)}(M_3')$} ;
    \draw [black,fill=black] (0.7,1.5) ellipse (0.03 and 0.03);
      \draw [blue, thick] (-2.3,0) -- (0,0) ;
     \draw [blue, thick, dashed] (0, 0) -- (0.7,0) ;
     \draw [blue, thick] (0.7,0) -- (3.6, 0) ;
    \node [blue] at (-1.5, 0.4) {$D_3^{(K)}(M_3')$} ;
    \node [blue] at (3, 0.4) {$D_3^{(I)}(M_3')$} ;
    \draw [black,fill=black] (0.7,0) ellipse (0.03 and 0.03);
    \end{scope}
\end{tikzpicture}
\caption{Action of the invertible, codimension-1, duality defects $D_6^{(J)}(M_6)$, on the dimension-3 defects $D_3^{(I)}(M_3')$, as computed in eqs. \eqref{eq:D3I_D6I} and \eqref{eq:D3I_D6J}. $(I,J,K)$ is any permutation of $(S,C,V)$.} \label{fig:D3_D6}
\end{figure}
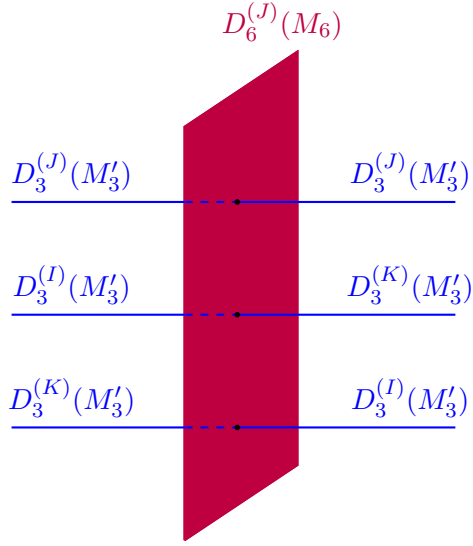

The action of $D_6^{(J)}(M_6)$ on the dimension-3 defects $D_3^{(I)}(M_3')$ for $I\neq J$ can be computed using \eqref{eq:D3_fusions}-\eqref{eq:quantum_torus}.
\begin{align}
    D_3^{(I)}(M_3')\,D_6^{(J)}(M_6)&= N\dw{M_6,\,\Z_2}\sum_{M_3\in H_3(M_6,\,\Z_2)} D_3^{(I)}(M_3')\,D_3^{(J)}(M_3)= \nn\\
        &= N\dw{M_6,\,\Z_2}\sum_{M_3\in H_3(M_6,\,\Z_2)} (-1)^{\langle M_3',M_3\rangle}D_3^{(J)}(M_3)\,D_3^{(I)}(M_3')=\nn\\
     &= N\dw{M_6,\,\Z_2}\sum_{M_3\in H_3(M_6,\,\Z_2)} (-1)^{\langle M_3',M_3\rangle}D_3^{(J)}(M_3)\,D_3^{(J)}(M_3')\,D_3^{(K)}(M_3')=\nn\\
     &= N\dw{M_6,\,\Z_2}\sum_{M_3\in H_3(M_6,\,\Z_2)} D_3^{(J)}(M_3+M_3')\,D_3^{(K)}(M_3')=\nn\\
    &= N\dw{M_6,\,\Z_2}\sum_{\Tilde{M}_3\in H_3(M_6,\,\Z_2)} D_3^{(J)}(\Tilde{M}_3)\,D_3^{(K)}(M_3')=\nn\\ 
    &=D_6^{(J)}(M_6)\,D_3^{(K)}(M_3') \label{eq:D3I_D6J}
\end{align}
where we used
\begin{equation}
    D_3^{(K)}(M_3')=D_3^{(J)}(M_3')\otimes D_3^{(I)}(M_3')\,.
\end{equation}
and replaced the sum over $M_3$ with one over $\Tilde{M_3}\equiv M_3+M_3'$.\\ 
Equations \eqref{eq:D3I_D6I} and \eqref{eq:D3I_D6J} are depicted in figure \ref{fig:D3_D6}.

\paragraph{Fusion of two GS2 defects.}
We have thus determined that the $D_6^{(I)}(M_6)$ defects, where $I\in\{S,C,V\}$ implement the order-2 Green-Schwarz automorphisms $\GS 2\up{I}$, equations \eqref{eq:GS2V}-\eqref{eq:GS2C}. Correspondingly, let us show that $D_6^{(I)}(M_6)$ is an invertible defect of order 2.  
\begin{align}
    D_6^{(I)}(M_6)\otimes D_6^{(I)}(M_6)&=\lb  N\dw{M_6,\,\Z_2}\rb ^2\sum_{M_3,M_3'\in H_3(M_6,\,\Z_2)} D_3^{(I)}(M_3)\, D_3^{(I)}(M_3')=\nn\\
    &=\lb  N\dw{M_6,\,\Z_2}\rb ^2\sum_{M_3,M_3'\in H_3(M_6,\,\Z_2)}(-1)^{\langle M_3,M_3'\rangle} D_3^{(I)}(M_3+M_3')=\nn\\
    &=\lb  N\dw{M_6,\,\Z_2}\rb ^2\sum_{M_3,\Tilde{M}_3'\in H_3(M_6,\,\Z_2)}(-1)^{\langle M_3, \Tilde{M}_3'\rangle} D_3^{(I)}(\Tilde{M}_3')=\nn\\
    &=\lb  N\dw{M_6,\,\Z_2}\rb ^2\sum_{\substack{\PD(M_3),\PD(\Tilde{M}_3')\\ \in H^3(M_6,\,\Z_2)}}\exp(i\pi\int_{M_6}\lb \PD(M_3)+c_3\up{I}\rb \cup \PD(\Tilde{M}_3'))
\end{align}
where we used \eqref{eq:quantum_torus} and then wrote the sum over $M_3,M_3'$ as a sum over their Poincaré duals $\PD(M_3),\PD(M_3')$. Integrating out $\PD(\Tilde{M}_3')$ sets $\PD(M_3)=-c_3^{(I)}$ (trivializing the integrand) and produces a factor of $|H^3(M_6,\,\Z_2)|$. \\
Recalling the definition of the normalization factor \eqref{eq:N_Z2_M6} and \cref{foot:chi},
we have:
\begin{equation}
     D_6^{(I)}(M_6)\otimes D_6^{(I)}(M_6)=\chiiM{}\,.
\end{equation}
showing that  $D_6^{(I)}(M_6)$ is an invertible defect of order 2 (the Euler counterterm could be removed by changing the normalization of $D_6^{(I)}(M_6)$ in \eqref{eq:D6I_def}).

\paragraph{Codimension-1 GS3 defects.} 
When fusing 
\begin{equation}
    D_6^{(I)}(M_6)\otimes D_6^{(J)}(M_6),\qquad I\neq J,
\end{equation}
we obtain new defects implementing the $\GS3$ automorphisms, which we describe below.
The defects obtained by higher-gauging the full $\Z_2^{(S)}\times\Z_2^{(C)}$ 3-form symmetry of the SymTFT on a codimension-1 manifold $M_6$ are:
\begin{align}
    T_6(M_6)&=D_6^{(S)}(M_6)\otimes D_6^{(C)}(M_6)=\lb  N\dw{M_6,\,\Z_2}\rb ^2\sum_{M_3,M_3'\in H_3(M_6,\,\Z_2)}D_3^{(S)}(M_3)\,D_3^{(C)}(M_3'), \label{eq:T6_def1}\\
    \ol{T}_6(M_6)&=D_6^{(C)}(M_6)\otimes D_6^{(S)}(M_6)=\lb  N\dw{M_6,\,\Z_2}\rb ^2\sum_{M_3,M_3'\in H_3(M_6,\,\Z_2)} D_3^{(C)}(M_3)\,D_3^{(S)}(M_3') \,, \label{eq:olT6_def1}
\end{align}
where the normalization factor is the square of \eqref{eq:N_Z2_M6}.

It is important to note that one can equivalently write:
\begin{align}
    T_6(M_6)&=D_6^{(C)}(M_6)\otimes D_6^{(V)}(M_6)=\lb  N\dw{M_6,\,\Z_2}\rb ^2\sum_{M_3,M_3'\in H_3(M_6,\,\Z_2)}D_3^{(C)}(M_3)\,D_3^{(V)}(M_3')=\nn\\
    &=D_6^{(V)}(M_6)\otimes D_6^{(S)}(M_6)=\lb  N\dw{M_6,\,\Z_2}\rb ^2\sum_{M_3,M_3'\in H_3(M_6,\,\Z_2)}
    D_3^{(V)}(M_3)\,D_3^{(S)}(M_3'),\label{eq:T6_def2}\\
    \ol{T}_6(M_6)
    &=D_6^{(S)}(M_6)\otimes D_6^{(V)}(M_6)=\lb  N\dw{M_6,\,\Z_2}\rb ^2\sum_{M_3,M_3'\in H_3(M_6,\,\Z_2)} 
    D_3^{(S)}(M_3)\,D_3^{(V)}(M_3') =\nn\\
    &=D_6^{(V)}(M_6)\otimes D_6^{(C)}(M_6)=\lb  N\dw{M_6,\,\Z_2}\rb ^2\sum_{M_3,M_3'\in H_3(M_6,\,\Z_2)} 
    D_3^{(V)}(M_3)\,D_3^{(C)}(M_3')\,. \label{eq:olT6_def2}
\end{align}
This follows from
\begin{align} \label{eq:D3prodrel1}
    D_3^{(S)}(M_3)\,D_3^{(C)}(M_3')&=D_3^{(C)}(M_3)\,D_3^{(V)}(M_3)\,D_3^{(C)}(M_3')=D_3^{(C)}(M_3+M_3')\,D_3^{(V)}(M_3)\,,\\
    D_3^{(C)}(M_3)\,D_3^{(V)}(M_3')&=D_3^{(V)}(M_3)\,D_3^{(S)}(M_3)\,D_3^{(V)}(M_3')=D_3^{(V)}(M_3+M_3')\,D_3^{(S)}(M_3)\,,
\end{align}
and similarly
\begin{align}\label{eq:D3prodrel2}
     D_3^{(C)}(M_3)\,D_3^{(S)}(M_3')&=D_3^{(S)}(M_3)\,D_3^{(V)}(M_3)\,D_3^{(S)}(M_3')=D_3^{(S)}(M_3+M_3')\,D_3^{(V)}(M_3)\,,\\
    D_3^{(S)}(M_3)\,D_3^{(V)}(M_3')&=D_3^{(V)}(M_3)\,D_3^{(C)}(M_3)\,D_3^{(V)}(M_3')=D_3^{(V)}(M_3+M_3')\,D_3^{(C)}(M_3)\,,
\end{align}
where we then redefine $M_3+M_3'\to M_3'$ and re-label the summed-over $M_3\leftrightarrow M_3'$.

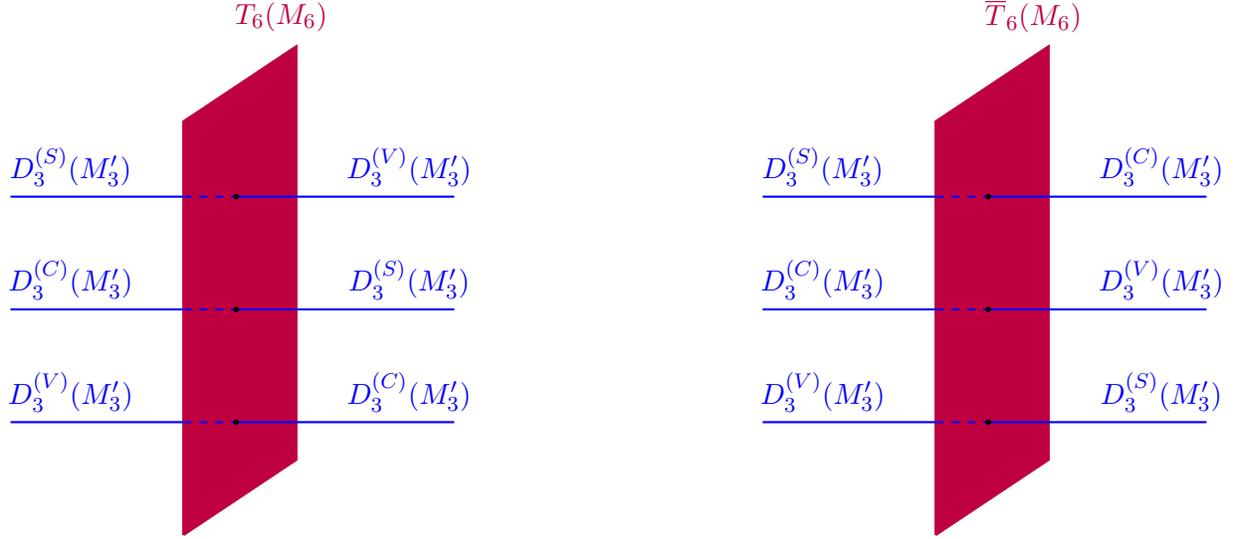
\begin{figure}
\centering
    \begin{tikzpicture}
    \begin{scope}
        \draw [purple, thick, fill=purple, opacity=1]
    (0,-1.5) -- (0,4) -- (1.5,5) -- (1.5,-0.5) -- (0,-1.5);
     \draw [purple, thick, fill=purple, opacity=1]
     (0,-1.5) -- (0,4) -- (1.5,5) -- (1.5,-0.5) -- (0,-1.5);
     \draw [blue, thick] (-2.3,3) -- (0, 3) ;
     \draw [blue, thick, dashed] (0, 3) -- (0.7,3) ;
     \draw [blue, thick] (0.7,3) -- (3.6, 3) ;
    \node [blue] at (-1.5, 3.4) {$D_3^{(S)}(M_3')$} ;
    \node[purple] at (1.3,5.4) {$T_6(M_6)$};
    \node [blue] at (3, 3.4) {$D_3^{(V)}(M_3')$} ;
    \draw [black,fill=black] (0.7,3) ellipse (0.03 and 0.03);
      \draw [blue, thick] (-2.3,1.5) -- (0, 1.5) ;
     \draw [blue, thick, dashed] (0, 1.5) -- (0.7,1.5) ;
     \draw [blue, thick] (0.7,1.5) -- (3.6, 1.5) ;
    \node [blue] at (-1.5, 1.9) {$D_3^{(C)}(M_3')$} ;
    \node [blue] at (3, 1.9) {$D_3^{(S)}(M_3')$} ;
    \draw [black,fill=black] (0.7,1.5) ellipse (0.03 and 0.03);
      \draw [blue, thick] (-2.3,0) -- (0,0) ;
     \draw [blue, thick, dashed] (0, 0) -- (0.7,0) ;
     \draw [blue, thick] (0.7,0) -- (3.6, 0) ;
    \node [blue] at (-1.5, 0.4) {$D_3^{(V)}(M_3')$} ;
    \node [blue] at (3, 0.4) {$D_3^{(C)}(M_3')$} ;
    \draw [black,fill=black] (0.7,0) ellipse (0.03 and 0.03);
    \end{scope}
     \begin{scope}[shift={(10,0)}]
        \draw [purple, thick, fill=purple, opacity=1]
    (0,-1.5) -- (0,4) -- (1.5,5) -- (1.5,-0.5) -- (0,-1.5);
     \draw [purple, thick, fill=purple, opacity=1]
     (0,-1.5) -- (0,4) -- (1.5,5) -- (1.5,-0.5) -- (0,-1.5);
     \draw [blue, thick] (-2.3,3) -- (0, 3) ;
     \draw [blue, thick, dashed] (0, 3) -- (0.7,3) ;
     \draw [blue, thick] (0.7,3) -- (3.6, 3) ;
    \node [blue] at (-1.5, 3.4) {$D_3^{(S)}(M_3')$} ;
    \node[purple] at (1.3,5.4) {$\ol{T}_6(M_6)$};
    \node [blue] at (3, 3.4) {$D_3^{(C)}(M_3')$} ;
    \draw [black,fill=black] (0.7,3) ellipse (0.03 and 0.03);
      \draw [blue, thick] (-2.3,1.5) -- (0, 1.5) ;
     \draw [blue, thick, dashed] (0, 1.5) -- (0.7,1.5) ;
     \draw [blue, thick] (0.7,1.5) -- (3.6, 1.5) ;
    \node [blue] at (-1.5, 1.9) {$D_3^{(C)}(M_3')$} ;
    \node [blue] at (3, 1.9) {$D_3^{(V)}(M_3')$} ;
    \draw [black,fill=black] (0.7,1.5) ellipse (0.03 and 0.03);
      \draw [blue, thick] (-2.3,0) -- (0,0) ;
     \draw [blue, thick, dashed] (0, 0) -- (0.7,0) ;
     \draw [blue, thick] (0.7,0) -- (3.6, 0) ;
    \node [blue] at (-1.5, 0.4) {$D_3^{(V)}(M_3')$} ;
    \node [blue] at (3, 0.4) {$D_3^{(S)}(M_3')$} ;
    \draw [black,fill=black] (0.7,0) ellipse (0.03 and 0.03);
    \end{scope}
\end{tikzpicture}
\caption{Action of the invertible, codimension-1, triality defects $T_6(M_6)$ and $\ol{T}_6(M_6)$, on the dimension-3 defects $D_3^{(I)}(M_3')$ for $I\in\{S,C,V\}$, as written in eqs. \eqref{eq:action_of_T6} and \eqref{eq:action_of_Tbar6}. \label{fig:D3_T6}}
\end{figure}

\paragraph{Action on dimension-3 defects.}
The action of $T_6(M_6)$ on the dimension-3 defects $D_3\up{I}(M_3)$ for $I\in\{S,C,V\}$, computed in appendix \ref{app:GS_SymTFT_fusions},
\be \label{eq:action_of_T6}
\ba
    D_3^{(S)}(N_3)\,T_6(M_6)&=T_6(M_6)\,D_3^{(V)}(N_3)\,, \\
     D_3^{(C)}(N_3)\,T_6(M_6)&=T_6(M_6)\,D_3^{(S)}(N_3)\,, \\
     D_3^{(V)}(N_3)\,T_6(M_6)&=T_6(M_6)\,D_3^{(C)}(N_3)\,,
\ea
\ee
shows that $T_6(M_6)$ implements the order-3 Green-Schwarz automorphism \eqref{eq:GS3}, as shown on the left of figure \ref{fig:D3_T6}.

The action of $\ol{T}_6(M_6)$ on the dimension-3 defects $D_3\up{I}(M_3)$ for $I\in\{S,C,V\}$ can be obtained by similar computations, or, equivalently, simply by exchanging $S\leftrightarrow C$ in the above equations, giving:
\be \label{eq:action_of_Tbar6}
\ba
    D_3^{(C)}(N_3)\,\ol{T}_6(M_6)&=\ol{T}_6(M_6)\,D_3^{(V)}(N_3)\,,\\
    D_3^{(S)}(N_3)\,\ol{T}_6(M_6)&=\ol{T}_6(M_6)\,D_3^{(C)}(N_3)\,,\\
    D_3^{(V)}(N_3)\,\ol{T}_6(M_6)&=\ol{T}_6(M_6)\,D_3^{(S)}(N_3)\,.
\ea
\ee
$\ol{T}_6(M_6)$ thus implements the order-3 Green-Schwarz automorphism \eqref{eq:GS3bar}, as shown on the right of figure \ref{fig:D3_T6}.

\paragraph{Fusion of GS2 and GS3 defects.} Using similar computations to the fusion of two duality GS2 defects, we compute the fusion between GS2 and GS3 defects, which are non-abelian. The calculations can be found in appendix \ref{app:GS_SymTFT_fusions}. We summarize the results below: 
\be \label{eq:GS2_GS3_fusions}
\ba
    D_6\up{K}(M_6)\otimes T_6(M_6)&=\chiiM{}\; D_6\up{I}(M_6)\,,\\
    D_6\up{K}(M_6)\otimes \ol{T}_6(M_6)&=\chiiM{}\, D_6\up{J}(M_6)\,,\\
    T_6(M_6)\otimes D_6\up{K}(M_6)&=\chiiM{}D_6\up{J}(M_6)\,,\\
    \ol{T}_6(M_6)\otimes D_6\up{K}(M_6)&=\chiiM{}D_6\up{I}(M_6)\,.
\ea
\ee
where $(I,J,K)$ is a cyclic permutation of $(S,C,V)$.

\paragraph{Fusion of two GS3 defects.} The detailed computations of the fusion of two GS3 defects can also be found in appendix \ref{app:GS_SymTFT_fusions}, the results are:
\be \label{eq:GS3_GS3_fusions}
\ba
    T_6(M_6)\otimes \ol{T}_6(M_6)&=\chiM{}^{-2}\,,\\
    \ol{T}_6(M_6)\otimes T_6(M_6)&=\chiM{}^{-2}\,,\\
    T_6(M_6)\otimes T_6(M_6)&=\chiiM{}\,\ol{T}_6(M_6)\,,\\
    \ol{T}_6(M_6)\otimes \ol{T}_6(M_6)&=\chiiM{}\,{T}_6(M_6)\,.
\ea
\ee

\paragraph{Codimension-1 defects and \tpdf{$S_3$}{S3}.}
The fusions discussed above reveal that the defects 
\begin{align}
    D_6\up{\id},\quad D_6\up{S},\quad D_6\up{C},\quad D_6\up{V},\quad T_6,\quad \ol{T}_6
\end{align}
form (up to Euler terms) the non-abelian group $S_3$
\begin{equation}
    S_3:\quad \langle a,b \;|\; a^3=b^2=1,\quad bab=a^2\rangle
\end{equation}
specifically, we have the following mapping
\begin{subequations}
    \begin{align}
    D_6\up{\id}&\leftrightarrow 1,& T_6&\leftrightarrow a, &\ol{T}_6&\leftrightarrow a^2 \\
    D_6\up{S}&\leftrightarrow b,& D_6\up{C}&\leftrightarrow a^2b,&
    D_6\up{V}&\leftrightarrow ab,&
\end{align}
\end{subequations}
with fusion corresponding to $S_3$ group multiplication (with Euler terms on the right hand side).

\subsubsection{Duality Twist Defects} \label{sec:duality_twist} By condensing a symmetry generator on a manifold with boundary one obtains a twist defect, as described in \cite{Kaidi:2022cpf} for duality twist interfaces in 3d and 5d SymTFTs.
For the 7d SymTFT case, we define the duality twist defect $D_6^{(I)}(M_6\gz,M_{5|0})$ for $I\in\{S,C,V\}$ by condensing $D_3^{(I)}(M_3)$ on a 6-manifold $M_6\gz$ with boundary $M_{5|0}=\de M_6\gz$ and imposing the Dirichlet boundary condition for $M_3$ on $M_{5|0}$:
\begin{equation} \label{eq:DI_twist}
    \begin{split}
        D_6^{(I)}(M_6\gz,M_{5|0})= N\dw{M_6\gz,\,\Z_2}\sum_{M_3\in H_3(M_6\gz,\,\Z_2)}D_3^{(I)}(M_3), \qquad I\in\{S,C,V\},
    \end{split}
\end{equation}
where the normalization factor was defined in \eqref{eq:N_Z2_M6gz}. 

\paragraph{Fusion of Duality Twist defects. \label{subsubsec:fusD6}} We will compute the fusion of two duality twist defects \eqref{eq:DI_twist}, located at $x=0$ and $x=\eps$ respectively, and then take $\eps\to0$. 
\begin{align}
    D_6^{(I)}(M_6\gz,M_{5|0})\otimes D_6^{(I)}(M_6\gep,M_{5|\eps})&= N\dw{M_6\gz,\,\Z_2} N\dw{M_6\gep,\,\Z_2}\sum_{\substack{M_3\in H_3(M_6\gz,\,\Z_2)\\M_3'\in H_3(M_6\gep,\,\Z_2)}}D_3^{(I)}(M_3) \, D_3^{(I)}(M_3')=\nn\\
    &= N\dw{M_6\gz,\,\Z_2} N\dw{M_6\gep,\,\Z_2}\sum_{\substack{M_3\in H_3(M_6\gz,\,\Z_2)\\M_3'\in H_3(M_6\gep,\,\Z_2)}} D_3^{(I)}(M_3+M_3')(-1)^{\langle M_3,M_3'\rangle}=
\end{align}
We now define $\Tilde{M}_3'=M_3+M_3'$ on $M_6\gep$, write the sum over 3-cycles $M_3,\Tilde{M}_3'$ as a sum over their Lefschetz duals $\LD(M_3),\,\LD(\Tilde{M}_3')$ and split the integral using $M_6\gz=M_6^{[0,\eps]}\cup M_6\gep$: 
\begin{align}
    = N\dw{M_6\gz,\,\Z_2} N\dw{M_6\gep,\,\Z_2}\sum_{\substack{\LD(M_3)\in H^3(M_6\gz,M_{5|0},\,\Z_2)\\
    \LD(\Tilde{M}_3')\in H^3(M_6\gep,M_{5|\eps},\,\Z_2)}}&\exp(i\pi\int_{M_6\ze}c_3\up{I}\cup \LD(M_3))\times\\
    \times&\exp(i\pi\int_{M_6\gep}\lb c_3\up{I}+\LD(M_3)\rb \cup \LD(\Tilde{M}_3'))=
\end{align}
Integrating out $\LD(\Tilde{M}_3')$ sets $\LD(M_3)|_{M_6\gep}=-c_3\up{I}$ and leaves the dynamical $\LD(M_3)$ only on $M_6\ze$, with Dirichlet boundary conditions on both boundaries:
\begin{align}
    &= N\dw{M_6\ze,\,\Z_2}\,\chiiM{\gep}\sum_{\substack{\LD(M_3)\in H^3(M_6\ze,M_{5|0}\cup M_{5|\eps},\,\Z_2)}}\exp(i\pi\int_{M_6\ze}c_3\up{I}\cup \LD(M_3))=\quad
\end{align}
We now use the isomorphism
\eqref{eq:iso_H^3_H^2} to map $\LD(M_3)\to b_2\in H^2(M_5,\,\Z_2)$:\footnote{For the normalization recall \cref{foot:norm2}.}
\begin{align}
    &=\chiiM{\gep}\,N\dw{M_5,\,\Z_2}\sum_{b_2\in H^2(M_5,\,\Z_2)}\exp(i\pi\int_{M_5}c_3\up{I}\cup b_2)\,.
\end{align}
Poincaré duality on $M_5$ converts the sum over $b_2\in H^2(M_5,\,\Z_2)$ to one over $M_3=\PD(b_2)\in H_3(M_5,\,\Z_2)$. We thus obtain, in the $\eps\to 0$ limit:
\begin{align} \label{eq:duality_twist_fusions}
   D_6^{(I)}(M_6\gz,M_{5|0})\otimes D_6^{(I)}(M_6\gep,M_{5|\eps})&=\chiiM{\gz}\;N\dw{M_5,\,\Z_2}\sum_{{M}_3\in H_3(M_5,\,\Z_2)}D_3^{(I)}({M}_3)
\end{align}
resulting in the condensation defect (with normalization factor given in eq. \eqref{eq:Norm_M5})
\begin{equation} 
    \cC_5\up{\Z_2\up{I}}(M_5)=\chiiM{\gz}\,N\dw{M_5,\,\Z_2}\sum_{M_3\in H_3(M_5,\,\Z_2)}\,\exp(i\pi\int_{M_3}c_3\up{I})\,.
\end{equation}
Therefore, the fusion is:
\begin{align} \label{eq:Dtwist_Dtwist}
    D_6^{(I)}(M_6,M_{5})\otimes D_6^{(I)}(M_6,M_{5})=\cC_5\up{\Z_2\up{I}}(M_5)\,.
\end{align}

\subsubsection{Triality Twist Defects}

We define triality twist defects by condensing $D_3^{(I)}(M_3)\,D_3^{(J)}(N_3)$ for $I\neq J\in\{S,C,V\}$ on a 6-manifold $M_6\gz$ with boundary $M_{5|0}=\de M_6\gz$ and imposing Dirichlet boundary conditions for $M_3,\,N_3$ on $M_{5|0}$:
\be \label{eq:T_twist}
\ba
    T_6(M_6\gz,M_{5|0})&=\lb  N\dw{M_6\gz,\,\Z_2}\rb ^2\sum_{M_3,N_3\in H_3(M_6\gz,\,\Z_2)}D_3^{(I)}(M_3)\,D_3^{(J)}(N_3), \\[2mm]
    \ol{T}_6(M_6\gz,M_{5|0})
    &=\lb  N\dw{M_6\gz,\,\Z_2}\rb ^2\sum_{M_3,N_3\in H_3(M_6\gz,\,\Z_2)} D_3^{(J)}(N_3)\,D_3^{(I)}(M_3) \,,
\ea
\ee
with normalization factor the square of \eqref{eq:N_Z2_M6gz}.

\paragraph{Fusion of Triality Twist Defects.} The detailed computations of triality twist defect fusions, obtained by similar methods to those used above, can be found in appendix \ref{app:twist_fusions}. The results are:
\begin{align} \label{eq:T_twist_fusions}
     T_6(M_6,M_{5})\otimes \ol{T}_6(M_6,M_{5})&=\chiM{\gz}^{-2}\;N\dw{M_5,\,\Z_2}^2\sum_{M_3,N_3\in H_3(M_5,\,\Z_2)}D_3^{(I)}(M_3)\,D_3^{(J)}(N_3)=\cC_5\up{\Z_2\up{I}\times\Z_2\up{J}}(M_5)\,,\nn\\
     \ol{T}_6(M_6,M_{5})\otimes {T}_6(M_6,M_{5})&=\chiM{\gz}^{-2}\;N\dw{M_5,\,\Z_2}^2\sum_{M_3,N_3\in H_3(M_5,\,\Z_2)}D_3^{(J)}(M_3)\,D_3^{(I)}(N_3)=\cC_5\up{\Z_2\up{J}\times\Z_2\up{I}}(M_5)\,, \nn\\
       T_6(M_6,M_{5})\otimes {T}_6(M_6,M_{5})&=\cC_5\up{\Z_2\up{I}\times\Z_2\up{J}}(M_5)\,   \ol{T}_6(M_6,M_{5})\,,\nn\\[3mm]
       \ol{T}_6(M_6,M_{5})\otimes \ol{T}_6(M_6,M_{5})&=\cC_5\up{\Z_2\up{J}\times\Z_2\up{I}}(M_5)\,   {T}_6(M_6,M_{5})\,,
\end{align}
and the expressions obtained from them by cyclically permuting $(I,J,K)$.

\paragraph{Fusion between Duality and Triality Twist defects.} Similarly:
\begin{align} \label{eq:D_T_twist_fusions}
    D_6\up{K}(M_6,M_5)\otimes T_6(M_6,M_5)&=\cC_5\up{\Z_2\up{K}}(M_5)\otimes  D_6\up{I}(M_6,M_5)\,,\nn\\
    D_6\up{K}(M_6,M_5)\otimes \ol{T}_6(M_6,M_5)&=\cC_5\up{\Z_2\up{K}}(M_5)\otimes D_6\up{J}(M_6,M_5)\,,\nn\\
    T_6(M_6,M_5)\otimes D_6\up{K}(M_6,M_5)&=\cC_5\up{\Z_2\up{I}\times \Z_2\up{J}}(M_5)\otimes D_6\up{J}(M_6,M_5)\,,\nn\\
    \ol{T}_6(M_6,M_5)\otimes D_6\up{K}(M_6,M_5)&=\cC_5\up{\Z_2\up{J}\times \Z_2\up{I}}(M_5)\otimes D_6\up{I}(M_6,M_5)\,.
\end{align}
Analogous equations follow by taking cyclic permutations of $(I,J,K)$.

\subsubsection{Gauging the \tpdf{$S_3$}{S3} Symmetry in the Bulk}
We now gauge the $S_3$ symmetry in the bulk. In order to do so, we generalize the approach in \cite{Bhardwaj:2022yxj, Kaidi:2022cpf, Antinucci:2022vyk, Antinucci:2022cdi} to 6d: first we take invariant combinations of defects in the pre-gauged theory, and then decorate with possible twist defects those that are invariant.

\paragraph{Topological operators from the ungauged theory:} We first need to consider $S_3$ invariant combinations of the operators of the ungauged theory
\begin{itemize}
    \item The $S_3$ invariant combination of $D_3^{(I)}$, that is 
    \begin{equation}
        D_3^{(S)} \oplus D_3^{(V)} \oplus D_3^{(C)} \,.
    \end{equation}
    \item Genuine codimension 2 defects are provided by  twist defects discussed in the previous section. The condensation of subgroups of $\mathbb{Z}_2 \times \mathbb{Z}_2$ on $M_6$ with $\partial M_6 =M_5$ becomes transparent. The topological operators are labeled by conjugacy classes $[g]$ of $S_3$, and representations of the stabilizer group of $g \in [g]$. For instance the ones labeled by the trivial representation are directly related to the twist defects presented in the previous section. 
    \begin{equation}\label{eq:gaugedops}
        D_5^{(S)} \oplus D_5^{(V)}  \oplus D_5^{(C)}, \qquad T_5  \oplus\overline{T}_5.
    \end{equation}
    where 
    \begin{equation} \label{eq:twisteval}
        D_5^{(I)} = D_6^{(I)}(M_6,M_5)|_{ M_5 \subset X_6 =\partial X_7}, \quad T_5= T_6(M_6,M_5)|_{ M_5 \subset X_6 =\partial X_7}, \quad \ol{T}_5= \ol{T}_6(M_6,M_5)|_{ M_5 \subset X_6 =\partial X_7}\,,
    \end{equation}
    where $X_7$ is the bulk and $X_6$ is the symmetry boundary. 
    In order to gauge the 0-form symmetry, $S_3$, we need to couple the bulk theory to $a\in H^1(M_7,S_3)$. The gauge field can have non-trivial holonomies around a twist defect supported on $M_5$, such that $\delta a = g$. Twist defects are not gauge invariant alone, but the combined configuration is gauge invariant. They are particularly important because they will generate the full $S_3$ symmetry when Dirichlet boundary conditions are specified for $a$ \cite{Antinucci:2022vyk, Antinucci:2022cdi}. 
\end{itemize}

\paragraph{Boundary conditions.} In the ungauged theory the b.c. come by specifying a Lagrangian algebra of the original ungauged theory, $\mathcal{L}$, or equivalently a polarization $(L_J, \ol{L}_I)$. When gauging $S_3$ we need to specify $S_3$ invariant boundary conditions. Therefore we take the $S_3$ orbit as follows
\begin{equation}
        | \mathcal{B}^{\rm sym} \rangle \rightarrow  \frac{1}{|\text{Stab}(\mathcal{B}^{\rm sym})
        |}  \sum_{g \in S_3} | \mathcal{B}^{\rm sym}_g \rangle \,,
\end{equation}
where $ | \mathcal{B}^{\rm sym}_g \rangle$ is given by the action of $g \in S_3$ on the polarization choice $(L_J, \ol{L}_I)$. We first need to understand the action when the twist defects end on the boundary conditions of the ungauged theory.\footnote{One can also study interfaces between two dual theories by looking at $S_3$ symmetry defects that end and form an L-shape on $ | \mathcal{B}^{\rm sym} \rangle$. This will be analogous to the 4d case studied in \cite{Antinucci:2022vyk}.} Since there is a one-to-one correspondence between polarizations of the relative theory and gapped boundary conditions of the ungauged bulk theory, we can label the $\mathcal{B}^{\rm sym}$ as follows,
\begin{equation}
        | \mathcal{B}^{\rm sym} \rangle = |L_J,\overline{L}_I\rangle\, .
\end{equation}
The $S_3$ orbit will span the full set of boundary conditions, therefore we have that 
\begin{equation} \label{eq:S3boundary}
        \frac{1}{|\text{Stab}(\mathcal{B}^{\rm sym})
        |}  \sum_{g \in S_3} | \mathcal{B}^{\rm sym}_g \rangle  = |L_V,\overline{L}_C\rangle +|L_C,\overline{L}_S\rangle+|L_S,\overline{L}_V\rangle+|L_V,\overline{L}_S\rangle+|L_C,\overline{L}_V\rangle+|L_S,\overline{L}_C\rangle\, .
\end{equation}
We need to evaluate now the defects $D_5^{(I)},T_5$ defined in \eqref{eq:twisteval} on the explicit symmetry boundaries, $|L_J,\overline{L}_I\rangle\ $. Since they are defined as condensations of the $D_3^{(I)}(M_3)$ surfaces, the boundary condition on $c_3\up{I}$ will dictate whether they trivialize or not on $|L_J,\overline{L}_I\rangle\ $. 
In particular we have that, when $c_3\up{I}$ has Dirichlet boundary conditions it is in relative cohomology: $c_3\up{I} \in H^3(X_6,X_7 ,\mathbb{Z}_2)$. When $M_5 = \partial M_6 \in X_6 = \partial X_7$ the 3-cycle $M_3$, where $c_3\up{I}$ with Dirichlet b.c. is integrated over, is such that
\begin{equation}
 M_3 \in  H_3(M_5,M_6,\mathbb{Z}_2) = H_3(M_5,\mathbb{Z}_2) \cap H_3(X_6,X_7 ,\mathbb{Z}_2).
\end{equation}
Since all $M_3 \in M_5$ must be in relative homology, it means that this intersection is trivial. Equivalently as stated in \cite{Kaidi:2022cpf}, the $M_3$ cycle is relative to itself. Therefore the Dirichlet boundary conditions on $c_3\up{I}$ will trivialize any $D_3^{(I)}$ on $M_5 = \partial M_6 \in X_6 = \partial X_7$. Neumann boundary conditions for $c_3\up{I}$ will instead project the $D_3^{(I)}$ entirely on $M_5 = \partial M_6 \in X_6 = \partial X_7$.

With this in hand let us evaluate $D_5^{(I)}$ and $T_5$ on $|L_J,\overline{L}_I\rangle\ $,
\be
\ba
        D_6^{(K)}(M_6,M_5 &=  \partial M_6 \in X_6 = \partial X_7)|_{(L_J,\overline{L}_I)} = D_5^{(K)}|_{(L_J,\overline{L}_I)}, \qquad \forall K\neq J\neq I \neq K  \\
         T_6(M_6,M_5 &=  \partial M_6 \in X_6 = \partial X_7)|_{(L_J,\overline{L}_I)}= T_5|_{(L_J,\overline{L}_I)},\qquad \forall I \neq J\\
         \overline{T}_6(M_6,M_5&=  \partial M_6 \in X_6 = \partial X_7)|_{(L_J,\overline{L}_I)}= \overline T_5|_{(L_J,\overline{L}_I)},\qquad \forall I \neq J
\ea    
\ee
where the evaluation of $T_5$ and $\overline T_5$ is non-trivial for any ${(L_J,\overline{L}_I)}$ due to the identities \eqref{eq:D3prodrel1} and \eqref{eq:D3prodrel2}.

\paragraph{Matching with Half-space Gauging Results.} The fusion rules will follow from \eqref{eq:duality_twist_fusions}, \eqref{eq:T_twist_fusions} and \eqref{eq:D_T_twist_fusions}. However, there are subtleties which we now discuss for the polarization $(L_J,\ol{L}_I)$.

When performing different topological manipulations in the absolute  6d\, $\so(8),\,\cN=(2,0)$ SCFTs involving gauging/stacking, as we did in section \ref{sec:half-space}, one can obtain the same global variant \emph{up to} stacking with the Arf--Kervaire 
invariant of the manifold, which is a $v_4$-fermionic SPT phase, as explained in section 6 of 
\cite{Gukov:2020btk}, and previously in \cite{Hsin:2021qiy}. 
Indeed, in our fusion results computed from half-space gauging and stacking, summarized in section \ref{sec:Bounty}, the factor $\cZ\dw{Y,M_6\gz}$, defined (up to normalization) in  \eqref{eq:ZY_def} is present on the right-hand side of the $\cT_5(M_5)\otimes \cT_5(M_5)$ and $\ol{\cT}_5(M_5)\otimes \ol{\cT}_5(M_5)$ fusions. 
{If we think in terms of the spins of the defects, condensing fermionic surface defects will give rise to fermionic codimension-one defects. Stacking them with a fermionic SPT provides codimension-one bosonic defects generating the $S_3$ symmetry, which is a bosonic symmetry as expected. 
For this reason we claim that the stacking is necessary. For instance, the stacking comes naturally if one computes codimension-one defect correlators with \eqref{eq:SIJ}, where $\bar{q}(C_3)=C_3 \cup X$, with $dX=v_4$.\footnote{See \cite{Hsin:2021qiy} for a more detailed discussion on the activation of the space-time background parametrized by the Wu class $v_4 \in H^4(BO(d),\mathbb{Z}_2)$, and the 3-group given by $1 \rightarrow \mathbb Z_2^{(3)} \rightarrow G^{(3)} \rightarrow O(d)\rightarrow 1$.}}

The SymTFT computations we have performed in the current section, instead, do not give rise to this factor and thus relate different global variants which do not differ by stacking with $\cZ\dw{Y,M_6\gz}$, (indeed, the $S_3$ bulk symmetry is not anomalous and can be gauged). 
To match with the half-space gauging results, we therefore define the symmetry defects on the symmetry boundary as given by the bulk twist defects stacked with $\cZ\dw{Y,M_6\gz}$: on the left-hand side of the fusion results it cancels since it is $\Z_2$-valued, but remains on the right-hand side when a non-trivial symmetry defect is present.

For each $|L_J,\overline{L}_I\rangle$ we will consider 
\be \label{eq:boundaryops}
\ba
&\cD_5^{\mathcal B^{\rm sym}}(M_5) = \cZ\dw{Y,M_6\gz}\, D_5^{(K)}(M_5)|_{(L_J,\overline{L}_I)}{{e^{i\pi\int_{M_6\gz}\bar{q}(C_3\up{I})}}}  \qquad \forall\, I \neq J \,,
\ea    
\ee
{where the last factor is present if the bulk field $c_3^{(I)}$ with Dirichlet boundary conditions is not fixed to zero on the boundary but to a background field $C_3^{(I)}$, recall equation \eqref{eq:S_boundary}.}
In addition, the triality twist defects act in the same way on each $|L_J,\overline{L}_I\rangle$, therefore we automatically have
\begin{equation} 
  \mathcal T_5^{\mathcal B^{\rm sym}}= \cZ\dw{Y,M_6\gz}\,T_5|_{(L_J,\overline{L}_I)}, \qquad  \overline{\mathcal T}_5^{\mathcal B^{\rm sym}}  =\cZ\dw{Y,M_6\gz}\,\overline T_5|_{(L_J,\overline{L}_I)} \qquad \forall\, I\neq J \,.
\end{equation}

The SymTFT duality twist defect fusion \eqref{eq:Dtwist_Dtwist} for $D_5\up{K}$, thus gives, on the symmetry boundary, the same result as \eqref{eq:6d_DD_res} if we identify $\cC_5\up{\Z_2\up{K}}|_{(L_J,\overline{L}_I)}$ with $  \cC_5\up{0}$. 
For triality fusions, the SymTFT triality twist fusions \eqref{eq:T_twist_fusions}, give the boundary results \eqref{eq:6d_TT1}-\eqref{eq:6d_TT4} if we take into account the aforementioned $\cZ\dw{Y,M_6\gz}$ factor and identify $\cC_5\up{\Z_2\up{I}\times\Z_2\up{J}}|_{(L_J,\overline{L}_I)}=\cC_5\up{\Z_2\up{J}\times\Z_2\up{K}}|_{(L_J,\overline{L}_I)}$ with $\cC_5\up{1}$ and $\cC_5\up{\Z_2\up{J}\times\Z_2\up{I}}$ with $\cC_5\up{0}$. Note that for these identifications, we choose the representative of the condensation defect with $c_3\up{J}$ on the left, since this is the dynamical field on the left SymTFT boundary $\Bsym$: the condensation of two dynamical fields in the first case gives the non-trivial twist whereas in the second case the $c_3\up{I}$ factor is frozen since $c_3\up{I}$ has Dirichlet boundary conditions and thus the twist is trivial. Summarizing this we have
\be
\ba
&\cC_5\up{0}(M_5)|_{\mathcal B^{\rm sym}}= \cC_5\up{\Z_2\up{K}}(M_5)|_{(L_J,\overline{L}_I)}=  \cC_5\up{\Z_2\up{J} \times \Z_2\up{I}}(M_5)|_{(L_J,\overline{L}_I)} \qquad \forall \,J \neq I\, , \\
&\cC_5\up{1}(M_5)|_{\mathcal B^{\rm sym}}=\cC_5\up{\Z_2\up{I} \times \Z_2\up{J}}(M_5)|_{(L_J,\overline{L}_I)} = \cC_5\up{\Z_2\up{J} \times \Z_2\up{K}}(M_5)|_{(L_J,\overline{L}_I)} \qquad \forall\, J \neq I, \; K\neq J \neq I \neq K\, .
\ea
\ee

Finally, the fusions between duality and triality defects follow from the SymTFT results \eqref{eq:D_T_twist_fusions} and match the boundary fusions \eqref{eq:6d_DT_res1}-\eqref{eq:6d_DT_res3}.
In summary the fusions are as follows:
\be
\ba
     \cD_5^{\mathcal B^{\rm sym}}(M_5)\otimes \cD_5^{\mathcal B^{\rm sym}}(M_5)&=\cC_5\up{0}(M_5)|_{\mathcal B^{\rm sym}}\\
    \ol{\cT}_5^{\mathcal B^{\rm sym}}(M_5)\otimes \cT_5^{\mathcal B^{\rm sym}}(M_5)&=\cC_5\up{0}(M_5)|_{\mathcal B^{\rm sym}}\\
     \cT_5^{\mathcal B^{\rm sym}}(M_5)\otimes \ol{\cT}_5^{\mathcal B^{\rm sym}}(M_5)&= \cC_5\up{1}(M_5)|_{\mathcal B^{\rm sym}}\\
      \cT_5^{\mathcal B^{\rm sym}}(M_5)\otimes \cT_5^{\mathcal B^{\rm sym}}(M_5)&=\cZ\dw{Y,M_6\gz}\, \cC_5\up{1}(M_5)|_{\mathcal B^{\rm sym}}\otimes \ol{\cT}_5^{\mathcal B^{\rm sym}}(M_5)\\
      \ol{\cT}_5^{\mathcal B^{\rm sym}}(M_5)\otimes \ol{\cT}_5^{\mathcal B^{\rm sym}}(M_5)&=
    \cZ\dw{Y,M_6\gz}\, \cC_5\up{0}(M_5)|_{\mathcal B^{\rm sym}}\otimes{\cT}_5^{\mathcal B^{\rm sym}}(M_5)\\
   \cD_5^{\mathcal B^{\rm sym}}(M_5)\otimes \ol{\cT}_5^{\mathcal B^{\rm sym}}(M_5)&=\cC_5\up{0}\;(M_5)|_{\mathcal B^{\rm sym}}\,{e^{i\pi\int_{M_6\gz}\bar{q}(C_3\up{I})}}\\
    \cT_5^{\mathcal B^{\rm sym}}(M_5)\otimes \cD_5^{\mathcal B^{\rm sym}}(M_5)&=\cC_5\up{1}(M_5)|_{\mathcal B^{\rm sym}}\,{e^{i\pi\int_{M_6\gz}\bar{q}(C_3\up{I})}}
    \\
     \cD_5^{\mathcal B^{\rm sym}}(M_5)\otimes {\cT}_5^{\mathcal B^{\rm sym}}(M_5)&=\ol{\cT}_5^{\mathcal B^{\rm sym}}(M_5)\otimes \cD_5^{\mathcal B^{\rm sym}}(M_5)\,.
\ea 
\ee

\paragraph{Comments on the $S_3$ Anomaly.} 
In order to discuss the possible 't Hooft anomaly of the $S_3$-ality symmetry we follow the strategy of \cite{Antinucci:2023ezl}. These are conditions on the SymTFT that come from gauging the symmetry $G$ of the bulk theory \eqref{eq:7dCSso8}. This is precisely our construction where $G=S_3$. 
\begin{enumerate}
\item The first condition is the absence of a $G$-invariant Lagrangian algebra, which implies that the symmetry is intrinsic, i.e. it cannot be related to an invertible symmetry by gauging \cite{gelaki2009centers, Kaidi:2022uux, Sun:2023xxv, Antinucci:2023ezl, Cordova:2023bja, Bashmakov:2022uek,Antinucci:2025fjp}.\footnote{Note that for more complicated starting categories, this condition does not always guarantee that the non-invertible symmetry is anomalous: there are examples of intrinsic non-invertible symmetries that are anomaly-free \cite{gelaki2009centers,Lu:2025yru}.}
\item The second condition is having a non-trivial anomaly functional for $S_3$ symmetry. This is usually related to the higher-dimensional generalization of the Frobenius-Schur indicator \cite{Antinucci:2023ezl}.
\end{enumerate}

The full $S_3$-ality symmetry and its $\mathbb Z_3$ triality subgroup are {intrinsic} by the first criterion: neither preserves a Lagrangian algebra, furthermore, since our starting higher-form symmetry (before gauging) was invertible abelian, the resulting $S_3$-ality symmetry and its $\Z_3$-ality subsymmetry are also anomalous. We therefore obtain the obstruction expected for a putative $G_2$ $(2,0)$ theory obtained by folding the $D_4$ theory by triality, since such a construction would require gauging the anomalous $\mathbb Z_3$ triality. Thus {a $\Z_3$-ality anomaly would be} consistent with, and {would provide} a field-theoretic explanation of, the absence of a $G_2$ $(2,0)$ theory from the $D_4$ folding. The $\mathbb Z_2$ subgroups are subtler. Each $\mathbb Z_2$ subgroup fixes one of the three Lagrangian subgroups (for example $\GS2\up{V}$ fixes $L_V$ while exchanging $L_S$ and $L_C$), so these duality subgroups do not satisfy the first criterion for an anomaly. It would be interesting to check whether the second criterion of \cite{Antinucci:2023ezl} applies to these $\mathbb Z_2$ dualities. If such a $\mathbb Z_2$ anomaly is present, it would likewise be consistent with the fact that the $B_3$ $(2,0)$ theory, which would arise by $\mathbb Z_2$ folding, has not been found. After coupling to a background $a_1\in H^1(Y_7,\mathbb Z_2)$, natural candidate anomaly functionals are a pure term $\int_{Y_7} a_1^7$ and mixed terms such as $\int_{Y_7} a_1\cup \bar q(c_3)$, where $c_3$ denotes the relevant dynamical 3-form field. From the SymTFT point of view, this check should be performed by explicitly gauging the $\mathbb Z_2$ symmetry in the bulk, equivalently by twisting the differential by $a_1$, and by tracking whether the chosen boundary condition captures fractionalization classes or counterterms that can shift the bulk contribution to the anomaly. We leave this more detailed analysis to future work.

{Let us also distinguish the present construction from the ordinary action on BPS strings. The action of GS automorphisms on the BPS string charge lattice was already analyzed in \cite{Apruzzi:2017iqe}; for the $D_4$ theory this is the standard $S_3$ permutation action. The new point here is that the $S_3$-ality defects act non-invertibly on surface defects of the 6d SCFT. Due to this non-invertible action on surface defects, we expect the associated Ward identities to constrain surface-operator correlation functions. In addition, non-invertible 0-form symmetries are known to constrain RG flows triggered by deformations of a given SCFT, and in particular the vacuum structure at the end of the flow \cite{Chang:2018iay,Damia:2023rg}. Although 6d SCFTs do not admit supersymmetry-preserving relevant deformations \cite{Cordova:2016xhm}, it would be interesting to use the $S_3$-ality symmetry to characterize RG flows triggered by local operator expectation values. Finally, upon ordinary $T^2$ compactification one expects 4d $\mathcal N=4$ $\mf{so}(8)$ SYM, or copies thereof. A natural future direction is to study compactifications twisted by the $S_3$-ality symmetry and identify the resulting lower-dimensional theories, in the spirit of \cite{Apruzzi:2017iqe,Kaidi:2022uux}.}

{A possible future direction is the study of interplays between the non-invertible duality, triality and $S_3$-ality internal symmetries constructed in the present work, with the superconformal symmetry of 6d SCFTs, which can also be anomalous \cite{Ohmori:2014kda,Cordova:2015fha,Cordova:2019wns}.\footnote{The 4d superconformal anomalies were given a topological formulation in \cite{Imbimbo:2023sph,Imbimbo:2025ffw}, including fermions and exact to all orders: extending this to 6d could provide the full 6d SCFT anomaly polynomials, completing the bosonic contributions of \cite{Ohmori:2014kda}.} A SymTFT formulation of spacetime symmetries \cite{Apruzzi:2025hvs} would provide a unified framework to study (non-invertible) internal and spacetime symmetries, anomalies and dualities, paving the way for investigations of possible intricate connections between them.}

\subsection*{Acknowledgments}

We thank Andrea Antinucci, Ho-Tat Lam, Craig Lawrie, {Daniele Migliorati,} Shu-Heng Shao, Xingyang Yu, Yunqin Zheng for many insightful discussions. 
The work of FA is partially supported by the University of Padua under the 2023 STARS Grants@Unipd programme (GENSYMSTR – Generalized Symmetries from Strings and Branes) and in part by the Italian MUR Departments of Excellence grant 2023-2027 ``Quantum Frontiers”.
The work of SSN and AW is supported by the UKRI Frontier Research Grant, underwriting the ERC Advanced Grant ``Generalized Symmetries in Quantum Field Theory and Quantum Gravity”.

\appendix
\section{Computation of Fusions of Triality and Condensation Defects}
\label{app:Twix}
In this appendix we provide the details for the fusions involving triality and condensation defects from half-space gauging, summarized in subsection \ref{sec:Bounty}.

\subsection{Fusion of two triality defects} 
\label{sec:triality_triality_fusions}
Triality fusion computations often involve the term
$\int_{M_6\ze}\bar{q}(C_3)$: since $C_3$ is a classical background, it is smooth across the boundaries of ${M_6\ze}$. In the $\eps\to0$ limit, the volume of the 6-manifold $M_6\ze\to0$ and, since there are no divergences or discontinuities, so does the integral of a quadratic function of a classical background evaluated on $M_6\ze$:
\begin{equation} \label{eq:q_0eps_to0}
    \int_{M_6\ze}\bar{q}(C_3)\to0 \qquad \text{as}\quad \eps\to0\,.
\end{equation}

\paragraph{$\bm{\ol{\cT}_5(M_5)\otimes {\cT}_5(M_5)}$.} To compute the fusion between two different triality defects $\ol{\cT}_5(M_5)$ and $\cT_5(M_5)$, we write schematically:
\begin{align} \label{eq:TbT_fusion_fig}
    &\;{\color{red}\ol{\cT}_5(M_{5|0})} \nn\\ 
     \cL_{\fT_J}[C_3\up{I}]  & {\color{red} \quad\bigg|_{c_3\up{J}|_{M_{5|0}}=0}} \quad \cL_{\fT_K}[c_3\up{J}] \,+\pi\lb c_3\up{J}\cup C_3\up{I}+\bar{q}(C_3\up{I})\rb\\[2mm]
    &&&\hspace{-4cm}\;{\color{red}{\cT}_5(M_{5|\eps})} \nn \\
    &&&\hspace{-4cm}{\color{red}\quad\bigg|_{\substack{c_3\up{K}|_{M_{5|\eps}}=0}}}\quad \cL_{\fT_J}[c_3\up{I}] \;+\pi\lb c_3\up{I}\cup c_3\up{K}+\cancel{2\,\bar{q}(c_3\up{K})}+c_3\up{K}\cup C_3\up{I}\rb\,.
\end{align}
From its definition in equation \eqref{eq:cT5_def_fus}, the  defect ${\cT}_5(M_{5|\eps})$ first acts with a Green-Schwarz interface $T_5$ at $M_{5|\eps}$, whose action on the gauge fields crossing it is:
\begin{subequations} \label{eq:T_M5eps_converts}
    \begin{align}
         c_3\up{J}|_{M_{5|\eps}}\,T_5(M_{5|\eps})&=T_5(M_{5|\eps})\,c_3\up{I}|_{M_{5|\eps}}\,,\\
     c_3\up{I}|_{M_{5|\eps}}\,T_5(M_{5|\eps})&=T_5(M_{5|\eps})\,c_3\up{K}|_{M_{5|\eps}}\,.
    \end{align}
\end{subequations}
We impose the Dirichlet boundary condition on $M_{5|\eps}$ for $c_3\up{K}$ when performing half-space twisted gauging: this, along with the invertibility of $T_5$, implies that we must also impose the Dirichlet boundary condition on $M_{5|\eps}$ for $c_3\up{I}$  and $c_3\up{J}$. In \eqref{eq:TbT_fusion_fig}, we used the fact that $\int \bar{q}$ is $\Z_2$-valued \eqref{eq:q_def} so $2\int \bar{q}(c_3\up{K})=0$. \\
The above picture translates to the equation:
\begin{align}
    Z_{\fT_J}[C_3\up{I};\, \ol{\cT}_5(M_{5|0}),\, \cT_5(M_{5|\eps})] =N\dw{M_6\ze,\,\Z_2} \sum_{c_3\up{J}\in H^3(M_6^{[0,\eps]},\,M_{5|0}\cup M_{5|\eps},\,\Z_2)} 
    \exp(i\pi\int_{M_6\ze}\lb c_3\up{J}\cup C_3\up{I}+\bar{q}(C_3\up{I})\rb)\,\times\nn\\
    \times N\dw{M_6\gep,\,\Z_2}^2\sum_{c_3\up{I},\,c_3\up{K} \in H^3(M_6^{\gs\eps},\,M_{5|\eps},\,\Z_2)}\,Z_{\fT_J}[c_3\up{I}]\,\exp(i\pi \int_{M_6\gep}\lb c_3\up{I}\cup c_3\up{K}+c_3\up{K}\cup C_3\up{I}\rb)\,.
\end{align}
Integrating out $c_3\up{K}$ enforces $c_3\up{I}=C_3\up{I}$, leaving the partition function $Z_{\fT_J}[C_3\up{I}]$ on $M_6\gep$: 
\begin{align}
    Z_{\fT_J}[C_3\up{I};\, \ol{\cT}_5(M_{5|0}),\, \cT_5(M_{5|\eps})] &=N\dw{M_6\ze,\,\Z_2} \sum_{c_3\up{J}\in H^3(M_6^{[0,\eps]},\,M_{5|0}\cup M_{5|\eps},\,\Z_2)} 
    \exp(i\pi\int_{M_6\ze}\lb c_3\up{J}\cup C_3\up{I}+\bar{q}(C_3\up{I})\rb)\,\times\nn\\
    &\times \,\chiiM{\gep}\,Z_{\fT_J}[ C_3\up{I}]\,.
\end{align}
As discussed before equation \eqref{eq:q_0eps_to0}, in the $\eps\to0$ limit the integral of a quadratic function of a classical background field vanishes.
We then use the isomorphism \eqref{eq:iso_H^3_H^2} and Poincaré duality on $M_5$ to map $c_3\up{J}\mapsto M_3\in H_3(M_5,\,\Z_2)$ and obtain, in the $\eps\to 0$ limit (for the normalization, we refer to \cref{foot:norm2}):
\begin{align} 
    \ol{\cT}_5(M_5)\otimes \cT_5(M_5)&=\chiiM{\gz}\;N\dw{M_5,\,\Z_2}\sum_{M_3\in H_3(M_5,\,\Z_2)}\,\exp(i\pi\int_{M_3}C_3\up{I})\,,
\end{align}
recalling the definition of the condensation defect \eqref{eq:C5I_def}, we therefore have: 
\begin{align}
    \ol{\cT}_5(M_5)\otimes \cT_5(M_5)&=\cC_5\up{0}(M_5)\,.
\end{align}

\paragraph{$\bm{\cT_5(M_5)\otimes \ol{\cT}_5(M_5)}$.} We depict the fusion in the opposite order to the one computed above as:
\begin{align} \label{eq:cT_cTbar_6d_fig}
    &\;{\color{red}\cT_5(M_{5|0})} \nn\\ 
     \cL_{\fT_J}[C_3\up{I}]  & {\color{red} \quad\bigg|_{c_3\up{K}|_{M_{5}}=0}} \quad \cL_{\fT_I}[c_3\up{K}] \,+\pi\lb  \bar{q}(c_3\up{K})+c_3\up{K}\cup C_3\up{I}\rb\\[2mm]
    &&&\hspace{-4cm}\;{\color{red}\ol{\cT}_5(M_{5|\eps})} \nn \\
    &&&\hspace{-4cm}{\color{red}\quad\bigg|_{\substack{c_3\up{J}|_{M_{5|\eps}}=0}}}\quad \cL_{\fT_J}[c_3\up{I}] \;+\pi\lb \bar{q}(c_3\up{I})+c_3\up{I}\cup c_3\up{J}+c_3\up{J}\cup C_3\up{I}+\bar{q}(C_3\up{I})\rb\,.
\end{align}
From its definition in equation \eqref{eq:cT5bar_def_fus}, the  defect $\ol{\cT}_5(M_{5|\eps})$ first acts with a Green-Schwarz interface $\ol{T}_5$ at $M_{5|\eps}$, whose action on the gauge fields crossing it is:
\begin{subequations} \label{eq:Tbar_M5eps_converts}
    \begin{align}
         c_3\up{K}|_{M_{5|\eps}}\, \ol{T}_5(M_{5|\eps})&=\ol{T}_5(M_{5|\eps})\,c_3\up{I}|_{M_{5|\eps}}\,, \\
         c_3\up{I}|_{M_{5|\eps}}\, \ol{T}_5(M_{5|\eps})&=\ol{T}_5(M_{5|\eps})\,c_3\up{J}|_{M_{5|\eps}}\,.
    \end{align}
\end{subequations}
Similarly to above, the Dirichlet boundary condition on $M_{5|\eps}$ for $c_3\up{J}$ and the invertibility of $\ol{T}_5$ imply the same boundary condition on $M_{5|\eps}$ also for $c_3\up{I},\,c_3\up{K}$. The equation corresponding to the above figure is:
\begin{align} \label{eq:cT_cTbar_6d_eq}
    Z_{\fT_J}[C_3\up{I};\, \cT_5(M_{5|0}),\, \ol{\cT}_5(M_{5|\eps})]=N\dw{M_6\ze,\,\Z_2} \sum_{c_3\up{K}\in H^3(M_6^{[0,\eps]},\,M_{5|0}\cup M_{5|\eps},\,\Z_2)} 
    \exp(i\pi\int_{M_6\ze}\lb  \bar{q}(c_3\up{K})+c_3\up{K}\cup C_3\up{I}\rb)\,\times\nn\\
    \times N\dw{M_6\gep,\,\Z_2}^2\sum_{c_3\up{I},\,c_3\up{J} \in H^3(M_6^{\gs\eps},\,M_{5|\eps},\,\Z_2)}\,Z_{\fT_J}[c_3\up{I}]\,\exp(i\pi\int_{M_6\gep}\lb \bar{q}(c_3\up{I})+c_3\up{I}\cup c_3\up{J}+c_3\up{J}\cup C_3\up{I}+\bar{q}(C_3\up{I})\rb)\,.
\end{align}
Integrating out $c_3\up{J}$  enforces $c_3\up{I}=C_3\up{I}$, leaving the partition function $Z_{\fT_J}[C_3\up{I}]$ on $M_6\gep$ (note that for $c_3\up{I}=C_3\up{I}$ we have $\int (\bar{q}(c_3\up{I})+\bar{q}(C_3\up{I}))=2\int \bar{q}(C_3\up{I})=0$ since $\int \bar{q}$ is $\Z_2$-valued), therefore: 
\begin{align}
    Z_{\fT_J}[C_3\up{I};\, \cT_5(M_{5|0}),\, \ol{\cT}_5(M_{5|\eps})]&=N\dw{M_6\ze,\,\Z_2} \sum_{c_3\up{K}\in H^3(M_6^{[0,\eps]},\,M_{5|0}\cup M_{5|\eps},\,\Z_2)} 
    \exp(i\pi\int_{M_6\ze}\lb  \bar{q}(c_3\up{K})+c_3\up{K}\cup C_3\up{I}\rb)\,\times\nn\\
    &\times \,\chiiM{\gep}\,Z_{\fT_J}[ C_3\up{I}]\,.
\end{align}
We then use the isomorphism \eqref{eq:iso_H^3_H^2} to map $c_3\up{K}\mapsto b_2\in H^2(M_5,\,\Z_2)$ which implies:
\begin{equation}
    \int_{M_6\ze}\bar{q}(c_3\up{K})\mapsto \int_{M_5}b_2\cup\beta(b_2)\,,
\end{equation}
where $\beta:H^2(M_5,\,\Z_2)\to H^{3}(M_5,\,\Z_2)$ is the Bockstein homomorphism associated to the short exact sequence $1\to\Z_2\to\Z_4\to\Z_2\to1$. Therefore, as $\eps\to0$ we have:
\begin{align}
    \cT_5(M_5)\otimes \ol{\cT}_5(M_5)=&\chiiM{\gz}\,N\dw{M_5,\,\Z_2}\sum_{b_2\in H^2(M_5,\,\Z_2)}\exp(i\pi\int_{M_5} b_2\cup\beta(b_2)+ b_2 \cup C_3\up{I})\,.
\end{align}
Using Poincaré duality on $M_5$ converts the sum over $b_2\in H^2(M_5,\,\Z_2)$ to one over $M_3=\PD(b_2)\in H_3(M_5,\,\Z_2)$. We can thus define the discrete torsion term:
\begin{align} \label{eq:QM5M3}
    Q(M_5,M_3)=\int_{M_5}\PD(M_3)\cup\beta(\PD(M_3))
\end{align}
and the resulting condensation defect with non-trivial discrete torsion:\footnote{We also include the Euler counterterm in its normalization.}
\begin{align} \label{eq:Cond_1_def}
    \cC_5\up{1}(M_5)= \chiiM{\gz}\,N\dw{M_5,\,\Z_2}\sum_{M_3\in H_3(M_5,\,\Z_2)}\,\exp(i\pi\,Q(M_5,M_3)+i\pi\int_{M_3}C_3\up{I})
\end{align}
where $N\dw{M_5,\,\Z_2}$ is given by equation \eqref{eq:Norm_M5}. The fusion is therefore:
\begin{align} \label{eq:cT_cTbar_6d_fusion}
    \cT_5(M_5)\otimes \ol{\cT}_5(M_5)&= \cC_5\up{1}(M_5)\,.
\end{align}

\paragraph{$\bm{\cT_5(M_5)\otimes \cT_5(M_5)}$.} Let us now depict the fusion between two triality defects $\cT_5(M_5)$:
\begin{align}
    &\;{\color{red}\cT_5(M_{5|0})} \nn\\ 
     \cL_{\fT_J}[C_3\up{I}]  &{\color{red} \quad\bigg|_{c_3\up{K}|_{M_{5}}=0}} \quad \cL_{\fT_I}[c_3\up{K}] \,+\pi\lb  \bar{q}(c_3\up{K})+c_3\up{K}\cup C_3\up{I}\rb
     \\[2mm]
    &&&\hspace{-6cm}\;{\color{red}\cT_5(M_{5|\eps})} \nn \\
    &&&\hspace{-6cm}{\color{red}\quad\bigg|_{\substack{c_3\up{K}|_{M_{5|\eps}}=0}}}\quad \cL_{\fT_K}[c_3\up{J}] \;+\pi\lb \bar{q}(c_3\up{J})+c_3\up{J}\cup c_3\up{K}+\bar{q}(c_3\up{K})+c_3\up{K}\cup C_3\up{I}\rb\,.
\end{align}
As before, ${\cT}_5(M_{5|\eps})$ first acts with a Green-Schwarz interface $T_5$ at $M_{5|\eps}$, recall equation \eqref{eq:T_M5eps_converts}, so we must also impose the Dirichlet boundary condition on $M_{5|\eps}$ for $c_3\up{J}$. We therefore write the equation:
\begin{align}
     Z_{\fT_J}[C_3\up{I};\, \cT_5(M_{5|0}),\, {\cT}_5(M_{5|\eps})]=N\dw{M_6\ze,\,\Z_2} \sum_{c_3\up{K}\in H^3(M_6^{[0,\eps]},\,M_{5|0}\cup M_{5|\eps},\,\Z_2)} 
    \exp(i\pi\int_{M_6\ze}\lb  \bar{q}(c_3\up{K})+c_3\up{K}\cup C_3\up{I}\rb)\,\times\nn\\
    \times N\dw{M_6\gep,\,\Z_2}^2\sum_{c_3\up{J},\,c_3\up{K} \in H^3(M_6^{\gs\eps},\,M_{5|\eps},\,\Z_2)}\,Z_{\fT_K}[c_3\up{J}]\,\exp(i\pi\int_{M_6\gep}\lb \bar{q}(c_3\up{J})+c_3\up{J}\cup c_3\up{K}+\bar{q}(c_3\up{K})+c_3\up{K}\cup C_3\up{I}\rb)\,.
\end{align}
Using the property of $\bar{q}$ written in equation \eqref{eq:q_prop}, we have:
\begin{align}
    Z_{\fT_J}[C_3\up{I};\, \cT_5(M_{5|0}),\, {\cT}_5(M_{5|\eps})]=N\dw{M_6\ze,\,\Z_2} \sum_{c_3\up{K}\in H^3(M_6^{[0,\eps]},\,M_{5|0}\cup M_{5|\eps},\,\Z_2)} 
    \exp(i\pi\int_{M_6\ze}\lb  \bar{q}(c_3\up{K})+c_3\up{K}\cup C_3\up{I}\rb)\,\times\nn\\
   \times N\dw{M_6\gep,\,\Z_2}^2\sum_{c_3\up{J},\,c_3\up{K} \in H^3(M_6^{\gs\eps},\,M_{5|\eps},\,\Z_2)}\,Z_{\fT_K}[c_3\up{J}]\,\exp(i\pi \int_{M_6\gep}\lb \bar{q}(c_3\up{J}+c_3\up{K}+C_3\up{I})+c_3\up{J}\cup C_3\up{I}+\bar{q}(C_3\up{I}) \rb)\,.
\end{align}
Replacing the sum over $c_3\up{K}$ on $M_6\gep$ with one over $\Tilde{c}_3\up{K}=c_3\up{J}+c_3\up{K}+C_3\up{I}$ gives:
\begin{align} \label{eq:cT5_squared_calc}
    Z_{\fT_J}[C_3\up{I};\, \cT_5(M_{5|0}),\, {\cT}_5(M_{5|\eps})]=N\dw{M_6\ze,\,\Z_2} \sum_{{c}_3\up{K}\in H^3(M_6^{[0,\eps]},\,M_{5|0}\cup M_{5|\eps},\,\Z_2)} 
    \exp(i\pi\int_{M_6\ze}\lb  \bar{q}(c_3\up{K})+c_3\up{K}\cup C_3\up{I}\rb)\,\times\nn\\
   \times N\dw{M_6\gep,\,\Z_2}^2\sum_{c_3\up{J},\,\Tilde{c}_3\up{K} \in H^3(M_6^{\gs\eps},\,M_{5|\eps},\,\Z_2)}\,Z_{\fT_K}[c_3\up{J}]\,\exp(i\pi \int_{M_6\gep}\lb \bar{q}(\Tilde{c}_3\up{K})+c_3\up{J}\cup C_3\up{I}+\bar{q}(C_3\up{I}) \rb)\,.
\end{align}
In the $\eps\to0$ limit, the term on $M_6\ze$ gives the condensation defect with non-trivial twist $\cC_5\up{1}(M_5)$, as discussed for the computation of $\cT_5(M_5)\otimes \ol{\cT}_5(M_5)$. The sum over $\Tilde{c}_3\up{K}$ produces a decoupled partition function
\begin{equation} \label{eq:Z_Ygz}
    \cZ\dw{Y,M_6\gz}=\chiM{\gz}\,N\dw{M_6\gz,\,\Z_2}\,\sum_{c_3\in H^3(M_6\gz,\,M_5,\,\Z_2)}\,e^{i\pi\, \int_{M_6\gz} \bar{q}(c_3)}
\end{equation}
that is analogous to equation \eqref{eq:Z_Y} but now defined on a manifold $M_6\gz$ with boundary $M_5$.\footnote{Note that since we include a factor of $\chiiM{\gz}$ in the normalization of the condensation defect, we have a compensating factor of $\chiM{\gz}$ in the normalization of \eqref{eq:Z_Ygz}.} Finally, we recognize from \eqref{eq:cT5bar_6d_eq} that the last terms are the insertion in the partition function of $\ol{\cT}_5(M_5)$, therefore:
\begin{align}
     \cT_5(M_5)\otimes \cT_5(M_5)=\cZ\dw{Y,M_6\gz}\, \cC_5\up{1}(M_5)\otimes \ol{\cT}_5(M_5)\,.
\end{align}
The fusion on the right-hand-side, involving the condensation defect, will be computed in appendix \ref{sec:cond_fusions}.

\paragraph{$\bm{\ol{\cT}_5(M_5)\otimes \ol{\cT}_5(M_5)}$.} We depict the fusion between two $\ol{\cT}_5(M_5)$ defects as:
\begin{align}
    &\;{\color{red}\ol{\cT}_5(M_{5|0})} \nn\\ 
     \cL_{\fT_J}[C_3\up{I}]  & {\color{red} \quad\bigg|_{c_3\up{J}|_{M_{5|0}}=0}} \quad \cL_{\fT_K}[c_3\up{J}] \,+\pi\lb c_3\up{J}\cup C_3\up{I}+\bar{q}(C_3\up{I})\rb\\[2mm]
    &&&\hspace{-6cm}\;{\color{red}\ol{\cT}_5(M_{5|\eps})} \nn \\
    &&&\hspace{-6cm}{\color{red}\quad\bigg|_{\substack{c_3\up{J}|_{M_{5|\eps}}=0}}}\quad \cL_{\fT_I}[c_3\up{K}] \;+\pi\lb c_3\up{K}\cup c_3\up{J}+ \bar{q}(c_3\up{J})+c_3\up{J}\cup C_3\up{I}+\bar{q}(C_3\up{I})\rb\,.
\end{align}
As before, $\ol{\cT}_5(M_{5|\eps})$ first acts with a Green-Schwarz interface $\ol{T}_5$ at $M_{5|\eps}$, recall equation \eqref{eq:Tbar_M5eps_converts}, so we must also impose the Dirichlet boundary condition on $M_{5|\eps}$ for $c_3\up{K}$.
The equation corresponding to the above figure is:
\begin{align}
    Z_{\fT_J}[C_3\up{I};\, \ol{\cT}_5(M_{5|0}),\, \ol{\cT}_5(M_{5|\eps})]=N\dw{M_6\ze,\,\Z_2} \sum_{c_3\up{J}\in H^3(M_6^{[0,\eps]},\,M_{5|0}\cup M_{5|\eps},\,\Z_2)}  
    \exp(i\pi\int_{M_6\ze}\lb c_3\up{J}\cup C_3\up{I}+\bar{q}(C_3\up{I})\rb)\,\times\nn\\
    \times N\dw{M_6\gep,\,\Z_2}^2\sum_{c_3\up{K},\,c_3\up{J} \in H^3(M_6^{\gs\eps},\,M_{5|\eps},\,\Z_2)}\,Z_{\fT_I}[c_3\up{K}]\,\exp(i\pi\int_{M_6\gep}\lb c_3\up{K}\cup c_3\up{J}+ \bar{q}(c_3\up{J})+c_3\up{J}\cup C_3\up{I}+\bar{q}(C_3\up{I})\rb)\,.
\end{align}
We recall that, as $\eps\to0$ we have $\int_{M_6\ze}\bar{q}(C_3\up{I})\to0$ and use the property of $\bar{q}$ in equation \eqref{eq:q_prop}, to write:
\begin{align}
    Z_{\fT_J}[C_3\up{I};\, \ol{\cT}_5(M_{5|0}),\, \ol{\cT}_5(M_{5|\eps})]=N\dw{M_6\ze,\,\Z_2} \sum_{c_3\up{J}\in H^3(M_6^{[0,\eps]},\,M_{5|0}\cup M_{5|\eps},\,\Z_2)}  
    \exp(i\pi\int_{M_6\ze}c_3\up{J}\cup C_3\up{I})\,\times\nn\\
    \times N\dw{M_6\gep,\,\Z_2}^2\sum_{c_3\up{K},\,c_3\up{J} \in H^3(M_6^{\gs\eps},\,M_{5|\eps},\,\Z_2)}\,Z_{\fT_I}[c_3\up{K}]\,\exp(i\pi\int_{M_6\gep}\lb \bar{q}(c_3\up{K}+ c_3\up{J}+C_3\up{I})+\bar{q}(c_3\up{K})+c_3\up{K}\cup C_3\up{I}\rb)\,.
\end{align}
We can replace the sum over $c_3\up{J}$ with one over $\Tilde{c}_3\up{J}=c_3\up{K}+c_3\up{J}+C_3\up{I}$:
\begin{align}
    Z_{\fT_J}[C_3\up{I};\, \ol{\cT}_5(M_{5|0}),\, \ol{\cT}_5(M_{5|\eps})]=N\dw{M_6\ze,\,\Z_2} \sum_{c_3\up{J}\in H^3(M_6^{[0,\eps]},\,M_{5|0}\cup M_{5|\eps},\,\Z_2)}  
    \exp(i\pi\int_{M_6\ze}c_3\up{J}\cup C_3\up{I})\,\times\nn\\
    \times N\dw{M_6\gep,\,\Z_2}^2\sum_{c_3\up{K},\,\Tilde{c}_3\up{J} \in H^3(M_6^{\gs\eps},\,M_{5|\eps},\,\Z_2)}\,Z_{\fT_I}[c_3\up{K}]\,\exp(i\pi\int_{M_6\gep}\lb \bar{q}(\Tilde{c}_3\up{J})+\bar{q}(c_3\up{K})+c_3\up{K}\cup C_3\up{I}\rb)\,.
\end{align}
In the $\eps\to0$ limit, the term on $M_6\ze$ gives the condensation defect with trivial twist $\cC_5\up{0}(M_5)$, as discussed for the computation of $\ol{\cT}_5(M_5)\otimes{\cT}_5(M_5)$. The sum over $\Tilde{c}_3\up{J}$ produces a decoupled $\cZ\dw{Y,M_6\gz}$, defined in equation \eqref{eq:Z_Ygz}. Finally, we recognize from \eqref{eq:cT5_6d_eq} that the last terms are the insertion in the partition function of ${\cT}_5(M_5)$, therefore:
\begin{equation}
    \ol{\cT}_5(M_5)\otimes \ol{\cT}_5(M_5)=
    \cZ\dw{Y,M_6\gz}\, \cC_5\up{0}(M_5)\otimes{\cT}_5(M_5)\,.
\end{equation}
The fusion on the right-hand-side, involving the condensation defect, will be computed in appendix \ref{sec:cond_fusions}.

\subsection{Fusions between duality and triality defects}
\paragraph{$\bm{\cD_5(M_5)\otimes \ol{\cT}_5(M_5)}$.} We can schematically depict the fusion of a duality defect $\cD_5(M_5)$ with a triality defect $\ol{\cT}_5(M_5)$ as:
\begin{align}
    &\;{\color{red}\cD_5(M_{5|0})} &&\;{\color{red}\ol{\cT}_5(M_{5|\eps})} \nonumber \\
    \cL_{\fT_J}[C_3\up{I}]  &{\color{red} \quad\bigg|_{c_3\up{J}|_{M_{5|0}}=0}}  \cL_{\fT_I}[c_3\up{J}]\;+\pi( c_3\up{J}\cup C_3\up{I})  &&{\color{red}\quad\bigg|_{\substack{c_3\up{J}|_{M_{5|\eps}}=0}}} \quad\cL_{\fT_J}[c_3\up{K}] \;+\pi\lb c_3\up{K}\cup c_3\up{J}+ c_3\up{J}\cup C_3\up{I}+\bar{q}(C_3\up{I})\rb\,.
\end{align}
The equation for the above picture is:
\begin{align}
    Z_{\fT_J}[C_3\up{I};\,\cD_5(M_{5|0}),\, \ol{\cT}_5(M_{5|\eps})]= N\dw{M_6\ze,\,\Z_2} \sum_{{c}_3\up{J} \in H^3(M_6^{[0,\eps]},\,M_{5|0}\cup M_{5|\eps},\,\Z_2)} 
    \exp(i\pi\int_{M_6\ze}{c}_3\up{J}\cup C_3\up{I})\,\times\nn\\
    \times N\dw{M_6\gep,\,\Z_2}^2\sum_{c_3\up{K},\,c_3\up{J}\in H^3(M_6^{\gs\eps},\,M_{5|\eps},\,\Z_2)}
    Z_{\fT_J}[c_3\up{K}]\,\exp(i\pi\int_{M_6\gep}\lb c_3\up{K}\cup c_3\up{J}+ c_3\up{J}\cup C_3\up{I}+\bar{q}(C_3\up{I})\rb)\,.
\end{align}
On $M_6\gep$, integrating out $c_3\up{J}$ sets $c_3\up{K}=C_3\up{I}$:
\begin{align}
    Z_{\fT_J}[C_3\up{I};\,\cD_5(M_{5|0}),\, \ol{\cT}_5(M_{5|\eps})]= N\dw{M_6\ze,\,\Z_2} \sum_{{c}_3\up{J} \in H^3(M_6^{[0,\eps]},\,M_{5|0}\cup M_{5|\eps},\,\Z_2)} 
    \exp(i\pi\int_{M_6\ze}{c}_3\up{J}\cup C_3\up{I})\,\times\nn\\
    \times\; \chiiM{\gep}\;
    Z_{\fT_J}[C_3\up{I}]\,\exp(i\pi\int_{M_6\gep}\bar{q}(C_3\up{I}))\,.
\end{align}
As for the fusion of two duality defects, described in detail in section \ref{sec:D5_D5_fusion}, in the $\eps\to0$ limit the term on $M_6\ze$ gives rise to a condensation defect $\cC_5\up{0}(M_5)$. There is, however, an extra term on $M_6\gz$: this is a 6d $\Z_2\up{I}$-SPT on a manifold $M_6\gep$ with boundary $M_5$:
\begin{align}
    \cD_5(M_5)\otimes \ol{\cT}_5(M_5)&=\cC_5\up{0}\;(M_5)\;e^{i\pi\int_{M_6\gz}\bar{q}(C_3\up{I})}\,.
\end{align}

\paragraph{$\bm{\cT_5(M_5)\otimes \cD_5(M_5)}$.} Fusing a triality defect $\cT_5(M_5)$ with a duality defect $\cD_5(M_5)$ is depicted as:
\begin{align}
    &\;{\color{red}\cT_5(M_{5|0})} &&\;{\color{red}\cD_5(M_{5|\eps})} \nonumber \\
    \cL_{\fT_J}[C_3\up{I}]  &{\color{red} \quad\bigg|_{c_3\up{K}|_{M_{5}}=0}}\hspace{-8mm} \cL_{\fT_I}[c_3\up{K}] \,+\pi\lb  \bar{q}(c_3\up{K})+c_3\up{K}\cup C_3\up{I}\rb&&{\color{red}\quad\bigg|_{c_3\up{J}|_{M_{5|\eps}}=0}} \hspace{-8mm} \cL_{\fT_J}[c_3\up{K}] \;+\pi\lb \bar{q}(c_3\up{K})+c_3\up{K}\cup c_3\up{J}+ c_3\up{J}\cup C_3\up{I}\rb\,.
\end{align}
with corresponding equation:
\begin{align}
    Z_{\fT_J}[C_3\up{I};\,\cT_5(M_{5|0}),\, \cD_5(M_{5|\eps})]= N\dw{M_6\gz,\,\Z_2} \sum_{c_3\up{K}\in H^3(M_6^{\gs0},\,M_{5|0},\,\Z_2)} 
    \exp(i\pi\int_{M_6\ze}\lb  \bar{q}(c_3\up{K})+c_3\up{K}\cup C_3\up{I}\rb)\,\times\nn\\
    \times N\dw{M_6\gep,\,\Z_2}\sum_{c_3\up{J}\in H^3(M_6^{\gs\eps},\,M_{5|\eps},\,\Z_2)}
    Z_{\fT_J}[c_3\up{K}]\,\exp(i\pi\int_{M_6\gep}\lb \bar{q}(c_3\up{K})+c_3\up{K}\cup c_3\up{J}+ c_3\up{J}\cup C_3\up{I}\rb)\,.
\end{align}
Integrating out $c_3\up{J}$ sets $c_3\up{K}|_{M_6\gep}=C_3\up{I}$, reducing the dynamical $c_3\up{K}$ on $M_6\ze$ with Dirichlet boundary conditions on both boundaries:
\begin{align}
     Z_{\fT_J}[C_3\up{I};\,\cT_5(M_{5|0}),\, \cD_5(M_{5|\eps})]= N\dw{M_6\ze,\,\Z_2} \sum_{c_3\up{K} \in H^3(M_6^{[0,\eps]},\,M_{5|0}\cup M_{5|\eps},\,\Z_2)} 
    \exp(i\pi\int_{M_6\ze}\lb  \bar{q}(c_3\up{K})+c_3\up{K}\cup C_3\up{I}\rb)\,\times\nn\\
    \times\; \chiiM{\gep}\;
    Z_{\fT_J}[C_3\up{I}]\,\exp(i\pi\int_{M_6\gep}\bar{q}(C_3\up{I}))\,.
\end{align}
As discussed for triality fusions in section \ref{sec:triality_triality_fusions}, the term on $M_6\ze$ gives rise to the condensation defect $\cC_5\up{1}(M_5)$, defined in \eqref{eq:Cond_1_def}. Like for the previous case, we also have a term on $M_6\gz$, so the result is:
\begin{align}
    \cT_5(M_5)\otimes \cD_5(M_5)&=\cC_5\up{1}(M_5)\;e^{i\pi\int_{M_6\gz}\bar{q}(C_3\up{I})}\,.
\end{align}

\paragraph{$\bm{\cD_5(M_5)\otimes {\cT}_5(M_5)}$ {and} $\bm{\ol{\cT}_5(M_5)\otimes \cD_5(M_5)}$. }
For  the fusion $\cD_5(M_5)\otimes {\cT}_5(M_5)$, we draw schematically:
\begin{align}
    &\;{\color{red}\cD_5(M_{5|0})} &&\;{\color{red}{\cT}_5(M_{5|\eps})} \nonumber \\
    \cL_{\fT_J}[C_3\up{I}]\quad  &{\color{red} \bigg|_{c_3\up{J}|_{M_{5|0}}=0}}  \cL_{\fT_I}[c_3\up{J}]\;+\pi( c_3\up{J}\cup C_3\up{I})  &&{\color{red}\quad\bigg|_{\substack{c_3\up{K}|_{M_{5|\eps}}=0}}} \cL_{\fT_K}[c_3\up{I}] \;+\pi\lb c_3\up{I}\cup c_3\up{K}+ \bar{q}(c_3\up{K})+c_3\up{K}\cup C_3\up{I}\rb\,.
\end{align}
The equation for the above picture is:
\begin{align}
    Z_{\fT_J}[C_3\up{I};\,\cD_5(M_{5|0}),\, \cT_5(M_{5|\eps})]= N\dw{M_6\ze,\,\Z_2} \sum_{c_3\up{J} \in H^3(M_6^{[0,\eps]},\,M_{5|0}\cup M_{5|\eps},\,\Z_2)} 
    \exp(i\pi\int_{M_6\ze}c_3\up{J}\cup C_3\up{I})\,\times\nn\\
    \times N\dw{M_6\gep,\,\Z_2}^2\sum_{c_3\up{K},\,c_3\up{I}\in H^3(M_6^{\gs\eps},\,M_{5|\eps},\,\Z_2)}
    Z_{\fT_K}[c_3\up{I}]\,\exp(i\pi\int_{M_6\gep}\lb c_3\up{I}\cup c_3\up{K}+ \bar{q}(c_3\up{K})+c_3\up{K}\cup C_3\up{I}\rb)\,.
\end{align}
Using the properties of $\bar{q}$, we can write:
\begin{align} 
      &Z_{\fT_J}[C_3\up{I};\,\cD_5(M_{5|0}),\, \cT_5(M_{5|\eps})]= N\dw{M_6\ze,\,\Z_2} \sum_{c_3\up{J} \in H^3(M_6^{[0,\eps]},\,M_{5|0}\cup M_{5|\eps},\,\Z_2)} 
    \exp(i\pi\int_{M_6\ze}c_3\up{J}\cup C_3\up{I})\,\times\nn\\
    &\times N\dw{M_6\gep,\,\Z_2}^2\sum_{c_3\up{K},\,c_3\up{I}\in H^3(M_6^{\gs\eps},\,M_{5|\eps},\,\Z_2)}
    Z_{\fT_K}[c_3\up{I}]\,\exp(i\pi\int_{M_6\gep}\lb \bar{q}(c_3\up{I}+c_3\up{K}+C_3\up{I})+\bar{q}(c_3\up{I})+c_3\up{I}\cup C_3\up{I}+\bar{q}(C_3\up{I})\rb)\,.
\end{align}
We can define $\Tilde{c}_3=c_3\up{I}+c_3\up{K}+C_3\up{I}$, to replace the sum over $c_3\up{K}$:
\begin{align} \label{eq:same1}
      &Z_{\fT_J}[C_3\up{I};\,\cD_5(M_{5|0}),\, \cT_5(M_{5|\eps})]= N\dw{M_6\ze,\,\Z_2} \sum_{c_3\up{J} \in H^3(M_6^{[0,\eps]},\,M_{5|0}\cup M_{5|\eps},\,\Z_2)} 
    \exp(i\pi\int_{M_6\ze}c_3\up{J}\cup C_3\up{I})\,\times\nn\\
    &\times N\dw{M_6\gep,\,\Z_2}^2\sum_{\Tilde{c}_3,\,c_3\up{I}\in H^3(M_6^{\gs\eps},\,M_{5|\eps},\,\Z_2)}
    Z_{\fT_K}[c_3\up{I}]\,\exp(i\pi\int_{M_6\gep}\lb \bar{q}(\Tilde{c}_3)+\bar{q}(c_3\up{I})+c_3\up{I}\cup C_3\up{I}+\bar{q}(C_3\up{I})\rb)\,.
\end{align}

For $\ol{\cT}_5(M_5)\otimes \cD_5(M_5)$, we instead draw:
\begin{align}
    &\;{\color{red}\ol{\cT}_5(M_{5|0})} &&\;{\color{red}\cD_5(M_{5|\eps})} \nonumber \\
    \cL_{\fT_J}[C_3\up{I}]  &{\color{red} \quad\bigg|_{c_3\up{J}|_{M_{5}}=0}}\hspace{-8mm} \cL_{\fT_K}[c_3\up{J}] \,+\pi\lb  c_3\up{J}\cup C_3\up{I}+\bar{q}(C_3\up{I})\rb&&{\color{red}\quad\bigg|_{c_3\up{J}|_{M_{5|\eps}}=0}} \hspace{-8mm} \cL_{\fT_K}[c_3\up{I}] \,+\pi\lb  c_3\up{I}\cup c_3\up{J}+\bar{q}(c_3\up{J})+c_3\up{J}\cup C_3\up{I}\rb\,.
\end{align}
with corresponding equation:
\begin{align}
    Z_{\fT_J}[C_3\up{I};\,\ol{\cT}_5(M_{5|0}),\, \cD_5(M_{5|\eps})]= N\dw{M_6\ze,\,\Z_2} \sum_{c_3\up{J}\in H^3(M_6\ze,\,M_{5|0}\cup M_{5|\eps},\,\Z_2)} 
    \exp(i\pi\int_{M_6\ze}\lb c_3\up{J}\cup C_3\up{I}+\bar{q}(C_3\up{I})\rb)\,\times\nn\\
    \times N\dw{M_6\gep,\,\Z_2}^2\sum_{c_3\up{J},\,c_3\up{I}\in H^3(M_6^{\gs\eps},\,M_{5|\eps},\,\Z_2)}
    Z_{\fT_K}[c_3\up{I}]\,\exp(i\pi\int_{M_6\gep}\lb c_3\up{I}\cup c_3\up{J}+\bar{q}(c_3\up{J})+c_3\up{J}\cup C_3\up{I}\rb)\,.
\end{align}
Recall that $\int_{M_6\ze}\bar{q}(C_3\up{I})\to0$ as $\eps\to0$. On $M_6\gep$ we use the properties of $\bar{q}$ and write:
\begin{align}
    Z_{\fT_J}[C_3\up{I};\,\ol{\cT}_5(M_{5|0}),\, \cD_5(M_{5|\eps})]= N\dw{M_6\ze,\,\Z_2} \sum_{c_3\up{J}\in H^3(M_6\ze,\,M_{5|0}\cup M_{5|\eps},\,\Z_2)} 
    \exp(i\pi\int_{M_6\ze} c_3\up{J}\cup C_3\up{I})\,\times\nn\\
    \times N\dw{M_6\gep,\,\Z_2}^2\sum_{c_3\up{J},\,c_3\up{I}\in H^3(M_6^{\gs\eps},\,M_{5|\eps},\,\Z_2)}
    Z_{\fT_K}[c_3\up{I}]\,\exp(i\pi\int_{M_6\gep}\lb \bar{q}(c_3\up{I}+c_3\up{J}+C_3\up{I})+\bar{q}(c_3\up{I})+c_3\up{I}\cup C_3\up{I}+\bar{q}(C_3\up{I})\rb)\,.
\end{align}
If we replace the sum over $c_3\up{J}$ with one over $\Tilde{c}_3=c_3\up{I}+c_3\up{J}+C_3\up{I}$, we have:
\begin{align} \label{eq:same2}
      &Z_{\fT_J}[C_3\up{I};\,\ol{\cT}_5(M_{5|0}),\, \cD_5(M_{5|\eps})]= N\dw{M_6\ze,\,\Z_2} \sum_{c_3\up{J} \in H^3(M_6^{[0,\eps]},\,M_{5|0}\cup M_{5|\eps},\,\Z_2)} 
    \exp(i\pi\int_{M_6\ze}c_3\up{J}\cup C_3\up{I})\,\times\nn\\
    &\times N\dw{M_6\gep,\,\Z_2}^2\sum_{\Tilde{c}_3,\,c_3\up{I}\in H^3(M_6^{\gs\eps},\,M_{5|\eps},\,\Z_2)}
    Z_{\fT_K}[c_3\up{I}]\,\exp(i\pi\int_{M_6\gep}\lb \bar{q}(\Tilde{c}_3)+\bar{q}(c_3\up{I})+c_3\up{I}\cup C_3\up{I}+\bar{q}(C_3\up{I})\rb)\,.
\end{align}
The right-hand sides of equations \eqref{eq:same1} and \eqref{eq:same2} are the same, showing that:
\begin{align}
    \cD_5(M_5)\otimes {\cT}_5(M_5)=\ol{\cT}_5(M_5)\otimes \cD_5(M_5)\,.
\end{align}

\subsection{Fusions with condensation defects} \label{sec:cond_fusions}
Equations \eqref{eq:cD_cD_6d_fig}-\eqref{eq:cD5_cD5_fusion_6d} and \eqref{eq:cT_cTbar_6d_fig}-\eqref{eq:cT_cTbar_6d_fusion} imply that we can write the insertion of the condensation defect $\cC_5\up{\ell},\,\ell\in\{0,1\}$ schematically as \cite{Choi:2022zal}
\begin{align} \label{eq:C5ell_6d_fig}
    &\;\; {\color{red}\cC_5\up{\ell}(M_5)} \; \nonumber \\
    \cL_{\fT_J}[C_3\up{I}]  & {\color{red} \quad\bigg|_{\substack{c_3\up{I}|_{M_{5}}=0\\
    c_3\up{J}|_{M_{5}}=0}}} \quad \cL_{\fT_J}[c_3\up{I}] \;+\pi\lb \ell\,\bar{q}(c_3\up{I})+c_3\up{I}\cup c_3\up{J}+c_3\up{J}\cup C_3\up{I}+\ell\,\bar{q}(C_3\up{I})\rb\,,
\end{align}
where both fields $c_3\up{I},\,c_3\up{J}$ are dynamical gauge fields with Dirichlet boundary conditions on the defect. The figure in \eqref{eq:C5ell_6d_fig} corresponds to the equation:
\begin{align} \label{eq:C5ell_6d_eq}
    Z_{\fT_J}[C_3\up{I};\,\cC_5\up{\ell}(M_5)]=&N\dw{M_6\gz,\,\Z_2}^2\sum_{\substack{c_3\up{I},\,c_3\up{J} \in H^3(M_6^{\gs 0},\,M_{5|0},\,\Z_2)}} 
    Z_{\fT_J}[c_3\up{I}]\,
    e^{i\pi\int_{M_6\gz}\lb \ell\,\bar{q}(c_3\up{I})+c_3\up{I}\cup c_3\up{J}+c_3\up{J}\cup C_3\up{I}+\ell\,\bar{q}(C_3\up{I})\rb}\,.
\end{align}

From the figures defining the symmetry defects $\cD_5(M_5)$, $\cT_5(M_5)$, $\ol{\cT}_5(M_5)$ in \eqref{eq:cD5_6d_fig}, \eqref{eq:cT5_6d_fig}, \eqref{eq:cT5bar_6d_fig}, we can generally denote them as $\cX_5(M_5)$ and depict:
\begin{equation} \label{eq:cX_6d_fig}
      \begin{split}
    &\;{\color{red}{\cX}_5(M_5)}\\
    \cL_{\fT_J}[C_3\up{I}]  &{\color{red} \quad\bigg|_{c_3|_{M_5}=0} } \quad  
   \cL[c_3]\,+\pi(\delta_{(\cX,\cT)}\,\bar{q}(c_3)+c_3\cup C_3\up{I}+\delta_{(\cX,\ol{\cT})}\,\bar{q}(C_3\up{I}))\,.
    \end{split}
\end{equation}
where $\delta_{(\cX,\cY)}$ is $1$ if $\cX=\cY$ and 0 otherwise.

The fusion $\cC_5\up{\ell}(M_5) \otimes \cX_5(M_5)$ gives a Dijkgraaf-Witten (DW) theory, (with twist if $\ell=1$) on $M_5$, multiplying $\cX_5(M_5)$. Indeed using the expressions \eqref{eq:C5ell_6d_fig}-\eqref{eq:C5ell_6d_eq} on $M_6\ze$ and the isomorphism \eqref{eq:iso_H^3_H^2}, we obtain the following $\Z_2$ TQFT on $M_5$ in the $\eps\to0$ limit:
\begin{align} \label{eq:Z2_DW}
    (\cZ_2)_{2\ell}\up{I}(M_5)&= \chiiM{\gz}
    \,N\dw{M_5,\,\Z_2}^2\,\sum_{b_2\up{I},\,b_2\up{J}\in H^2(M_5,\,\Z_2)}\exp(i\pi \int_{M_5}\ell \,b_2\up{I}\cup\beta(b_2\up{I})+b_2\up{J}\cup(\delta\,b_2\up{I}-C_3\up{I}))\,.
\end{align}
On $M_6\gep$ we have integrated out $c_3\up{J}$ appearing in the definition of the condensation defect \eqref{eq:C5ell_6d_fig}-\eqref{eq:C5ell_6d_eq}: this sets $c_3\up{I}=C_3\up{I}$ and leaves on $M_6\gep$ the insertion of the defect $\cX_5$:\footnote{It also produces a factor of $|H^3(M_6\gz,\Z_2)|$ which combined with $N\dw{M_6\gz,\,\Z_2}^2$ gives the Euler term $\chiiM{\gz}$ which we have included in the normalization of \eqref{eq:Z2_DW}.}
\begin{equation}
    \cC_5\up{\ell}(M_5) \otimes \cX_5(M_5)=(\cZ_2)_{2\ell}\up{I}(M_5)\;\cX_5(M_5)\,.
\end{equation}

For the fusion $\cX_5(M_5)\otimes \cC_5\up{\ell}(M_5)$, we can again integrate out $c_3\up{J}$ appearing in the definition of the condensation defect, that leaves us with $\cX_5$ on $M_6\gep$. On $M_6\ze$ we use the isomorphism \eqref{eq:iso_H^3_H^2} and obtain the condensation defect on $M_5$ $\cC_5\up{\delta_{(\cX,\cT)}}$: the twist is zero except when $\cX=\cT$ (owing to the $\bar{q}(c_3)$ term in its definition). Therefore:
\begin{equation}
    \cX_5(M_5)\otimes \cC_5\up{\ell}(M_5)=\cC_5\up{\delta_{(\cX,\cT)}}(M_5)\otimes  \cX_5(M_5)=(\cZ_2)_{2 \delta_{(\cX,\cT)}}\up{I}(M_5)\;\cX_5(M_5).
\end{equation}
Analogously, when fusing two condensation defects, we obtain:
\begin{align}
    \cC_5\up{\ell}(M_5)\otimes \cC_5\up{\ell'}(M_5)&=(\cZ_2)_{2\ell}\up{I}(M_5)\;\cC_5\up{\ell'}(M_5)\,.
\end{align}

\section{Details on SymTFT computations}
\subsection{Invertible GS defects} \label{app:GS_SymTFT_fusions}
In this subsection, we present the derivations of results in section \ref{sec:SymTFT-defects}.

\paragraph{Action of $T_6(M_6)$ on dimension-3 defects.} Below is the derivation of equation \eqref{eq:action_of_T6} using properties \eqref{eq:D3_fusions}-\eqref{eq:quantum_torus}:
\begin{align}
     D_3^{(S)}(N_3)\,T_6(M_6)&=\lb  N\dw{M_6,\,\Z_2}\rb ^2\sum_{M_3,M_3'\in H_3(M_6,\,\Z_2)}D_3^{(S)}(N_3)\,D_3^{(S)}(M_3)\,D_3^{(C)}(M_3')=\nn\\
     &=\lb  N\dw{M_6,\,\Z_2}\rb ^2\sum_{M_3,M_3'\in H_3(M_6,\,\Z_2)}(-1)^{\langle N_3,M_3'\rangle}D_3^{(S)}(M_3)\,D_3^{(C)}(M_3')\,D_3^{(S)}(N_3)=\nn\\
     &=\lb  N\dw{M_6,\,\Z_2}\rb ^2\sum_{M_3,M_3'\in H_3(M_6,\,\Z_2)}D_3^{(S)}(M_3)\,D_3^{(C)}(M_3'+N_3)\,D_3^{(V)}(N_3)=\nn\\
     &=\lb  N\dw{M_6,\,\Z_2}\rb ^2\sum_{M_3,\Tilde{M}_3'\in H_3(M_6,\,\Z_2)}D_3^{(S)}(M_3)\,D_3^{(C)}(\Tilde{M}_3')\,D_3^{(V)}(N_3)=\nn\\
     &=T_6(M_6)\,D_3^{(V)}(N_3)\,,
\end{align}
\begin{align}
     D_3^{(C)}(N_3)\,T_6(M_6)&=\lb  N\dw{M_6,\,\Z_2}\rb ^2\sum_{M_3,M_3'\in H_3(M_6,\,\Z_2)}D_3^{(C)}(N_3)\,D_3^{(S)}(M_3)\,D_3^{(C)}(M_3')=\nn\\
     &=\lb  N\dw{M_6,\,\Z_2}\rb ^2\sum_{M_3,M_3'\in H_3(M_6,\,\Z_2)}D_3^{(V)}(N_3)\,D_3^{(S)}(N_3)\,D_3^{(S)}(M_3)\,D_3^{(C)}(M_3')=\nn\\
     &=\lb  N\dw{M_6,\,\Z_2}\rb ^2\sum_{M_3,M_3'\in H_3(M_6,\,\Z_2)}D_3^{(S)}(M_3+N_3)D_3^{(S)}(N_3)\,D_3^{(C)}(N_3)\,D_3^{(C)}(M_3')=\nn\\
     &=\lb  N\dw{M_6,\,\Z_2}\rb ^2\sum_{M_3,M_3'\in H_3(M_6,\,\Z_2)}D_3^{(S)}(M_3+N_3)\,D_3^{(C)}(M_3'+N_3)\,D_3^{(S)}(N_3)=\nn\\
     &=\lb  N\dw{M_6,\,\Z_2}\rb ^2\sum_{\Tilde{M}_3,\Tilde{M}_3'\in H_3(M_6,\,\Z_2)}D_3^{(S)}(\Tilde{M}_3)\,D_3^{(C)}(\Tilde{M}_3')\,D_3^{(S)}(N_3)=\nn\\
     &=T_6(M_6)\,D_3^{(S)}(N_3)\,,
\end{align}
\begin{align}
     D_3^{(V)}(N_3)\,T_6(M_6)&=\lb  N\dw{M_6,\,\Z_2}\rb ^2\sum_{M_3,M_3'\in H_3(M_6,\,\Z_2)}D_3^{(V)}(N_3)\,D_3^{(S)}(M_3)\,D_3^{(C)}(M_3')=\nn\\
     &=\lb  N\dw{M_6,\,\Z_2}\rb ^2\sum_{M_3,M_3'\in H_3(M_6,\,\Z_2)}D_3^{(C)}(N_3)\,D_3^{(S)}(N_3)\,D_3^{(S)}(M_3)\,D_3^{(C)}(M_3')=\nn\\
    &=\lb  N\dw{M_6,\,\Z_2}\rb ^2\sum_{M_3,M_3'\in H_3(M_6,\,\Z_2)}D_3^{(S)}(M_3+N_3)\,D_3^{(C)}(M_3')\,D_3^{(C)}(N_3)=\nn\\
    &=\lb  N\dw{M_6,\,\Z_2}\rb ^2\sum_{\Tilde{M}_3,M_3'\in H_3(M_6,\,\Z_2)}D_3^{(S)}(\Tilde{M}_3)\,D_3^{(C)}(M_3')\,D_3^{(C)}(N_3)=\nn\\
     &=T_6(M_6)\,D_3^{(C)}(N_3)\,,
\end{align}

\paragraph{Fusion of GS2 and GS3 defects.} Let $(I,J,K)$ be a cyclic permutation of $(S,C,V)$, and recall definitions \eqref{eq:T6_def1}-\eqref{eq:olT6_def2} of $T_6(M_6)$ and $\ol{T}_6(M_6)$. We compute:
\begin{align}
    D_6\up{K}(M_6)\otimes T_6(M_6)&=\lb  N\dw{M_6,\,\Z_2}\rb ^3\sum_{M_3,N_3',M_3'\in H_3(M_6,\,\Z_2)}D_3^{(K)}(M_3)\,D_3^{(K)}(N_3')\,D_3^{(I)}(M_3')=\nn\\
    &=\lb  N\dw{M_6,\,\Z_2}\rb ^3\sum_{M_3,N_3',M_3'\in H_3(M_6,\,\Z_2)}(-1)^{\langle M_3,N_3'\rangle}D_3^{(K)}(M_3+N_3')\,D_3^{(I)}(M_3')=\nn\\ 
    &=\lb  N\dw{M_6,\,\Z_2}\rb ^3\sum_{M_3,\Tilde{N}_3',M_3'\in H_3(M_6,\,\Z_2)}(-1)^{\langle M_3,\Tilde{N}_3'\rangle}D_3^{(K)}(\Tilde{N}_3')\,D_3^{(I)}(M_3')=\nn\\ 
    &=\lb  N\dw{M_6,\,\Z_2}\rb ^2\sum_{\substack{\PD(M_3),\PD(\Tilde{N}_3')\\ \in H^3(M_6,\,\Z_2)}}\exp(i\pi\int_{M_6}\lb \PD(M_3)+c_3\up{K}\rb \cup \PD(\Tilde{N}_3'))\times\nn\\
    &\qquad\times  N\dw{M_6,\,\Z_2} \sum_{\substack{M_3'\in H_3(M_6,\,\Z_2)}}D_3^{(I)}(M_3')=\nn\\[2mm]
    &=\chiiM{}\; D_6\up{I}(M_6)
\end{align}
After writing the sum over $M_3,\Tilde{N}_3'$ as a sum over their Poincaré duals $\PD(M_3),\PD(\Tilde{N}_3')$, we integrated out $\PD(\Tilde{N}_3')$, trivializing the first integrand\footnote{It also produces a factor of $|H^3(M_6,\,\Z_2)|$, which, combined with $\lb  N\dw{M_6,\,\Z_2}\rb ^2$ gives $\chiiM{}$.} and leaving (up to an Euler term) the defect $D_6\up{I}(M_6)$. Similarly, exchanging $I\leftrightarrow J$ in the above calculation, we have:
\begin{align}
     D_6\up{K}(M_6)\otimes \ol{T}_6(M_6)=\chiiM{}\, D_6\up{J}(M_6)\,.
\end{align}
The fusions are non-abelian as can be seen by computing
\begin{align}
    T_6(M_6)\otimes D_6\up{K}(M_6)&=\lb  N\dw{M_6,\,\Z_2}\rb ^3\sum_{N_3,M_3,M_3'\in H_3(M_6,\,\Z_2)}D_3^{(J)}(N_3)\,D_3^{(K)}(M_3)\,D_3^{(K)}(M_3')=\nn\\
    &=\lb  N\dw{M_6,\,\Z_2}\rb ^3\sum_{N_3,M_3,M_3'\in H_3(M_6,\,\Z_2)}D_3^{(J)}(N_3)\,(-1)^{\langle M_3,M_3'\rangle}D_3^{(K)}(M_3+M_3')\,=\nn\\ 
    &=\lb  N\dw{M_6,\,\Z_2}\rb ^3\sum_{N_3,M_3,\Tilde{M}_3'\in H_3(M_6,\,\Z_2)}D_3^{(J)}(N_3)\,(-1)^{\langle M_3,\Tilde{M}_3'\rangle}D_3^{(K)}(\Tilde{M}_3')=\nn\\ 
    &= N\dw{M_6,\,\Z_2} \sum_{\substack{N_3\in H_3(M_6,\,\Z_2)}}D_3^{(J)}(N_3)\;\times\nn\\
    &\;\;\times\lb  N\dw{M_6,\,\Z_2}\rb ^2\sum_{\substack{\PD(M_3),\PD(\Tilde{M}_3')\\ \in H^3(M_6,\,\Z_2)}}\exp(i\pi\int_{M_6}\lb \PD(M_3)+c_3\up{K}\rb \cup \PD(\Tilde{M}_3'))\nn\\
    &=\chiiM{}\; D_6\up{J}(M_6)
\end{align}
where, similarly to above, after writing the sum over $M_3,\Tilde{M}_3'$ as a sum over their Poincaré duals $\PD(M_3),\PD(\Tilde{M}_3')$, we integrated out $\PD(\Tilde{M}_3')$, trivializing the integrand and leaving (up to an Euler term) the defect $D_6\up{J}(M_6)$. 

By exchanging $J\leftrightarrow I$ in the above, we also obtain:
\begin{align}
    \ol{T}_6(M_6)\otimes D_6\up{K}(M_6)&=\chiiM{}D_6\up{I}(M_6)\,.
\end{align}
We have therefore shown that equations \eqref{eq:GS2_GS3_fusions} hold.

\paragraph{Fusion of two GS3 defects.}
\begin{align}
    T_6(M_6)\otimes \ol{T}_6(M_6)&=\lb  N\dw{M_6,\,\Z_2}\rb ^4\sum_{\substack{M_3,N_3,N_3',M_3'\\\in H_3(M_6,\,\Z_2)}}D_3^{(I)}(M_3)\,D_3^{(J)}(N_3)\,D_3^{(J)}(N_3')\,D_3^{(I)}(M_3')=\nn\\
    &=\lb  N\dw{M_6,\,\Z_2}\rb ^4\sum_{\substack{M_3,N_3,N_3',M_3'\\\in H_3(M_6,\,\Z_2)}}(-1)^{\langle N_3,N_3' \rangle+\langle M_3,N_3+N_3'+M_3' \rangle}D_3^{(J)}(N_3+N_3')\,D_3^{(I)}(M_3+M_3')=\nn\\
    &=\lb  N\dw{M_6,\,\Z_2}\rb ^4\sum_{\substack{M_3,N_3,\Tilde{N}_3',\Tilde{M}_3'\\\in H_3(M_6,\,\Z_2)}}(-1)^{\langle N_3+M_3,\Tilde{N}_3'\rangle}D_3^{(J)}(\Tilde{N}_3')\,(-1)^{\langle M_3,\Tilde{M}_3'\rangle}D_3^{(I)}(\Tilde{M}_3')=
    \end{align}
we can redefine $N_3+M_3\to N_3$ and use Poincaré duality to write
\begin{align}
    &=\lb  N\dw{M_6,\,\Z_2}\rb ^2\sum_{\substack{\PD(N_3),\PD(\Tilde{N}_3')\\ \in H^3(M_6,\,\Z_2)}}\exp(i\pi\int_{M_6}\lb \PD(N_3)+c_3\up{J}\rb \cup \PD(\Tilde{N}_3'))\times\nn\\
    &\times\lb  N\dw{M_6,\,\Z_2}\rb ^2\sum_{\substack{\PD(M_3),\PD(\Tilde{M}_3')\\ \in H^3(M_6,\,\Z_2)}}\exp(i\pi\int_{M_6}\lb \PD(M_3)+c_3\up{I}\rb \cup \PD(\Tilde{M}_3'))\,.
\end{align}
For each factor, we now proceed similarly to the fusion of two GS2 defects, i.e. we integrate out $\PD(\Tilde{M}_3')$ and $\PD(\Tilde{N}_3')$: this trivializes the integrands and produces a factor $|H^3(M_6,\,\Z_2)|^2$, which, when combined with $\lb  N\dw{M_6,\,\Z_2}\rb ^4$, produces an Euler counterterm $\chiM{}^{-2}$ (this could be eliminated by redefining the normalization of the GS3 defects) giving:
\begin{align}
     T_6(M_6)\otimes \ol{T}_6(M_6)=\chiM{}^{-2}\,.
\end{align}
The fusion $\ol{T}_6(M_6)\otimes T_6(M_6)$ can be obtained by simply exchanging the labels $I\leftrightarrow J$ in the above computation and yields the same result:
\begin{align}
    \ol{T}_6(M_6)\otimes T_6(M_6)=\chiM{}^{-2}\,.
\end{align}
Let us now compute the fusion of $T_6(M_6)$ with itself:
\begin{align}
    T_6(M_6)\otimes T_6(M_6)&=\lb  N\dw{M_6,\,\Z_2}\rb ^4\sum_{\substack{M_3,N_3,M_3',N_3'\\\in H_3(M_6,\,\Z_2)}}D_3^{(I)}(M_3)\,D_3^{(J)}(N_3)\,D_3^{(I)}(M_3')\,D_3^{(J)}(N_3')= \nn\\
    &=\lb  N\dw{M_6,\,\Z_2}\rb ^4\sum_{\substack{M_3,N_3,M_3',N_3'\\\in H_3(M_6,\,\Z_2)}}(-1)^{\langle M_3+N_3,M_3'\rangle+\langle N_3,N_3'\rangle}D_3^{(I)}(M_3+M_3')\,D_3^{(J)}(N_3+N_3')= \nn\\
    &=\lb  N\dw{M_6,\,\Z_2}\rb ^4\sum_{\substack{M_3,N_3,\Tilde{M}_3',\Tilde{N}_3'\\\in H_3(M_6,\,\Z_2)}}(-1)^{\langle N_3,M_3\rangle+\langle M_3+N_3,\Tilde{M}_3'\rangle+\langle N_3,\Tilde{N}_3'\rangle}D_3^{(I)}(\Tilde{M}_3')\,D_3^{(J)}(\Tilde{N}_3')=
\end{align}
redefining $M_3+N_3\to M_3$ and using Poincaré duality gives:
\begin{align}
    =\lb  N\dw{M_6,\,\Z_2}\rb ^4\sum_{\substack{\PD(M_3),\PD(N_3),\\\PD(\Tilde{M}_3'),\PD(\Tilde{N}_3') \in H^3(M_6,\,\Z_2)}}&\exp(i\pi\int_{M_6}\PD(N_3)\cup\PD(M_3))\times\nn\\
    \times&\exp(i\pi\int_{M_6}\lb \PD(M_3)+c_3\up{I}\rb \cup \PD(\Tilde{M}_3'))\times\nn\\
    \times&\exp(i\pi\int_{M_6}\lb \PD(N_3)+c_3\up{J}\rb \cup \PD(\Tilde{N}_3'))\,.
\end{align}
Integrating out $\PD(\Tilde{M}_3')$ sets $\PD(M_3)=-c_3\up{I}$. Replacing this into the above expression, and using again Poincaré duality, we obtain:
\begin{align}
    T_6(M_6)\otimes T_6(M_6)&=\chiiM{}\,\lb  N\dw{M_6,\,\Z_2}\rb ^2\sum_{N_3,\Tilde{N}_3'\in H_3(M_6,\,\Z_2)}D_3^{(I)}(N_3)\,D_3^{(J)}(\Tilde{N}_3')(-1)^{\langle N_3,\Tilde{N}_3'\rangle}=\nn\\
    &=\chiiM{}\,\lb  N\dw{M_6,\,\Z_2}\rb ^2\sum_{N_3,\Tilde{N}_3'\in H_3(M_6,\,\Z_2)}D_3^{(J)}(\Tilde{N}_3')\,D_3^{(I)}(N_3)
\end{align}
showing that:
\begin{align}
    T_6(M_6)\otimes T_6(M_6)=\chiiM{}\,\ol{T}_6(M_6)\,.
\end{align}
The fusion of $\ol{T}_6(M_6)$ with itself can be obtained by exchanging $I\leftrightarrow J$ in the above computation:
\begin{align}
    \ol{T}_6(M_6)\otimes \ol{T}_6(M_6)=\chiiM{}\,{T}_6(M_6)\,.
\end{align}

\subsection{Triality twist defect fusions} \label{app:twist_fusions}
We will compute the fusion of two triality twist defects \eqref{eq:T_twist}, located at $x=0$ and $x=\eps$ respectively, and then take $\eps\to0$. The results are summarized in equation \eqref{eq:T_twist_fusions}. We first compute the fusions of two different triality twist defects:
\begin{align}
    T_6(M_6\gz,M_{5|0})\otimes \ol{T}_6(M_6\gep,M_{5|\eps})&= \lb N\dw{M_6\gz,\,\Z_2} N\dw{M_6\gep,\,\Z_2}\rb^2\times\nn\\[2mm]
    &\times\sum_{\substack{M_3,N_3\in H_3(M_6\gz,\,\Z_2)\\N_3',M_3'\in H_3(M_6\gep,\,\Z_2)}}D_3^{(I)}(M_3)\,D_3^{(J)}(N_3)\,D_3^{(J)}(N_3')\,D_3^{(I)}(M_3')=\nn\\
    &= \lb N\dw{M_6\gz,\,\Z_2} N\dw{M_6\gep,\,\Z_2}\rb^2\times\nn\\[2mm]
    &\times\sum_{\substack{M_3,N_3\in H_3(M_6\gz,\,\Z_2)\\N_3',M_3'\in H_3(M_6\gep,\,\Z_2)}}(-1)^{\langle N_3,N_3' \rangle+\langle M_3,N_3+N_3'+M_3' \rangle}D_3^{(J)}(N_3+N_3')\,D_3^{(I)}(M_3+M_3')=\nn\\
\end{align}
We now set, on $M_6\gep$, $\Tilde{N}_3'=N_3+N_3'$, $\Tilde{M}_3'=M_3+M_3'$, and split the homologies using $M_6\gz=M_6\ze\cup M_6\gep$:
\begin{align}
    &= \lb N\dw{M_6\gz,\,\Z_2} N\dw{M_6\gep,\,\Z_2}\rb^2\,\lb\sum_{\substack{M_3,N_3\in H_3(M_6\ze,\,\Z_2)}}(-1)^{\langle M_3,N_3\rangle} D_3\up{J}(N_3)\,D_3\up{I}(M_3)\rb\,\times\nn\\[2mm]
    &\times\sum_{\substack{M_3,N_3,\Tilde{N}_3',\Tilde{M}_3'\in H_3(M_6\gep,\,\Z_2)}}(-1)^{\langle N_3+M_3,\Tilde{N}_3'\rangle}D_3^{(J)}(\Tilde{N}_3')\,(-1)^{\langle M_3,\Tilde{M}_3'\rangle}D_3^{(I)}(\Tilde{M}_3')=\nn\\
    &= \lb N\dw{M_6\gz,\,\Z_2} N\dw{M_6\gep,\,\Z_2}\rb^2\,\lb\sum_{\substack{M_3,N_3\in H_3(M_6\ze,\,\Z_2)}} D_3\up{I}(N_3)\,D_3\up{J}(M_3)\rb\,\times\nn\\[2mm]
    &\times\sum_{\substack{M_3,N_3,\Tilde{N}_3',\Tilde{M}_3'\in H_3(M_6\gep,\,\Z_2)}}(-1)^{\langle N_3+M_3,\Tilde{N}_3'\rangle}D_3^{(J)}(\Tilde{N}_3')\,(-1)^{\langle M_3,\Tilde{M}_3'\rangle}D_3^{(I)}(\Tilde{M}_3')=
\end{align}
We write the sum over 3-cycles as a sum over their Lefschetz duals  then integrate out $\LD(\Tilde{M}_3')$ and $\LD(\Tilde{N}_3')$, which reduce the dynamical $\LD(M_3)$ and $\LD(N_3)$ to $M_6\ze$ with Dirichlet boundary conditions on both boundaries.  Using then the isomorphism \eqref{eq:iso_H^3_H^2} and Poincaré duality on $M_5$, we obtain in the $\eps\to0$ limit:
\begin{align}
    T_6(M_6,M_{5})\otimes \ol{T}_6(M_6,M_{5})&=\chiM{\gz}^{-2}\;N\dw{M_5,\,\Z_2}^2\sum_{M_3,N_3\in H_3(M_5,\,\Z_2)}D_3^{(I)}(M_3)\,D_3^{(J)}(N_3)\,.
\end{align}
By exchanging $I\leftrightarrow J$, we similarly have:
\begin{align}
    \ol{T}_6(M_6,M_{5})\otimes {T}_6(M_6,M_{5})&=\chiM{\gz}^{-2}\;N\dw{M_5,\,\Z_2}^2\sum_{M_3,N_3\in H_3(M_5,\,\Z_2)}D_3^{(J)}(M_3)\,D_3^{(I)}(N_3)\,.
\end{align}
We now consider the fusion of a triality twist defect with itself:
\begin{align}
    T_6(M_6\gz,M_{5|0})\otimes {T}_6(M_6\gep,M_{5|\eps})&= \lb N\dw{M_6\gz,\,\Z_2} N\dw{M_6\gep,\,\Z_2}\rb^2\times\nn\\[2mm]
    &\times\sum_{\substack{M_3,N_3\in H_3(M_6\gz,\,\Z_2)\\M_3',N_3'\in H_3(M_6\gep,\,\Z_2)}}\,D_3^{(I)}(M_3)\,D_3^{(J)}(N_3)\,D_3^{(I)}(M_3')\,D_3^{(J)}(N_3')= \nn\\
    &= \lb N\dw{M_6\gz,\,\Z_2} N\dw{M_6\gep,\,\Z_2}\rb^2\times\nn\\[2mm]
    &\times\sum_{\substack{M_3,N_3\in H_3(M_6\gz,\,\Z_2)\\M_3',N_3'\in H_3(M_6\gep,\,\Z_2)}}(-1)^{\langle M_3+N_3,M_3'\rangle+\langle N_3,N_3'\rangle}D_3^{(I)}(M_3+M_3')\,D_3^{(J)}(N_3+N_3')= \nn\\
\end{align}
Like for the previous case, we now set, on $M_6\gep$, $\Tilde{N}_3'=N_3+N_3'$, $\Tilde{M}_3'=M_3+M_3'$, and split the homologies using $M_6\gz=M_6\ze\cup M_6\gep$:
\begin{align}
    &= \lb N\dw{M_6\gz,\,\Z_2} N\dw{M_6\gep,\,\Z_2}\rb^2\,\lb\sum_{\substack{M_3,N_3\in H_3(M_6\ze,\,\Z_2)}} D_3\up{I}(N_3)\,D_3\up{J}(M_3)\rb\,\times\nn\\[2mm]
     &\times\sum_{\substack{M_3,N_3,\Tilde{N}_3',\Tilde{M}_3'\in H_3(M_6\gep,\,\Z_2)}}(-1)^{\langle N_3,M_3\rangle+\langle M_3+N_3,\Tilde{M}_3'\rangle+\langle N_3,\Tilde{N}_3'\rangle}D_3^{(I)}(\Tilde{M}_3')\,D_3^{(J)}(\Tilde{N}_3')=\nn\\
     &= \lb N\dw{M_6\gz,\,\Z_2} N\dw{M_6\gep,\,\Z_2}\rb^2\,\lb\sum_{\substack{M_3,N_3\in H_3(M_6\ze,\,\Z_2)}} D_3\up{I}(N_3)\,D_3\up{J}(M_3)\rb\,\times\nn\\[2mm]
     &\times\sum_{\substack{M_3,N_3,\Tilde{N}_3',\Tilde{M}_3'\in H_3(M_6\gep,\,\Z_2)}}(-1)^{\langle N_3,M_3\rangle+\langle M_3,\Tilde{M}_3'\rangle+\langle N_3,\Tilde{N}_3'\rangle}D_3^{(I)}(\Tilde{M}_3')\,D_3^{(J)}(\Tilde{N}_3')=    
\end{align}
where, in the last step, we re-defined $M_3+N_3\to M_3$ on $M_6\gep$. We write the sum over 3-cycles as a sum over their Lefschetz duals, then integrate out $\LD(\Tilde{M}_3')$, which sets 
$\LD(M_3)|_{M_6\gep}=-c_3\up{I}$, which we then replace in the expression:
\begin{align}
     &= \lb N\dw{M_6\gz,\,\Z_2} N\dw{M_6\gep,\,\Z_2}\rb^2\,\lb\sum_{\substack{M_3,N_3\in H_3(M_6\ze,\,\Z_2)}} D_3\up{I}(N_3)\,D_3\up{J}(M_3)\rb\,\times\nn\\[2mm]
     &\times\sum_{\substack{N_3,\Tilde{N}_3'\in H_3(M_6\gep,\,\Z_2)}}(-1)^{\langle N_3,\Tilde{N}_3'\rangle}D_3^{(I)}(N_3)\,D_3^{(J)}(\Tilde{N}_3')= \nn\\
      &= \lb N\dw{M_6\gz,\,\Z_2} N\dw{M_6\gep,\,\Z_2}\rb^2\,\lb\sum_{\substack{M_3,N_3\in H_3(M_6\ze,\,\Z_2)}} D_3\up{I}(N_3)\,D_3\up{J}(M_3)\rb\,\times\nn\\[2mm]
     &\times\sum_{\substack{N_3,\Tilde{N}_3'\in H_3(M_6\gep,\,\Z_2)}}D_3^{(J)}(\Tilde{N}_3')\,D_3^{(I)}(N_3)=      
\end{align}

After using then the isomorphism \eqref{eq:iso_H^3_H^2} and Poincaré duality on $M_5$, and taking the $\eps\to0$ limit, we obtain:
\begin{align}
    T_6(M_6,M_{5})\otimes {T}_6(M_6,M_{5})&=\lb \chiM{\gz}^{-2}\;N\dw{M_5,\,\Z_2}^2\sum_{M_3,N_3\in H_3(M_5,\,\Z_2)}D_3^{(I)}(M_3)\,D_3^{(J)}(N_3)\rb\,\ol{T}_6(M_6,M_{5}).
\end{align}
Exchanging $I\leftrightarrow J$ gives:
\begin{align}
    \ol{T}_6(M_6,M_{5})\otimes \ol{T}_6(M_6,M_{5})&=\lb \chiM{\gz}^{-2}\;N\dw{M_5,\,\Z_2}^2\sum_{M_3,N_3\in H_3(M_5,\,\Z_2)}D_3^{(J)}(M_3)\,D_3^{(I)}(N_3)\rb\,{T}_6(M_6,M_{5}).
\end{align}
These results are written in a more concise manner in equation \eqref{eq:T_twist_fusions}.

The fusions between duality and triality twist defects can be computed very similarly, following the derivation for duality twist defect fusions in section \ref{sec:duality_twist}, the fusions of GS2 and GS3 defects in appendix \ref{app:GS_SymTFT_fusions} and the above triality twist defect fusions. The results are provided in equation \eqref{eq:D_T_twist_fusions}.

\section{Bosonic SymTFT and Interfaces \label{app:bosint}}
We have learnt that in the $K_{\mathfrak{so}(8)}$ Chern-Simons theory the 3-dimensional surface operators are fermionic. We can start with an absolute theory $A[\mathbb Z_2\up{I}]$, where $\Z_2\up{I}$ indicates the 2-form symmetry of the theory. To describe the SymTFT of this theory the standard procedure entails the gauging in the bulk via the Dijkgraaf-Witten (DW) discrete gauge theory for this symmetry,
\begin{equation} \label{eq:7dDWth}
    S_{\rm 7d \, DW} = \pi \int_{M_7} c_3^{(I)} \cup \delta c_3^{(J)} 
\end{equation}
in this case the matrix $K$ reads, 
\begin{equation}
    K = \begin{pmatrix}
        0 & 2 \\
        2 & 0
    \end{pmatrix}
\end{equation}
and the spins of the operators (which are defined in $d=2k+1$ dimensions with odd $k$)
\begin{equation}
    U_3\up{\lambda_I}(M_3)=\exp(i\int_{M_3}\lambda_I\,c_3\up{I})\,\qquad\quad \lambda_I\in\bD= \mathbb{Z}_2\up{I} \times \mathbb{Z}_2\up{J}
\end{equation}
are determined by \eqref{eq:spin_braiding_CS}, and read
\begin{equation}
    s((1,0))= 0, \qquad s((0,1))=0, \qquad s((1,1))= \frac{1}{2}.
\end{equation}
So we notice that $U^{(I)}, U^{(J)}$ are bosonic whereas $U^{(K)}$ is fermionic. Due to different spins of the $U$ operators, the theory does not have manifest $S_3$ symmetry. In addition it does not contain all possible topological manipulations of the boundary $\cA[\mathbb Z_2^I]$. For instance the topological manipulation, $\tau$, that is stacking with an SPT given by the quadratic refinement $\int_{M_6}\bar{q}(c_3^{I})$ changes the spins of the $U$ operators, while gauging remains a boundary manipulation captured by the DW theory. This is in contrast with the $K_{\mathfrak{so}(8)}$ Chern-Simons where the $U$ defects are all fermionic and the $S_3$ symmetry is manifest. 

On the other hand we can still proceed with the theory \eqref{eq:7dDWth} and construct the $S_3$ bulk topological operators by fusing interfaces. For instance the SPT stacking operation, $\tau$, much like stacking in the 4d Yang-Mills theory, provides interfaces, $\tau, \sigma\tau, \tau\sigma, \tau\sigma\tau $, 
\be
\ba
&  &\quad {\color{red}\tau} \\
& \pi \, c_3^{(I)} \cup \delta c_3^{(J)}  &{\color{red} \quad\bigg|} &\quad &\pi \, c_3^{(K)} \cup \delta c_3^{(J)} \\
& &\quad {\color{red}\sigma\tau} \\
& \pi \, c_3^{(I)} \cup \delta c_3^{(J)}  &{\color{red} \quad\bigg|} &\quad &\pi \, c_3^{(K)} \cup \delta c_3^{(I)} \\
& &\quad {\color{red}\tau\sigma} \\
& \pi \, c_3^{(I)} \cup \delta c_3^{(J)}  &{\color{red} \quad\bigg|} &\quad &\pi \, c_3^{(J)} \cup \delta c_3^{(K)} \\
& &\quad {\color{red}\tau\sigma\tau} \\
& \pi \, c_3^{(I)} \cup \delta c_3^{(J)}  &{\color{red} \quad\bigg|} &\quad &\pi \, c_3^{(I)} \cup \delta c_3^{(K)} 
\ea
\ee

Other interfaces are also provided by the codimension-one condensation defects $D_6^{(I)},T_6, \overline{T}_6$ defined in \eqref{eq:D6I_def}, \eqref{eq:T6_def1} and \eqref{eq:olT6_def1}. This can be argued from the action of the condensation defects on the $U^{(I)}$ surface defects, where we denote by $(I,J,K)$ a cyclic permutation of $(S,C,V)$,
\be
\ba
&  & {\color{blue}D_6^{(I)}} \\
& \pi \, c_3^{(I)} \cup \delta c_3^{(J)}  &{\color{blue} \quad\bigg|} &\quad &\pi \, c_3^{(K)} \cup \delta c_3^{(J)} \\
& &  {\color{blue}D_6^{(J)}} \\
& \pi \, c_3^{(I)} \cup \delta c_3^{(J)}  &{\color{blue} \quad\bigg|} &\quad &\pi \, c_3^{(I)} \cup \delta c_3^{(K)} \\
& & {\color{blue}D_6^{(K)}} \\
& \pi \, c_3^{(I)} \cup \delta c_3^{(J)}  &{\color{blue} \quad\bigg|} &\quad &\pi \, c_3^{(J)} \cup \delta c_3^{(I)} \\
& & {\color{green}T_6} \\
& \pi \, c_3^{(I)} \cup \delta c_3^{(J)}  &{\color{green} \quad\bigg|} &\quad &\pi \, c_3^{(K)} \cup \delta c_3^{(I)} \\
& & {\color{green}\overline{T}_6} \\
& \pi \, c_3^{(I)} \cup \delta c_3^{(J)}  &{\color{green} \quad\bigg|} &\quad &\pi \, c_3^{(J)} \cup \delta c_3^{(K)} 
\ea
\ee
where only $D_6^{(K)}$ provides a $\mathbb{Z}_2$ 0-form symmetry of the 7d DW theory, and therefore we can construct duality defects of the boundary theory as twist defects, which become genuine after we gauge $\mathbb{Z}_2$. The action of these interfaces on gapped boundary conditions follows from the action on the $c_3^{(I)}$, 
\be
\ba
& \tau:  &\; (L_J,\overline{L}_I) \; &\mapsto \; (L_J,\overline{L}_K)\\ 
& \sigma\tau:  &\; (L_J,\overline{L}_I) \; &\mapsto \; (L_I,\overline{L}_K)\\ 
& \tau\sigma:  &\; (L_J,\overline{L}_I) \; &\mapsto \; (L_K,\overline{L}_J)\\ 
& \tau\sigma\tau:  &\; (L_J,\overline{L}_I) \; &\mapsto \; (L_K,\overline{L}_I)\\ 
& D_6^{(I)}:  &\; (L_J,\overline{L}_I) \; &\mapsto \; (L_J,\overline{L}_K)\\ 
& D_6^{(J)}:  &\; (L_J,\overline{L}_I) \; &\mapsto \; (L_K,\overline{L}_I)\\ 
&  D_6^{(K)}:  &\; (L_J,\overline{L}_I) \; &\mapsto \; (L_I,\overline{L}_J)\\ 
& T_6:  &\; (L_J,\overline{L}_I) \;  &\mapsto \; (L_I,\overline{L}_K) \\ 
& \overline T_6:  &\; (L_J,\overline{L}_I) \;&\mapsto \; (L_K,\overline{L}_J)
\ea
\ee

By combining these two types of interfaces we can now construct codimension one bulk symmetry defects for the full $S_3$, that are
\begin{equation}
    D_6^{(K)}\;, \; D_6^{(I)}\tau\;, D_6^{(J)} \tau\sigma \tau\; , T_6 \tau\sigma\; , \overline{T}_6\sigma\tau .
\end{equation}
The non-invertible $S_3$-ality boundary defect operators can now be constructed by considering twist defects ending on the symmetry boundary \eqref{eq:S3boundary}.

\bibliographystyle{ytphys}
\small 
\baselineskip=.94\baselineskip
\let\bbb\bibitem\def\bibitem{\itemsep4pt\bbb}
\bibliography{ref}

\end{document}